\documentclass[namedreferences]{SolarPhysics}
\usepackage[optionalrh]{spr-sola-addons}
\usepackage{graphicx}
\usepackage{url}             
\usepackage{color}
\usepackage{lscape}

\usepackage{solaheader}
\newcommand{\solareceiveddate}{6 May 2012}
\newcommand{\solaaccepteddate}{24 September 2012}
\newcommand{\doi}{012-0149-8}

\renewcommand{\TODAY}{11 October 2012}

\newcommand{\grad}{\mbox{\boldmath$\nabla$}}
\newcommand{\vdot}{{\mathbf{\cdot}}}
\newcommand{\B}{\mathbf{B}}

\newcommand{\bxi}{B_x^{\rm i}}
\newcommand{\byi}{B_y^{\rm i}}
\newcommand{\bzi}{B_z^{\rm i}}
\newcommand{\bxh}{B_x^{\rm h}}
\newcommand{\byh}{B_y^{\rm h}}
\newcommand{\bzh}{B_z^{\rm h}}
\newcommand{\jxh}{J_x^{\rm h}}
\newcommand{\jyh}{J_y^{\rm h}}
\newcommand{\jzh}{J_z^{\rm h}}

\newcommand{\xuh}{x^{\rm h}}
\newcommand{\yuh}{y^{\rm h}}
\newcommand{\zuh}{z^{\rm h}}
\newcommand{\xui}{x^{\rm i}}
\newcommand{\yui}{y^{\rm i}}
\newcommand{\zui}{z^{\rm i}}
\newcommand{\myie}{{\it i.e., }}
\newcommand{\myeg}{{\it e.g., }}
\newcommand{\cbl}{{Paper~I}}
\newcommand{\st}{S} 

\newcommand{\aap}{    {\it Astron. Astrophys.}}

\newcommand{\apj}{    {\it Astrophys. J.}}
\newcommand{\apjl}{    {\it Astrophys. J. Lett.}}
\newcommand{\apjs}{    {\it Astrophys. J. Suppl.}}

\newcommand{\cpc}{    {\it Comput. Phys. Commun.}}

\newcommand{\jcp}{    {\it J. Chem. Phys.}}

\newcommand{\sjna}{  {\it SIAM J. Numeric. Anal.}}
\newcommand{\solphys}{{\it Solar Phys.}}

\begin{document}
\begin{article}
\begin{opening}

\title{Resolving the Azimuthal Ambiguity in Vector Magnetogram Data with the Divergence-Free Condition: the Effects of Noise and Limited Spatial Resolution}

\author{A.D.~\surname{Crouch}}

\runningauthor{A.D.~Crouch}
\runningtitle{Resolving the Azimuthal Ambiguity in Vector Magnetogram Data}

\institute{A.D.~Crouch\\NorthWest Research Associates, 3380 Mitchell Lane, Boulder, CO 80301, USA\\email: \url{ash@cora.nwra.com}}


\begin{abstract}
We investigate how the azimuthal ambiguity in solar vector magnetogram data can be resolved by using the divergence-free property of magnetic fields.
In a previous article, by \citeauthor{2009SoPh..260..271C}~(\textit{Solar Phys.} {\bf 260}, 271, \citeyear{2009SoPh..260..271C}), error-free synthetic data were used to test several methods that each make a different assumption about how the divergence-free property can be used to resolve the ambiguity.
In this paper this testing is continued with an examination of the effects of Poisson photon noise and limited instrumental spatial resolution.
We find that all currently available methods based on the divergence-free property can produce undesirable results when photon noise or unresolved structure are present in the data.
We perform a series of experiments aimed at improving the performance of the global minimisation method, which is the most promising of the methods.
We present a two-step approach that produces reasonable results in tests using synthetic data.
The first step of this approach involves the global minimisation of a combination of the absolute value of the approximation for the divergence and a smoothness constraint, which is designed to minimise the difference between the magnetic field in neighbouring pixels.
In the second step, pixels with measurements known to be strongly affected by photon noise are revisited with a smoothing algorithm that also seeks to minimise the difference between the magnetic field in neighbouring pixels.
\end{abstract}


\end{opening}

\section{Introduction}

Knowledge of solar magnetic structures and solar activity in general is greatly enhanced by measurements of the vector magnetic field in the solar atmosphere.
However, there is an ambiguity of 180$^\circ$ in the direction of the inferred component of the magnetic field perpendicular to the line-of-sight, when it is derived from the linear polarisation of magnetically sensitive spectral lines \cite{1969PhDT.........3H}.
Many algorithms are available to resolve this ambiguity in single-height vector magnetogram data, and while some methods perform better than others (for overviews and comparisons see \opencite{2006SoPh..237..267M}; \opencite{2009SoPh..260...83L}), it is apparent that each method involves at least one major assumption or approximation that may not be valid for solar magnetic fields in general.

A class of methods that can potentially lessen the need for additional assumptions resolve the azimuthal ambiguity by using the divergence-free property of magnetic fields and multiple-height vector magnetogram data (\myeg \opencite{1990AcApS..10..371W}; \opencite{1993A+A...278..279C}; \opencite{1993A+A...279..214L}; \opencite{1999A+A...347.1005B}; \opencite{2007ApJ...654..675L}; \opencite{2008SoPh..247...25C}; \opencite{2009SoPh..260..271C}).
In this context, multiple-height vector magnetogram data include information regarding the variation of the magnetic field in the horizontal heliographic directions and along the line-of-sight direction (the latter of which can be inferred from observations, see, \myeg \opencite{1992ApJ...398..375R}; \opencite{1994A+A...291..622C}; \opencite{1995ApJ...439..474M}; \opencite{1996SoPh..164..169D}; \opencite{1996SoPh..169...79L}; \opencite{1998ApJ...494..453W}, \citeyear{2001ApJ...547.1130W}; \opencite{2000ApJ...530..977S}; \opencite{2002A+A...381..290E}; \opencite{2003SoPh..212..361L}; \opencite{2005ApJ...631L.167S}, \citeyear{2007ApJS..169..439S}).
It is important to emphasise that the horizontal heliographic directions are not orthogonal to the line-of-sight direction, except at disk centre.
The problem of computing the divergence in terms of horizontal heliographic and line-of-sight derivatives, for any position on the solar disk, was addressed by \inlinecite{2008SoPh..247...25C} and \citeauthor{2009SoPh..260..271C}~(\citeyear{2009SoPh..260..271C}, henceforth \cbl{}).

In \cbl{}, using error-free synthetic data with measurements available at two discrete heights, three methods were tested that each make different assumptions about how to use the divergence-free property to resolve the ambiguity.
The synthetic data employed in \cbl{} did not account for noise that is typical of solar observational data.
In this article we use more realistic synthetic data to test how the currently available methods respond to the effects of noise.
Following the procedure of \inlinecite{2009SoPh..260...83L} two types of noise are included in the synthetic data: noise to simulate Poisson photon noise in the observed polarization spectra, and a spatial binning to simulate the effects of limited instrumental spatial resolution.

The outline of this article is as follows.
In Section~\ref{sec_divb} we re-iterate the derivation of the divergence-free condition for any position on the solar disk in terms of observable quantities, as described in \inlinecite{2008SoPh..247...25C} and \cbl{}.
In Section~\ref{sec_synth} we describe the synthetic data and metrics used to test the performance of the ambiguity-resolution algorithms.
In Section~\ref{sec_now} we test the various methods based on the divergence-free property that were examined in \cbl{}.
We find that all currently available methods can produce undesirable results when the data contain either photon noise or unresolved structure.
Nevertheless, we find the same general trend that was found in \cbl{}, in that the global minimisation method is more robust than both \citeauthor{1990AcApS..10..371W}'s (\citeyear{1990AcApS..10..371W}) criterion and the sequential minimisation method \cite{1999A+A...347.1005B}.
Consequently, in Section~\ref{sec_extra} we continue the development of the global minimisation method by implementing and testing several techniques to minimise the effects of noise on the disambiguation results.
In Section~\ref{sec_conc} we draw conclusions.

\section{The Divergence-Free Condition}
\label{sec_divb}

In this section we re-iterate how the divergence of the field can be expressed in terms of observable quantities for any position on the solar disk, as described by \inlinecite{2008SoPh..247...25C} and \cbl{}.
Within a limited field of view we assume that a layer of constant optical depth can be represented by the heliographic plane, which is tangent to the solar surface at the position given by the central meridian angle, the latitude and the $P$- and $B_0$-angles.
On the heliographic plane, the image and heliographic components of the magnetic field are related by

\begin{eqnarray}
\bxh & = & a_{11} \bxi + a_{12} \byi + a_{13} \bzi \, , \nonumber \\
\byh & = & a_{21} \bxi + a_{22} \byi + a_{23} \bzi \, , \label{B_h} \\
\bzh & = & a_{31} \bxi + a_{32} \byi + a_{33} \bzi \, ,  \nonumber
\end{eqnarray}

\noindent
where the coefficients $a_{ij}$ are given by Equation~(1) of \inlinecite{1990SoPh..126...21G},
the superscipts h and i label heliographic and image components, respectively,
$\bzi = B_\|$ is the line-of-sight component of the field, and the image components of the field in the directions perpendicular to the line-of-sight direction are

\begin{equation}
\bxi = B_\perp \cos \xi \, , \qquad \mbox{and} \qquad \byi = B_\perp \sin \xi \, ,
\label{btrans}
\end{equation}

\noindent
where $B_\perp$ is the magnitude of the transverse component of the magnetic field (perpendicular to the line-of-sight) and $\xi$ is the azimuthal angle measured counterclockwise from the $\xui$-axis to the transverse component of the field (where $\xui$ and $\yui$ are coordinates on the plane perpendicular to the line-of-sight).
In solar vector magnetogram data, the azimuthal ambiguity occurs because \(\xi\) can only be inferred within the range $0 \leq \xi < 180^\circ$, using the linear polarisation of magnetically sensitive spectral lines.

When written in heliographic coordinates, the calculation of the divergence requires the derivative of the magnetic field in the direction perpendicular to the heliographic plane, $\partial / \partial \zuh$, which cannot be directly measured away from disk centre. 
However, the derivative of the magnetic field in the line-of-sight direction, $\partial / \partial \zui$, can be measured for any position on the solar disk in principle (\myeg \opencite{1992ApJ...398..375R}; \opencite{1994A+A...291..622C}; \opencite{1995ApJ...439..474M}; \opencite{1996SoPh..164..169D}; \opencite{1996SoPh..169...79L}; \opencite{1998ApJ...494..453W}, \citeyear{2001ApJ...547.1130W}; \opencite{2000ApJ...530..977S}; \opencite{2002A+A...381..290E}; \opencite{2003SoPh..212..361L}; \opencite{2005ApJ...631L.167S}, \citeyear{2007ApJS..169..439S}). 
For more detailed discussion of the various methods for inferring the line-of-sight variation of the magnetic field see \inlinecite{2008SoPh..247...25C} and \cbl{}.
For all of the available methods the magnetic field is inferred at discrete heights (in the line-of-sight direction) and, therefore, the line-of-sight derivatives of the magnetic field must be approximated.
The derivatives  $\partial / \partial \zuh$ and  $\partial / \partial \zui$ are related by

\begin{equation}
\frac{\partial f}{\partial \zui} = a_{13} \frac{\partial f}{ \partial \xuh} + a_{23} \frac{\partial f}{ \partial \yuh} + a_{33} \frac{\partial f}{\partial \zuh} \, ,
\label{dlos}
\end{equation}

\noindent
where $f$ is any differentiable function, and $\xuh$ and $\yuh$ are coordinates on the heliographic plane.
The derivatives $\partial / \partial \xuh$ and $\partial / \partial \yuh$ can be approximated using discrete measurements over the heliographic plane.
We use line-of-sight distance $\zui$, but it should be noted that the magnetic field may be inferred as a function of optical depth $\tau$; in such cases, to determine the correspondence between line-of-sight distance and optical depth, an atmosphere model would need to be employed (\myeg \opencite{1986ApJ...306..284M}; \opencite{1981ApJS...45..635V}; \opencite{1994A+A...291..622C}; \opencite{2007ApJS..169..439S}).

Using Equations~(\ref{B_h}) and (\ref{dlos}),  the divergence of the magnetic field can be expressed in terms of observable quantities (\myie derivatives of the image components of the magnetic field with respect to $\xuh$, $\yuh$ and $\zui$),

\begin{equation}
a_{33} \grad \vdot \B  = D_a + a_{33} \frac{\partial  \bzi}{\partial \zui} \, , 
\label{divb2}
\end{equation}

\noindent
where

\begin{eqnarray}
D_a & = & a_{31} \frac{\partial \bxi}{\partial \zui} + a_{32} \frac{\partial \byi}{\partial \zui}
+ \left( a_{11} a_{33} - a_{13} a_{31} \right) \frac{\partial \bxi}{\partial \xuh}
+ \left( a_{12} a_{33} - a_{13} a_{32} \right) \frac{\partial \byi}{\partial \xuh} \nonumber \\
& + & \left( a_{21} a_{33} - a_{23} a_{31} \right) \frac{\partial \bxi}{\partial \yuh}
+ \left( a_{22} a_{33} - a_{23} a_{32} \right) \frac{\partial \byi}{\partial \yuh} \, , \label{da}
\end{eqnarray}

\noindent
which demonstrates that the calculation of the divergence generally involves the line-of-sight derivatives of all three image components of the magnetic field, except at disk centre where $a_{ij} = \delta_{ij}$ and only \( \partial  \bzi / \partial \zui \) is required (see \inlinecite{2008SoPh..247...25C} for further discussion).

It is important to emphasise that the line-of-sight derivatives of the transverse image components of the field, $\partial \bxi / \partial \zui$ and $\partial \byi / \partial \zui$, depend on the ambiguity resolution at all heights used to approximate these derivatives.
Therefore, the ambiguity resolution at one height can affect the ambiguity resolution at other heights, when the ambiguity is resolved using an algorithm based on the divergence for data positioned away from disk centre.
On the other hand, for data positioned at disk centre and algorithms based on the divergence, each height can be disambiguated independently if the approximation for horizontal heliographic derivatives involves only measurements at one height.

\section{Synthetic Data and Performance Metrics}
\label{sec_synth}

\inlinecite{2009SoPh..260...83L} compared several techniques for disambiguating single-height magnetogram data using synthetic data for which information regarding the variation of the magnetic field out of the heliographic plane was not available.
They constructed two families of synthetic data that included two different types of noise that are expected in solar observational data: noise to simulate Poisson photon noise in the observed polarization spectra, and a spatial binning to simulate the effects of limited instrumental spatial resolution in the directions perpendicular to the line-of-sight.
Methods based on the divergence require magnetic field data from at least two heights in order to approximate the line-of-sight derivatives of the field.
To test the performance of the various algorithms based on the divergence we use the same magnetic field configurations and the same types of noise as were considered by \inlinecite{2009SoPh..260...83L} with magnetic field data available at two heights.
The pixels at each height are aligned in the line-of-sight direction and separated by a line-of-sight distance \( \Delta \zui \).
All of the synthetic datasets used in this article were provided by K.D. Leka and Graham Barnes (2012, private communication). 
In the following paragraphs we present a brief summary of the procedures used to generate the synthetic data.

To simulate the influence of noise on the observed spectra the two families of synthetic data are generated using the same general approach, with the following steps (complete details are provided in \opencite{2009SoPh..260...83L}):
(1) At each pixel at each measurement height an analytic magnetic field is computed (described below).
(2) Unno-Rachkovsky-Stokes profiles are generated using a Milne-Eddington atmosphere, with parameters consistent with the \(\lambda\)630.25~nm Fe {\rm I} line, and subsequently  convolved with an instrumental response function.
(3) The resulting Stokes profiles are modified according to the specifics of the type of noise being tested.
(4) The modified ``noise-added'' Stokes profiles are inverted using the same Milne-Eddington algorithm used in step~(2).

\begin{figure}[ht]
\begin{center}
\begin{tabular}{c@{\hspace{0.005\textwidth}}c}
\includegraphics[width=0.45\textwidth]{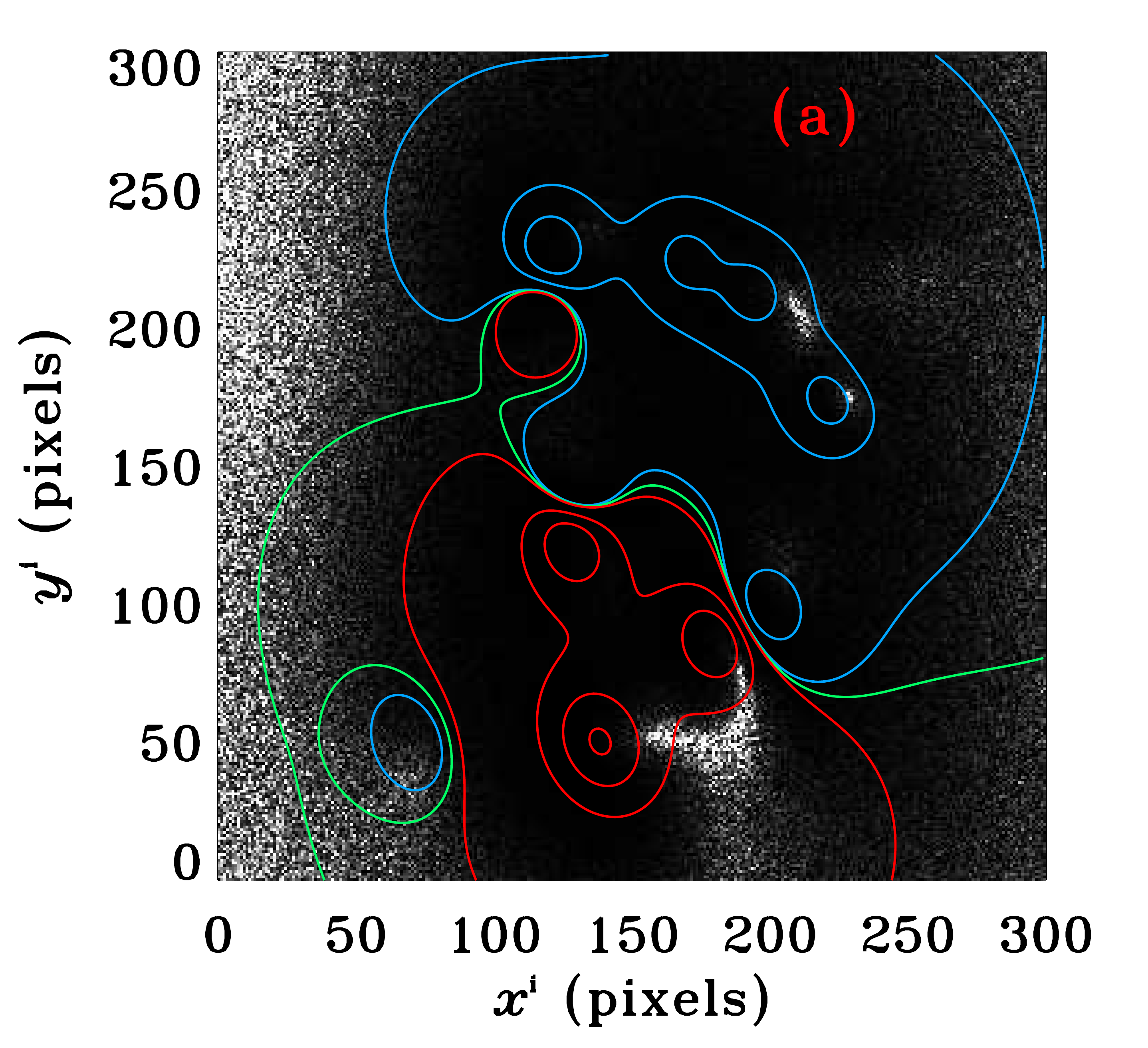} & \includegraphics[width=0.45\textwidth]{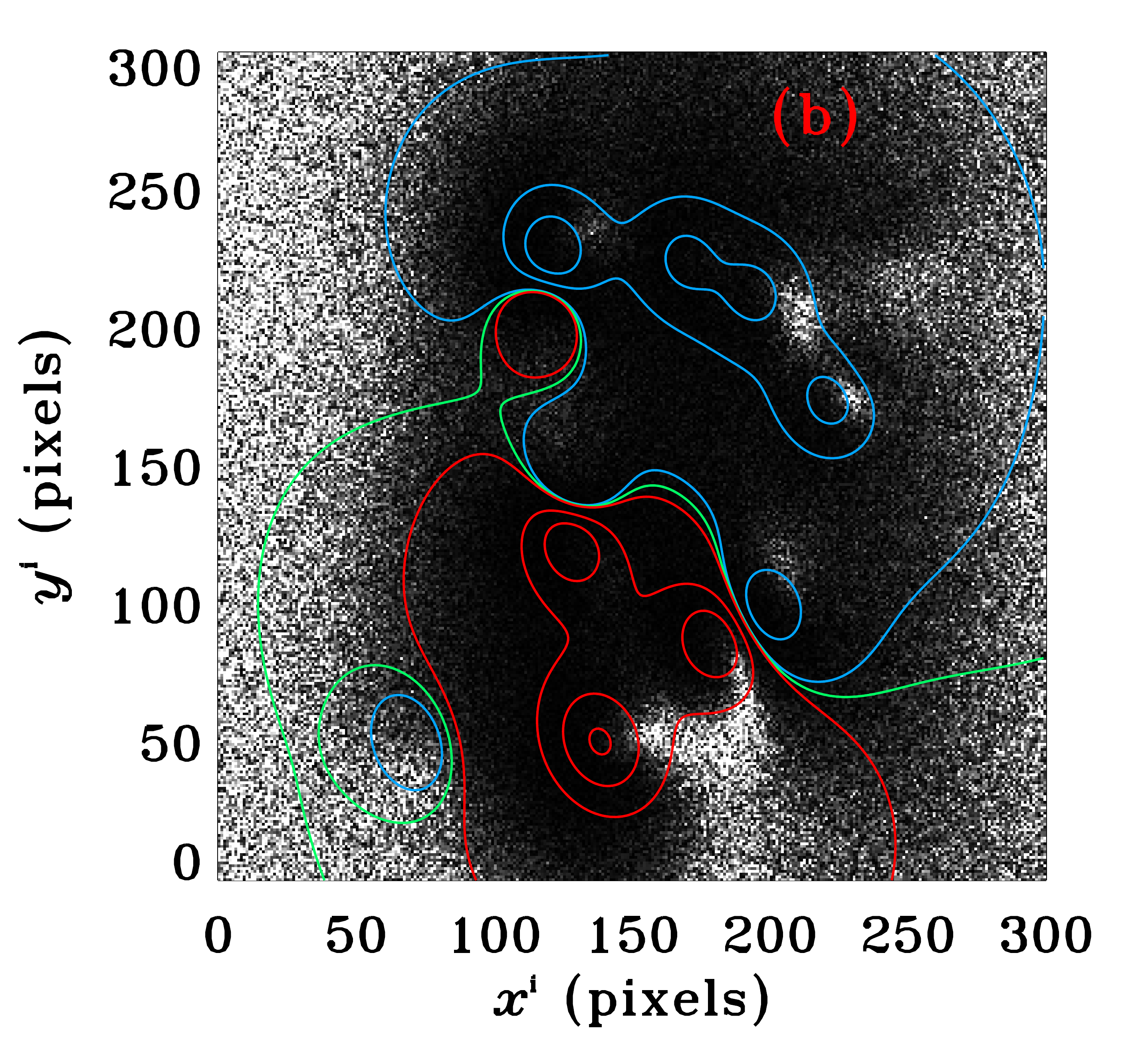} 
\end{tabular}
\end{center}
\caption{
(a) Angular difference between the transverse components of the magnetic field for the no-noise and low-noise cases.
For the low-noise case, the ambiguity is resolved by taking  the realisation of the transverse component of the magnetic field that is closest to the transverse component of the no-noise magnetic field.
The grey-scale image is designed to saturate at \(45^\circ\).
Positive/negative vertical magnetic flux for the no-noise case is indicated by red/blue contours at 50, 700, 1500 and 2800~G, and the corresponding magnetic neutral line is indicated by the green contour.
This plot is similar to that in the right-hand panel of Figure~3 of Leka {\it et al.} (2009).
(b) Same as (a) except for the high-noise case.
}
\label{adiff}
\end{figure}

For the case with photon noise, the magnetic field is constructed from a collection of point sources located on a plane below the heliographic plane.
The magnetic field is linear force-free and the contribution to the field from each source is determined by a Green's function (\myeg \opencite{1977ApJ...212..873C}).
This magnetogram is located away from disk centre, centred at ($9^\circ$S, $36^\circ$E).
To mimic the effect of noise in photon-counting instruments, randomly generated Poisson-distributed noise is incorporated into the Stokes profiles at step~(3).
Three different levels of noise are added: one no-noise case (skipping steps~(2) through (4)) and two noise-added cases to simulate noise levels expected in commonly used instruments.
The separation of the two measurement heights in the line-of-sight direction, \( \Delta \zui \), is equal to the pixel size.
The addition of this photon noise causes the inferred magnetic field to deviate from the original, no-noise, analytic magnetic field.
For the two noise-added cases, the ambiguity-resolved ``answer'' at a given location is taken as the realisation of the transverse component of the noise-added magnetic field that is closest to the transverse component of the original, no-noise, analytic magnetic field.
Figure~\ref{adiff} highlights the effects of this photon noise by showing maps of the angular difference between the ambiguity-resolved, transverse components of the magnetic field for the no-noise and noise-added cases.
Further discussion of this test case can be found in Section~3.1 of \inlinecite{2009SoPh..260...83L}.

For the case of limited spatial resolution, a current-free (potential) magnetic field is constructed that has fine-scale structure with properties similar to both penumbrae and plage.
The model field is computed on a grid with a pixel size of \(0.03''\).
To simulate the effects of imperfect instrumental spatial resolution the Stokes profiles at step~(3) are spatially averaged by factors of 5, 10, and 30, resulting in effective pixel sizes of \(0.15'', 0.3''\), and \(0.9''\), respectively.
Some of the fine-scale features present in the original magnetic field, that are fully spatially resolved on the \(0.03''\) grid, are unresolved at lower spatial resolution.
The separation of the two measurement heights in the line-of-sight direction, \( \Delta \zui \), is \(0.18''\) for all pixel sizes.
In this case, the ambiguity-resolved ``answer'' at a given location is taken as the realisation of the transverse component of the magnetic field that is closest to transverse component of the directly binned (spatially averaged), original, analytic magnetic field.
For some additional discussion of this test case see Section~3.2 of \inlinecite{2009SoPh..260...83L}, and \inlinecite{2012SoPh..277...89L}, \inlinecite{2012SoPh..276..423G} and \inlinecite{2012SoPh..276..441L}.

For each synthetic dataset, the magnetic field is sampled at (exact) discrete spatial locations in the line-of-sight direction. 
Thus, the effects of limited spatial resolution in the line-of-sight direction are not included in these synthetic data.
The separation of the pixels in the line-of-sight direction, \( \Delta \zui \), 
is assumed to be constant over the field of view, although this may not be true for solar observations depending on how the solar atmosphere varies in the directions perpendicular to the line-of-sight.
The synthetic datasets include the effects of curvature; however, the layer where the magnetic field is measured is assumed to be at constant geometrical height, which may also not be accurate for solar observations.
Clearly, the synthetic datasets employed in this article are designed to isolate some, but not all, sources of noise and/or uncertainty that are expected in solar observations.
The sources of noise that are not modelled in these synthetic datasets may be potentially significant for disambiguation methods that rely on line-of-sight derivatives of the magnetic field and, therefore, further investigation is required to quantify these effects.

To quantify the performance of the various algorithms we use the following metrics:
the fraction of pixels correctly disambiguated \( \mathcal{M}_{\rm area} \), 
and the fraction of transverse field above a specified threshold \( \mathcal{T} \) correctly disambiguated \( \mathcal{M}_{B_\perp  > \mathcal{T}} \), for which 
we present results for two thresholds \( \mathcal{T} \): 100~G (gauss) and 500~G \cite{2009SoPh..260...83L}.

\section{The Currently Available Algorithms}
\label{sec_now}

In this section we test the three disambiguation methods that were examined in \cbl{}, with each algorithm implemented exactly as described therein.
The algorithms employ different assumptions and implementation details regarding how to use the divergence-free condition to resolve the azimuthal ambiguity.

\subsection{The Wu and Ai (1990) Criterion}
\label{sec_wuai}

\inlinecite{1990AcApS..10..371W} outlined an approach that was later tested on synthetic data by \inlinecite{1993A+A...278..279C}, \inlinecite{1993A+A...279..214L}, \inlinecite{2007ApJ...654..675L} and \cbl{}.
Following \cbl{}, for any position on the solar disk (except the limb), the divergence-free condition can be written as an inequality, \myie{}

\begin{equation}
a_{33} D_a \frac{\partial \bzi}{\partial \zui} = - \left( a_{33} \frac{\partial \bzi }{\partial \zui} \right)^2 \le 0 \, .
\label{divb1}
\end{equation}

\noindent
In principle, Equation~(\ref{divb1}) can be used to resolve the ambiguity by choosing the realisation of the azimuthal angle such that \( D_a \) satisfies the inequality, given only the sign of \( \partial \bzi / \partial \zui \), not its magnitude.
A couple of limitations of this approach were demonstrated in \cbl{}; these include
(1) the results are sensitive to the smoothness of the initial configuration of azimuthal angles and the order in which pixels are visited; and
(2) the underlying assumption made by this method can be incorrect over a significant fraction of the field of view when the derivatives are approximated by discrete measurements (in that the sign of  \( D_a \) can be  incorrect for the correct choice of azimuthal angle, or the sign of  \( D_a \) can be the same for both the correct and the incorrect choices of azimuthal angle).

We examine the case where the initial configuration of azimuthal angles is such that \( \byi > 0 \).
Results for the noise-added test cases are provided in Figure~\ref{tpd1} and Table~\ref{tpd_tab}, and those for the tests of limited spatial resolution are provided in Figure~\ref{flowers1} and  Table~\ref{flowers_tab}.
In the interest of saving space we only show results for the lower height of each magnetogram; however, it is worth noting that we find the same trend found in \cbl{}, where the results retrieved by the \citeauthor{1990AcApS..10..371W} criterion at the upper height tend to be worse than those at the lower height for data positioned away from disk centre (this is because the upper height is disambiguated after the lower height).

For the multipole field configuration without noise, the results for the \citeauthor{1990AcApS..10..371W} criterion (see Figure~\ref{tpd1} and Table~\ref{tpd_tab}) are similar to those found in \cbl{}. 
We find  that the assumption made by this method is violated over a significant fraction of the field of view (results not shown, but see Figures~4 and 5 of \cbl{} for demonstrations of this issue).
Since the underlying assumption is not tested during the disambiguation procedure this method appears to do quite well for the case without photon noise.
However, as the level of photon noise increases the results produced by this method deteriorate significantly.

Regarding the tests of limited spatial resolution (Figure~\ref{flowers1} and  Table~\ref{flowers_tab}), for the higher resolution cases the solutions retrieved by the \citeauthor{1990AcApS..10..371W} criterion  appear to be reasonable; however, we again find that the assumption made by this method is violated over a significant fraction of the field of view (for all pixel sizes).
For the cases with pixel sizes \(  0.15'' \) and \( 0.3'' \) the  quality of the solutions decreases gradually as the spatial resolution degrades. 
On the other hand, the solution retrieved for the lowest resolution case (pixel size \( 0.9'' \)) is noticeably worse than the other cases.

\subsection{The Sequential Minimisation Method}

The sequential minimisation method resolves the azimuthal ambiguity by assuming that the approximation for the magnitude of the divergence at each pixel is minimised for the set of pixels used in the approximation.
This assumption is applied sequentially at each pixel in the field of view.
This method was first proposed by \inlinecite{1999A+A...347.1005B} and was examined in \cbl{}.
For the test cases examined in \cbl{} it was shown that the sequential minimisation method is more robust than the \citeauthor{1990AcApS..10..371W} criterion but it can produce errors when the underlying assumption is violated.
The underlying assumption is that the correct choice of azimuthal angle will be retrieved for the set of pixels used to approximate the divergence, provided that any previously resolved pixels have the correct choice of azimuthal angle.
Moreover, in \cbl{} it was demonstrated that the sequential minimisation method can propagate erroneous solutions into regions that do not violate the underlying assumption.
This method is not sensitive to the initial configuration of azimuthal angles although it is sensitive to the order in which pixels are examined.

Results for the sequential minimisation method are provided in Tables~\ref{tpd_tab} and \ref{flowers_tab} and Figures~\ref{tpd1} and \ref{flowers1}.
For the multipole field configuration without noise the results are worse than the similar case examined in \cbl{}. This is primarily due to the propagation of erroneous solutions.
As the degree of photon noise increases many more pixels violate the underlying assumption (not shown but see Figures~6(b) and 7(c) and 7(d) of \cbl{} for a demonstration of the issue). 
Consequently, the solutions retrieved by this method get progressively worse as the level of noise increases.
As in \cbl{}, for data positioned away from disk centre we find that the results retrieved by this method at the upper height (not shown) are very similar to those found at the lower height.
For the tests of limited spatial resolution (Figure~\ref{flowers1} and  Table~\ref{flowers_tab}) we find again that the underlying assumption made by this method is violated for a small fraction of pixels (for all pixel sizes), and propagation of erroneous solutions accounts for a large fraction of the incorrect results retrieved by this method.

\subsection{The Global Minimisation Method}
\label{sec_global}

The global minimisation method resolves the azimuthal ambiguity by assuming that the correct configuration of azimuthal angles corresponds to the minimum of

\begin{equation}
E = \sum_{i=1}^{n_x}  \sum_{j=1}^{n_y} \left( | ( \grad \vdot \B )_{i,j,k} | +  | ( \grad \vdot \B )_{i,j,{k+1}} | \right)
\label{eglobal}
\end{equation}

\noindent
where \( ( \grad \vdot \B )_{i,j,k} \) is the approximation for the divergence at pixel $(i, j, k)$. 
The indices $i$, $j$ and $k$ label pixels in the $\xui$-, $\yui$- and $\zui$-directions, respectively, and $n_x$ and $n_y$ are the number of pixels in the $\xui$- and $\yui$-directions, respectively.
Simulated annealing (\myeg \opencite{1953JChPh..21.1087M}; \opencite{1983Sci...220..671K}; \opencite{1992nrfa.book.....P}) is used to search for the configuration of azimuthal angles that  corresponds to the global minimum of $E$.
For the error-free synthetic data used in \cbl{} it was demonstrated that the global minimisation method generally produced better results than both the \citeauthor{1990AcApS..10..371W} criterion and the sequential minimisation method.
However, it was shown in \cbl{} that the assumption made by the global minimisation method (that the global minimum of \(E\) does indeed correspond to the correct configuration of azimuthal angles) can be violated when the divergence is approximated from discrete measurements, depending on how the magnetic field varies and how it is sampled by the observations.

The simulated annealing algorithm is implemented as described in \cbl{} with the following modifications.
Here, we use a ``temperature'' that can be position--dependent.
By using a temperature that is local this modification improves the efficiency of the simulated annealing algorithm in comparison to one with a uniform temperature.
Moreover, an algorithm with a position-dependent temperature tends to retrieve solutions with lower values for \(E\) (when noise is present) than an algorithm with a uniform temperature (all other things equal).
Starting with the initial configuration of azimuthal angles we examine \( 1000 n_x n_y \) reconfigurations of randomly chosen pixels at each height that is disambiguated (which are all accepted) and take two times the maximum value of \( | \Delta E | \) in a small neighbourhood as the initial temperature at each pixel, where \( \Delta E \) is the change in \(E\) (Equation~(\ref{eglobal})) caused by changing the azimuthal angle.

Because simulated annealing is a stochastic algorithm, we disambiguate each synthetic magnetogram 20 times with different sequences of random numbers for each disambiguation.
We select the solution that produces the lowest value of \(E\) for further examination; results are provided in Figures~\ref{tpd1} and \ref{flowers1} and Tables~\ref{tpd_tab} and \ref{flowers_tab}.

For the multipole field positioned away from disk centre without photon noise, the global minimisation method retrieves the correct solution at every pixel over both heights, consistent with the findings of \cbl{} (for cases with comparable values for $\Delta \zui$).
As the level of photon noise increases the quality of the results produced by the global minimisation method decreases substantially (Figure~\ref{tpd1} and Table~\ref{tpd_tab}).
As found in \cbl{}, areas with the incorrect ambiguity resolution tend to occur at similar locations at both heights for data positioned away from disk centre.
Photon noise tends to have the greatest influence on the results retrieved by the global minimisation method in regions where the component of the magnetic field transverse to the line-of-sight is weak \cite{2009SoPh..260...83L}: Note the trend in \( \mathcal{M}_{B_\perp  > 100~{\rm G}} \) for the different noise levels compared to that in \( \mathcal{M}_{B_\perp  > 500~{\rm G}} \) and see Figure~\ref{adiff}.
For both noise-added cases, $E$ for the solution retrieved by the global minimisation algorithm is less than that for the answer: For the low-noise  case  the ratio of $E$ for the retrieved solution to that for the answer is \(0.94\),  whereas for the high-noise case the ratio is \(0.84\).
This demonstrates that the assumption made by the global minimisation method is violated for these test cases.
Moreover, the results produced by the global minimisation method, as implemented here, are relatively poor when compared with the results produced by the better performing single-height algorithms tested in \inlinecite{2009SoPh..260...83L}.

\begin{figure}[ht]
\begin{center}
\includegraphics[width=0.5\textwidth]{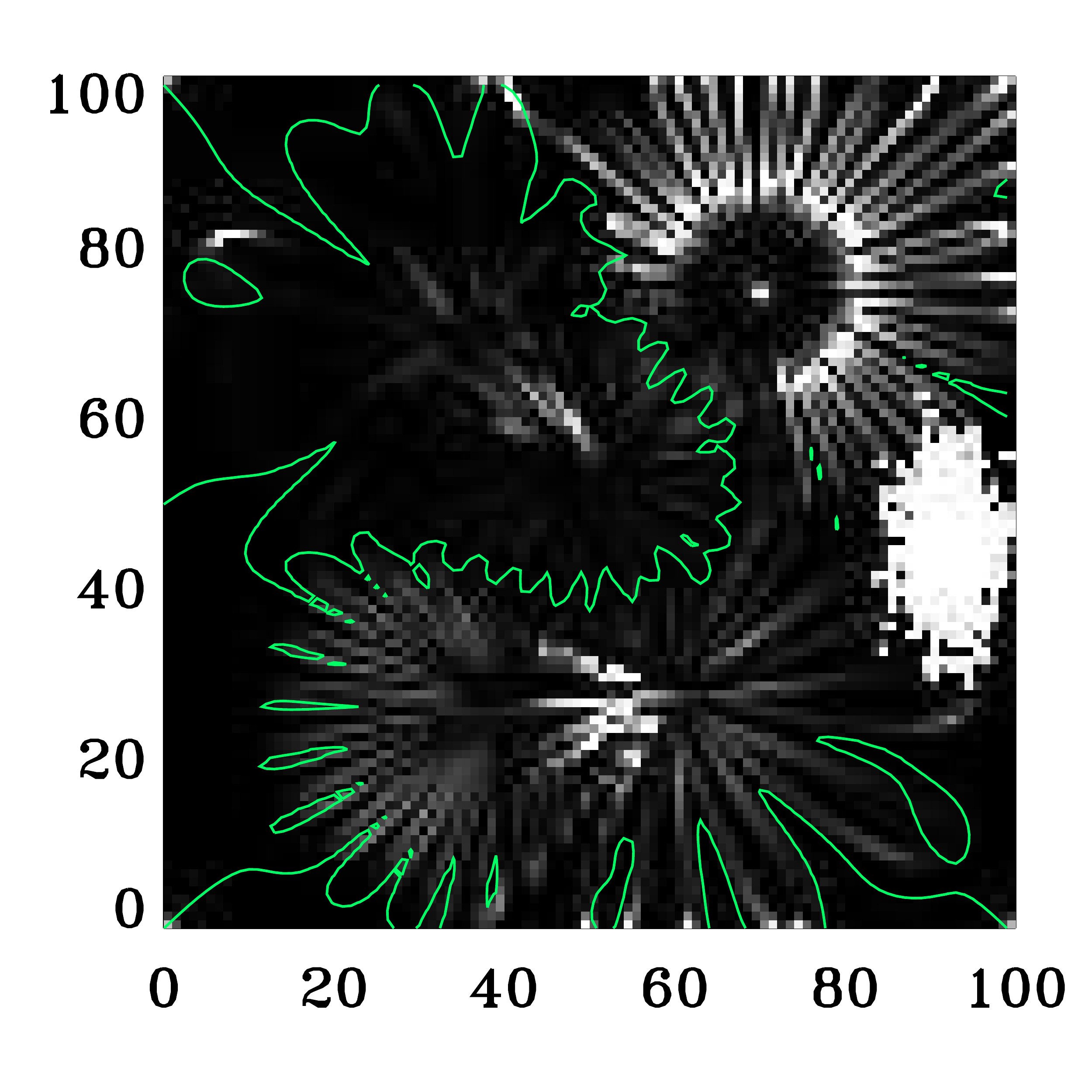}
\end{center}
\caption{An estimate of the degree of unresolved structure in each \(0.9''\) pixel as a function of \(\xuh\) and \(\yuh\).
For each \(0.9''\) pixel we examine the underlying  original model magnetic field projected onto the image plane ( \( \bxi \), \( \byi \) ), as sampled on the \( 0.03'' \) grid.
Shown is the cosine of the largest angular separation between any two  vectors ( \( \bxi \), \( \byi \) ) that contribute to each \(0.9''\) pixel.
Light/dark shaded pixels correspond to areas with a large/small range of angles; the range shown is \( [-1,1] \). 
Pixels with large values of this quantity are a reasonable predictor of the locations where the global minimisation method fails to retrieve the correct azimuthal angle.
The magnetic neutral line is indicated by the green contour.
}
\label{fbin}
\end{figure}

For the tests of limited spatial resolution (Figure~\ref{flowers1} and  Table~\ref{flowers_tab}) we again find that the assumption made by the global minimisation method is violated, in that $E$ for the  retrieved solutions  is less than that for the answers.
The ratios of $E$ for the retrieved solutions to those for the answers are 0.89, 0.94 and 0.98 for pixel sizes of \( 0.15'', 0.3''\) and  \(0.9''\), respectively.
Despite this, for the higher spatial resolution cases the solutions retrieved by the global minimisation method are quite reasonable for most of the field of view.
The exception to this is the plage region (centre, right of the field of view), which evidently poses a challenge for this method even at the higher spatial resolution.
The plage region consists of the collection of randomly distributed small-scale azimuth centres with a significant horizontal magnetic field. The strength of the vertical component of the magnetic field of each small-scale structure is proportional to \(\exp ( -r^2/a^2 ) \), where \(a\) is \(0.156''\) and \(r\) is the distance from the azimuth centre \cite{2009SoPh..260...83L}.
Thus, the small-scale structures in the plage region are spatially resolved on the \(0.03''\) grid, but are spatially unresolved at lower spatial resolution.

For the lowest resolution case (with pixel size \(0.9''\)) incorrect solutions are retrieved by the global minimisation method in and around regions with significant unresolved structure (see Figure~\ref{fbin}).
These regions include the plage region, the area between the two flux concentrations in the bottom, centre of the field of view, and around the flux concentration at the top, right of the field of view (where there is a ring of small-scale azimuth centres that are spatially unresolved at lower spatial resolution, see \myeg Figures~5 and 7 of \inlinecite{2009SoPh..260...83L}). 
In these locations disambiguation methods are not expected to retrieve good solutions as the magnetic field vectors underlying each \(0.9''\) pixel have a large range of angular differences (see Figure~\ref{fbin}) and this information is lost at low spatial resolution.

\begin{landscape}
\begin{table}
\caption{Performance metrics for the noise-added cases for the various ambiguity-resolution algorithms. The lower height is evaluated.}
\label{tpd_tab}
\begin{tabular}{llll|lll|lll}
\hline
& \multicolumn{3}{c}{\( \mathcal{M}_{\rm area} \)}
& \multicolumn{3}{c}{\( \mathcal{M}_{B_\perp  > 100~\rm{G}} \)}
& \multicolumn{3}{c}{\( \mathcal{M}_{B_\perp  > 500~\rm{G}} \)}
\\
Noise level: & None & Low & High & None & Low & High & None & Low & High \\ 
\hline
\citeauthor{1990AcApS..10..371W}                           & 0.91  & 0.61 & 0.54 & 0.95 & 0.71 & 0.57 & 0.97 & 0.80 & 0.61 \\ 
Sequential minimisation                         & 0.55 & 0.57 & 0.52 & 0.54 & 0.56 & 0.52 & 0.55 & 0.54 & 0.53 \\ 
Global minimisation, \( | \grad \vdot \B | \) (Equation~(\ref{eglobal}))        & 1.00 & 0.91 & 0.78 & 1.00 & 0.98 & 0.90 & 1.00 & 1.00 & 1.00 \\ 
Global minimisation, \( | \grad \vdot \B | \) (Equation~(\ref{eglobal})) with smoothed data        & 1.00 & 0.92 & 0.83 & 1.00 & 0.98 & 0.91 & 1.00 & 1.00 & 0.97 \\ 
Global minimisation, \( | \grad \vdot \B | + |J_z| \) (Equation~(\ref{eds2}))        & 1.00 & 0.97 & 0.88 & 1.00 & 1.00 & 0.96 & 1.00 & 1.00 & 1.00 \\ 
Global minimisation, \( | \grad \vdot \B | + |J| \) (Equation~(\ref{eds6}))        & 1.00 & 0.96 & 0.87 & 1.00 & 0.99 & 0.95 & 1.00 & 1.00 & 1.00 \\ 
Global minimisation, \( | \grad \vdot \B | + \st \) (Equation~(\ref{eds3}))        & 1.00 & 0.99 & 0.92 & 1.00 & 1.00 & 0.97 & 1.00 & 1.00 & 1.00 \\ 
Hybrid                                                                             & 1.00 & 1.00 & 0.99 & 1.00 & 1.00 & 1.00 & 1.00 & 1.00 & 1.00 \\ 
\hline
\end{tabular}
\end{table}
\begin{table}
\caption{Performance metrics for the limited resolution cases for the various ambiguity-resolution algorithms. The lower height is evaluated.}
\label{flowers_tab}
\begin{tabular}{llll|lll|lll}
\hline
& \multicolumn{3}{c}{\( \mathcal{M}_{\rm area} \)} 
& \multicolumn{3}{c}{\( \mathcal{M}_{B_\perp  > 100~\rm{G}} \)} 
& \multicolumn{3}{c}{\( \mathcal{M}_{B_\perp  > 500~\rm{G}} \)}
\\
Pixel size: &   $0.15''$ & $0.3''$ & $0.9''$ 
            &   $0.15''$ & $0.3''$ & $0.9''$ 
            &   $0.15''$ & $0.3''$ & $0.9''$ \\
\hline
\citeauthor{1990AcApS..10..371W}                                                              &     0.95 & 0.92 & 0.81 &   0.95 & 0.92 & 0.79  &   0.95 & 0.92 & 0.79 \\ 
Sequential minimisation                                                                       &     0.90 & 0.88 & 0.80 &   0.89 & 0.87 & 0.77  &   0.89 & 0.87 & 0.77 \\ 
Global minimisation, \( | \grad \vdot \B | \) (Equation~(\ref{eglobal}))                      &     0.99 & 0.99 & 0.97 &   1.00 & 1.00 & 0.99  &   1.00 & 1.00 & 0.99 \\ 
Global minimisation, \( | \grad \vdot \B | \) (Equation~(\ref{eglobal})) with smoothed data   &     0.99 & 0.98 & 0.94 &   0.99 & 0.99 & 0.96  &   1.00 & 0.99 & 0.97 \\ 
Global minimisation, \( | \grad \vdot \B | + |J_z| \) (Equation~(\ref{eds2}))                 &     0.99 & 0.99 & 0.99 &   1.00 & 1.00 & 1.00  &   1.00 & 1.00 & 1.00 \\ 
Global minimisation, \( | \grad \vdot \B | + |J| \) (Equation~(\ref{eds6}))                   &     1.00 & 0.99 & 0.99 &   1.00 & 1.00 & 1.00  &   1.00 & 1.00 & 1.00 \\ 
Global minimisation, \( | \grad \vdot \B | + \st \) (Equation~(\ref{eds3}))                   &     0.99 & 0.99 & 0.99 &   1.00 & 1.00 & 1.00  &   1.00 & 1.00 & 1.00 \\ 
Hybrid                                                                                        &     0.99 & 0.99 & 0.99 &   1.00 & 1.00 & 1.00  &   1.00 & 1.00 & 1.00 \\ 
\hline
\end{tabular}
\end{table}
\end{landscape}

\begin{figure}[ht]
\begin{center}
\begin{tabular}{c@{\hspace{0.005\textwidth}}c@{\hspace{0.005\textwidth}}c}
\includegraphics[width=0.3\textwidth]{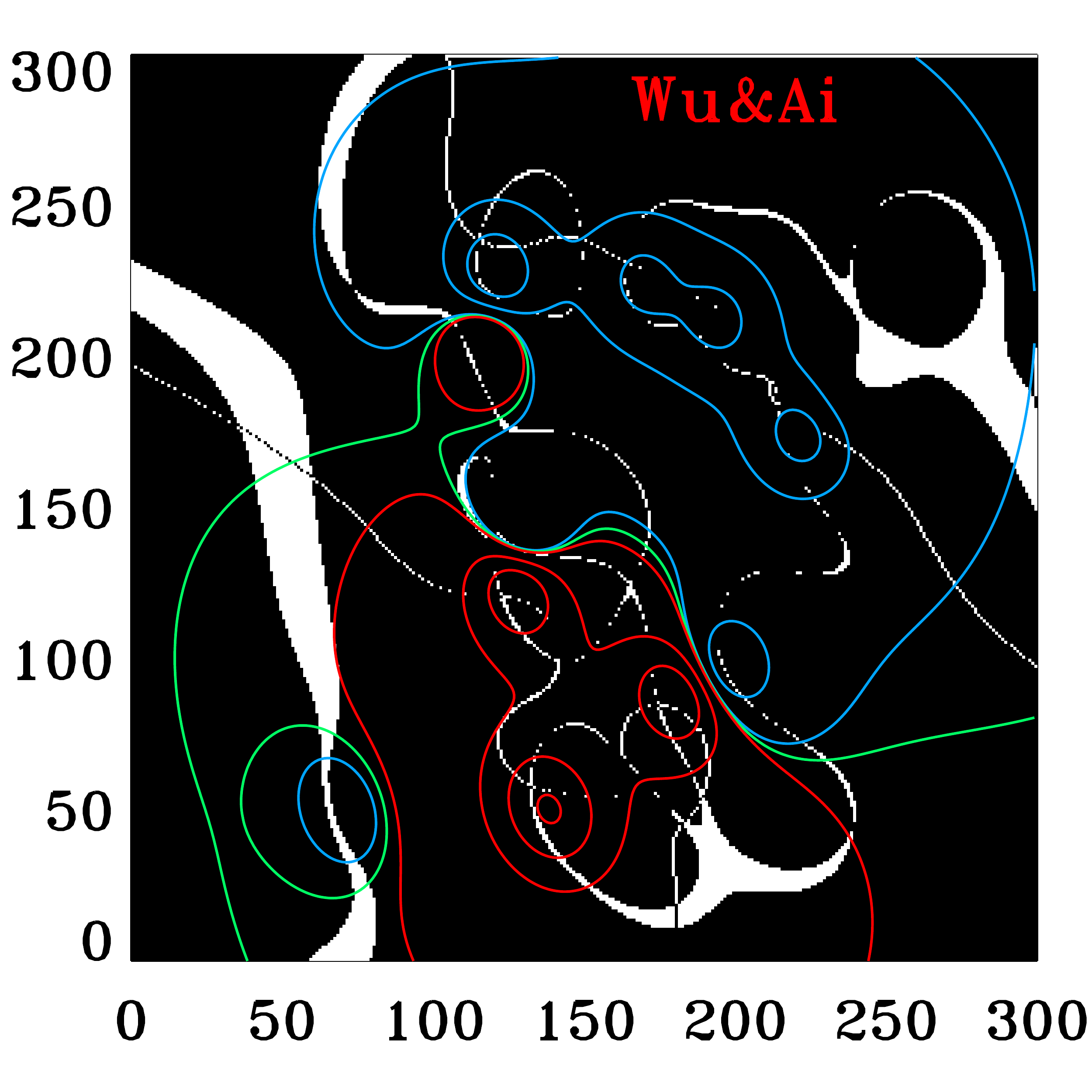} &
\includegraphics[width=0.3\textwidth]{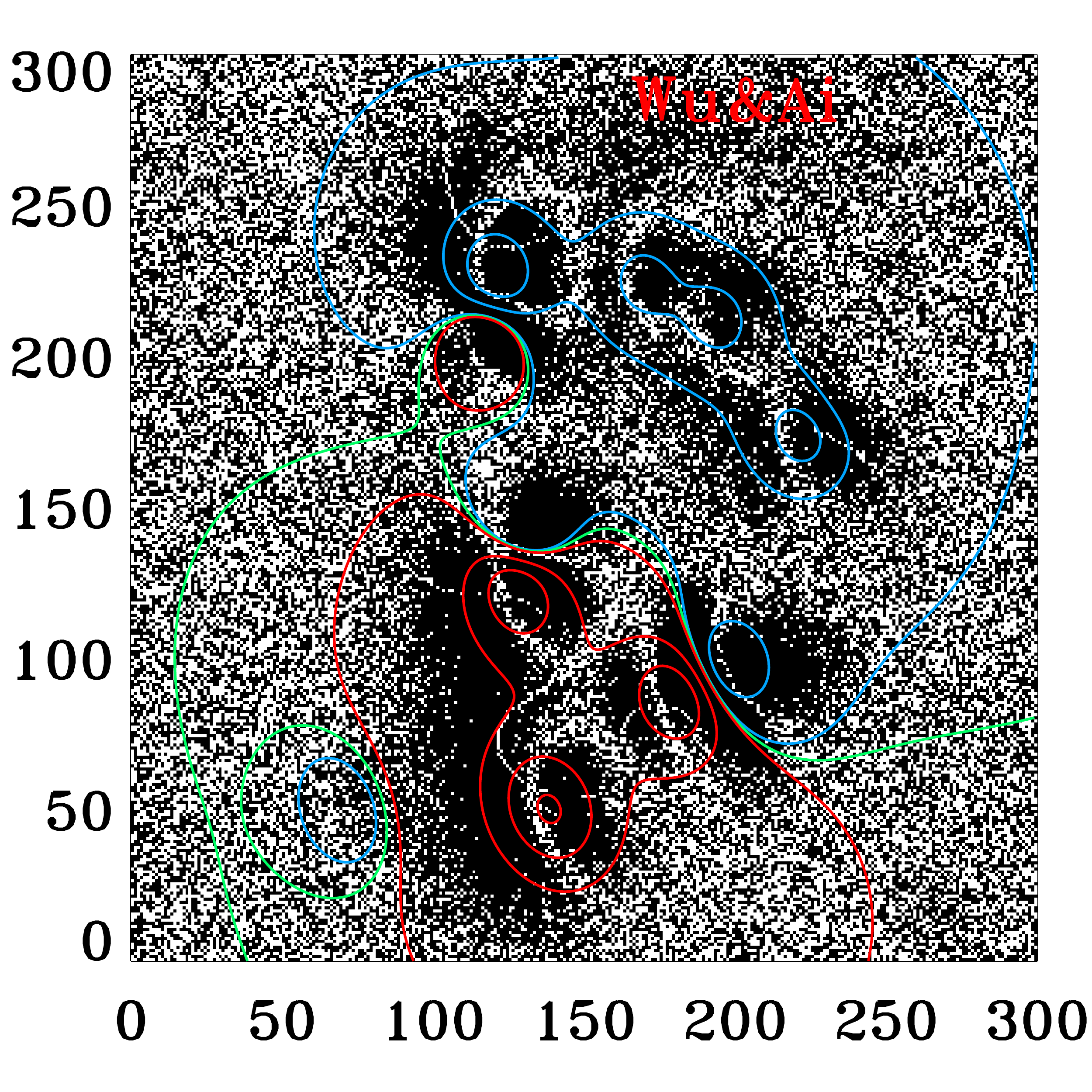} &
\includegraphics[width=0.3\textwidth]{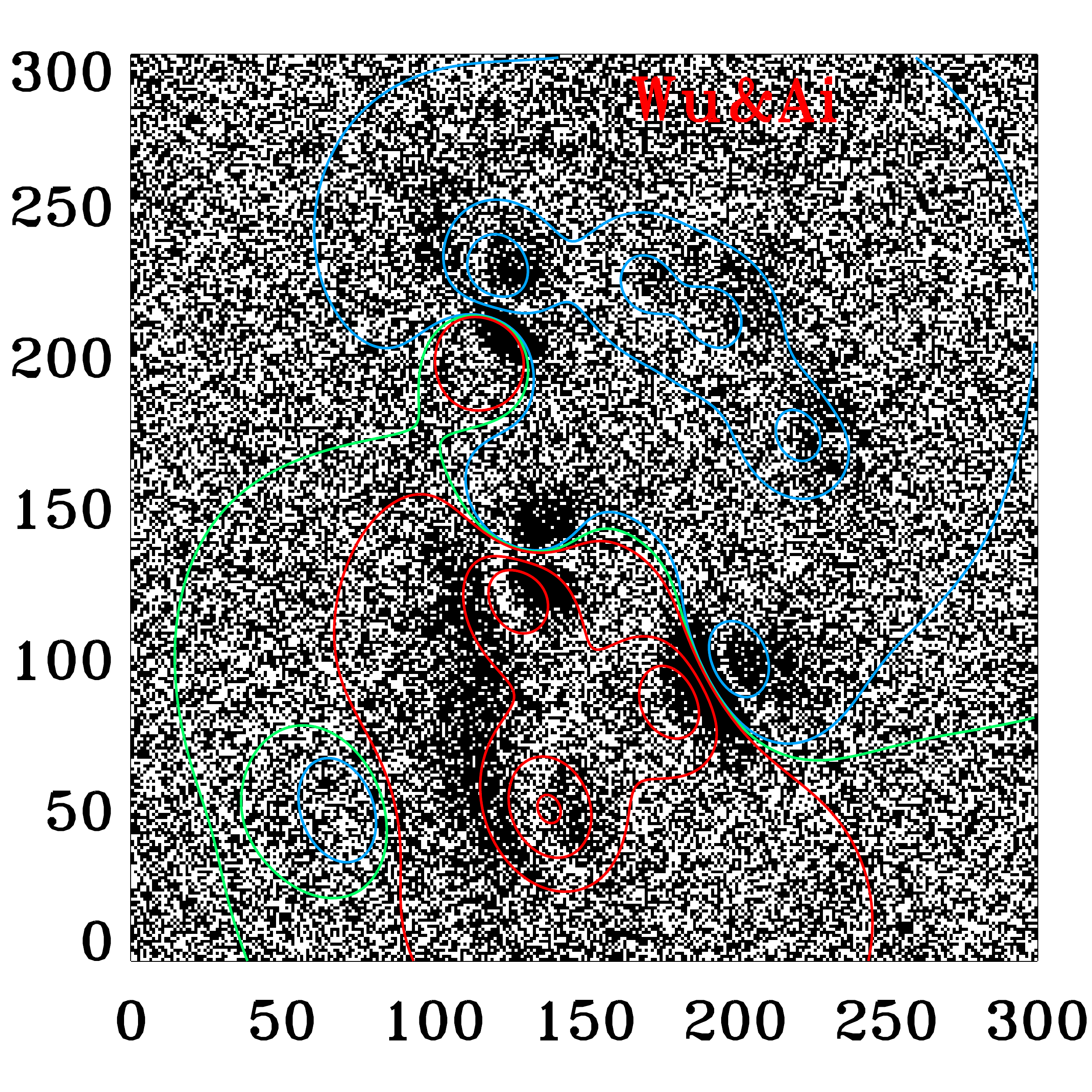} \\[-0.01\textwidth]
\includegraphics[width=0.3\textwidth]{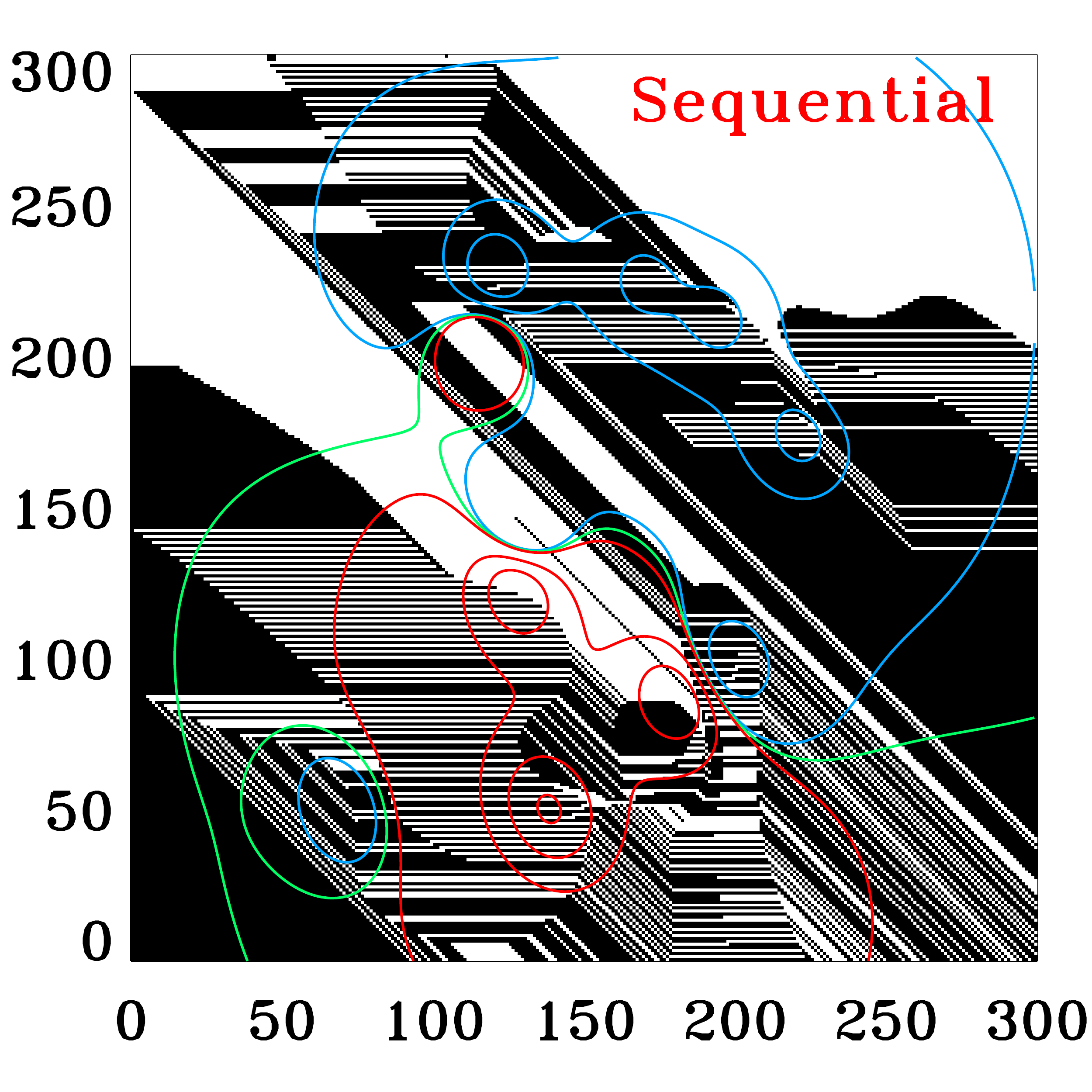} &
\includegraphics[width=0.3\textwidth]{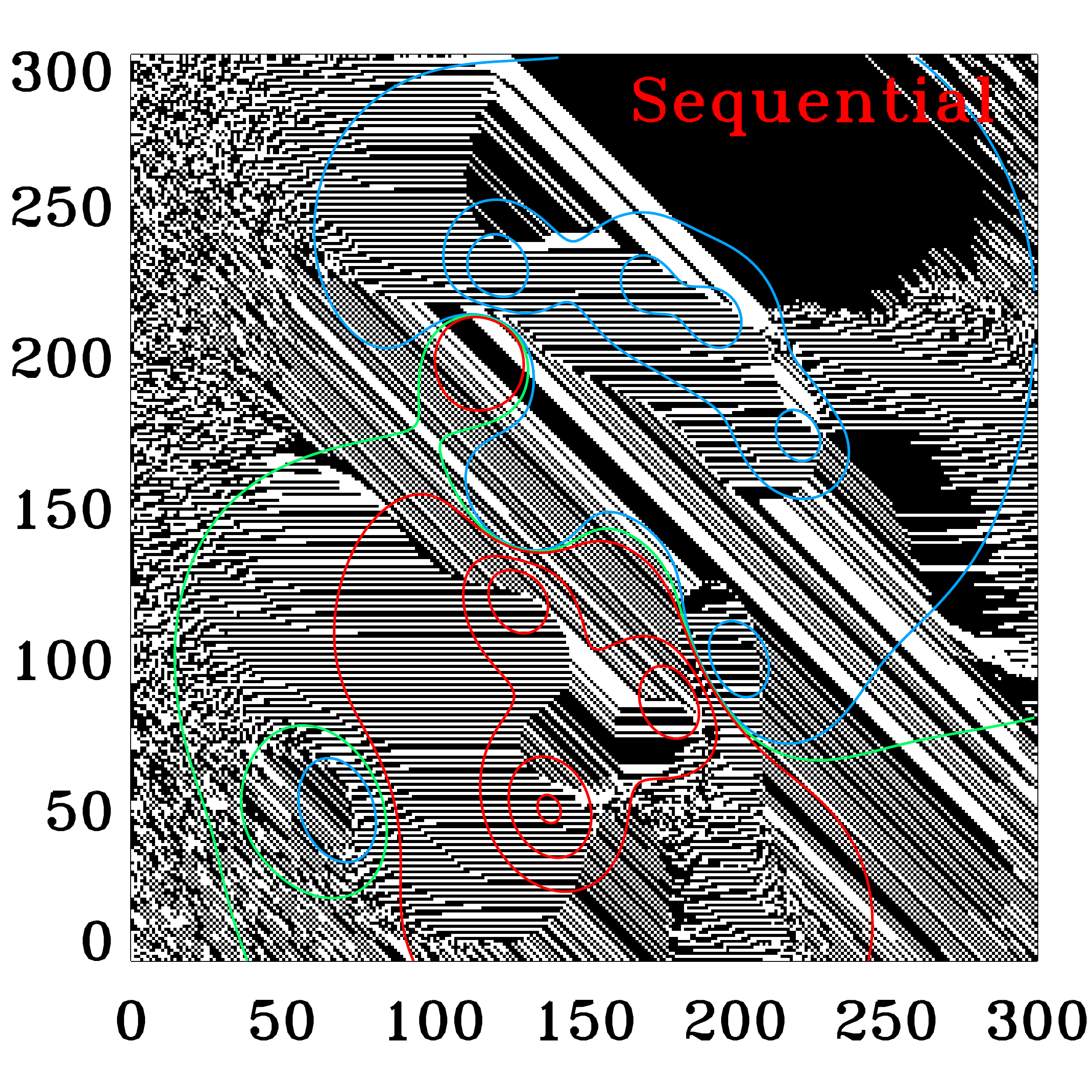} &
\includegraphics[width=0.3\textwidth]{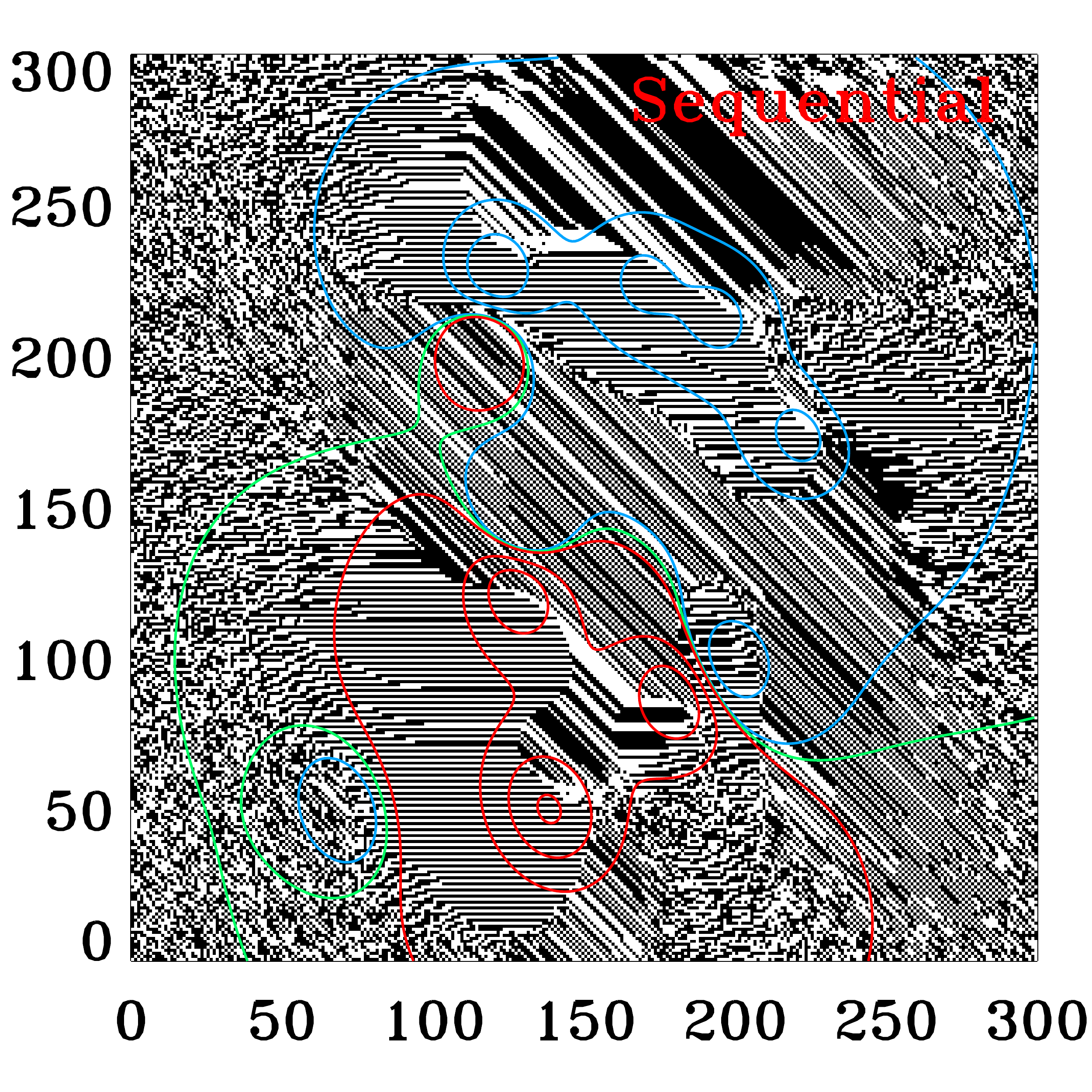} \\[-0.01\textwidth]
\includegraphics[width=0.3\textwidth]{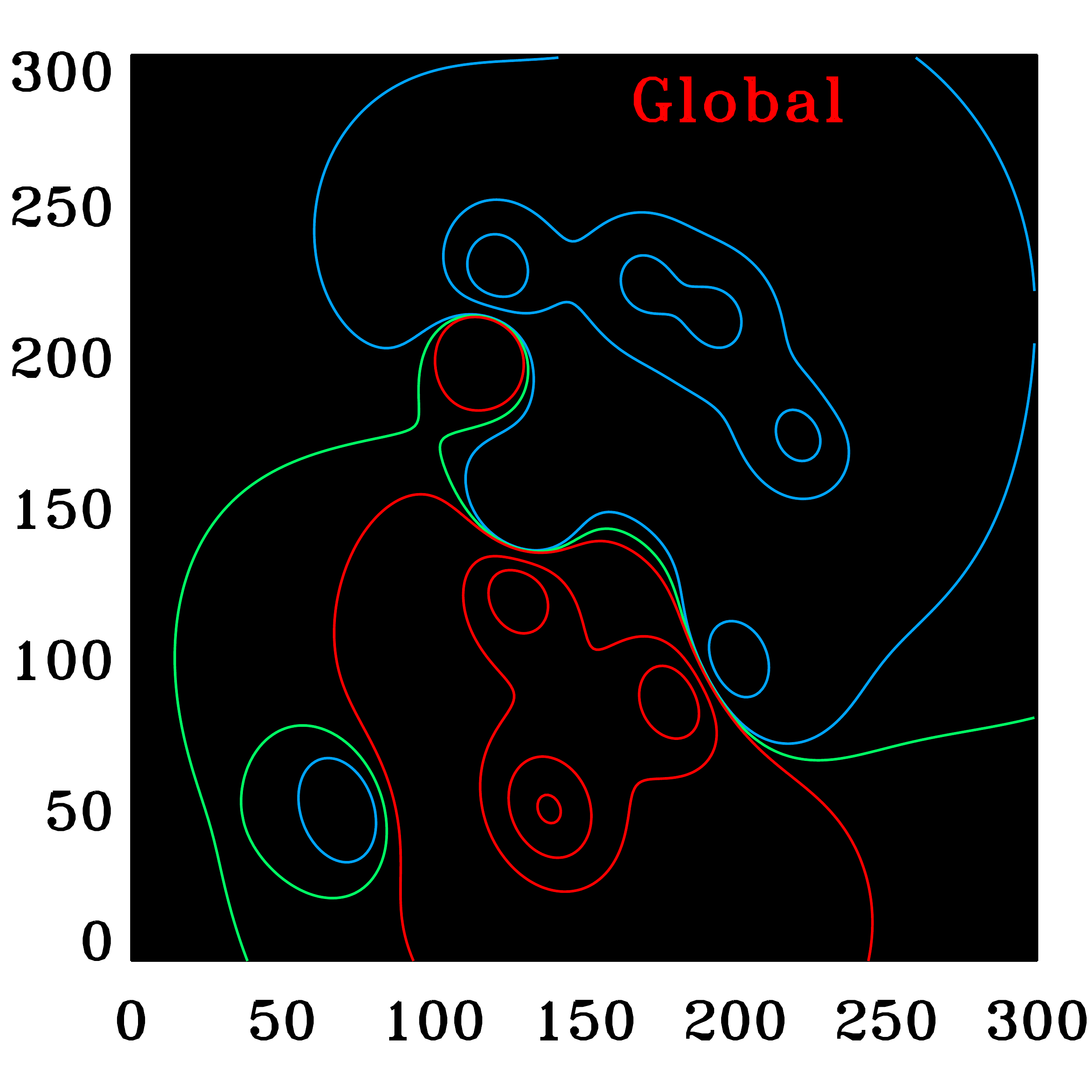} &
\includegraphics[width=0.3\textwidth]{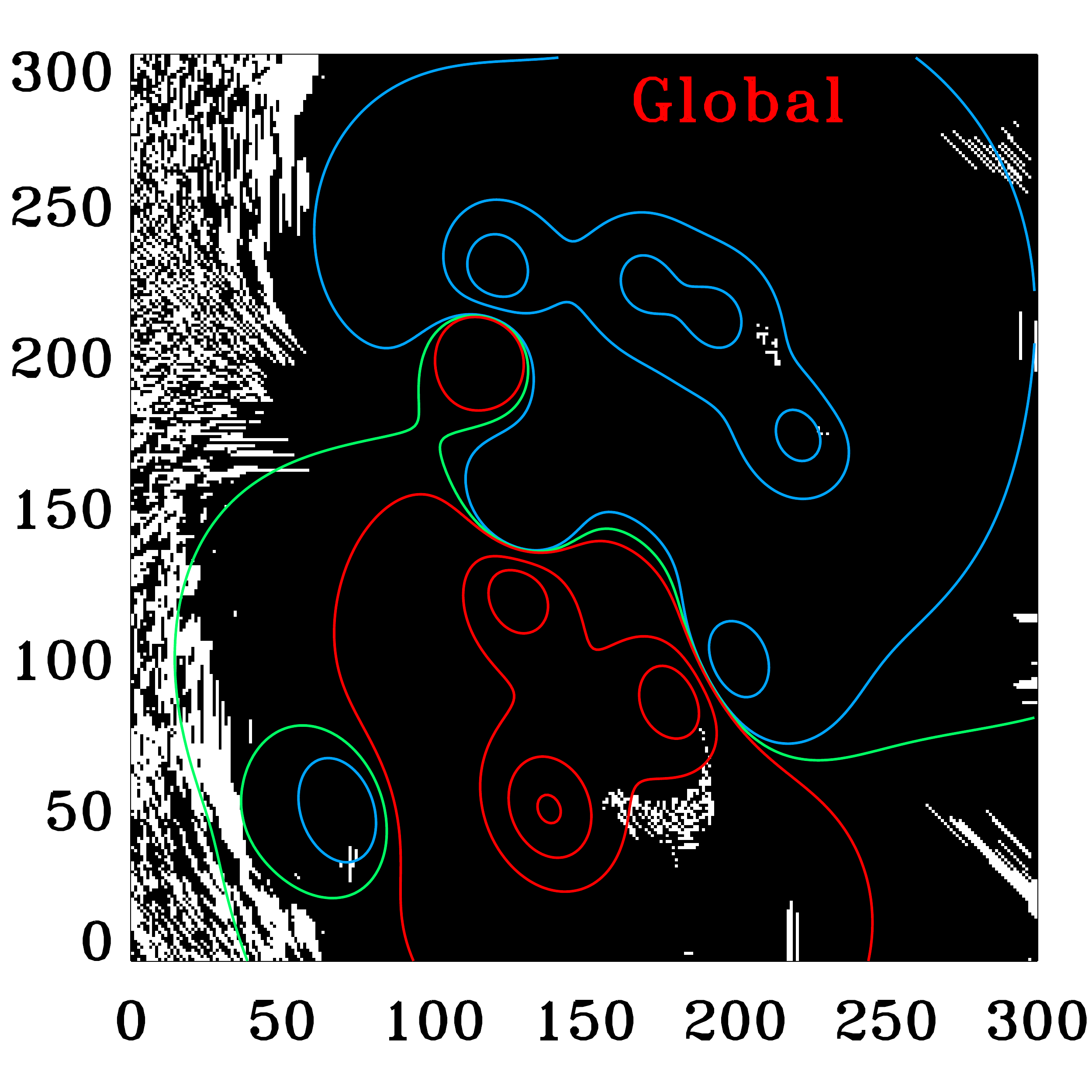} &
\includegraphics[width=0.3\textwidth]{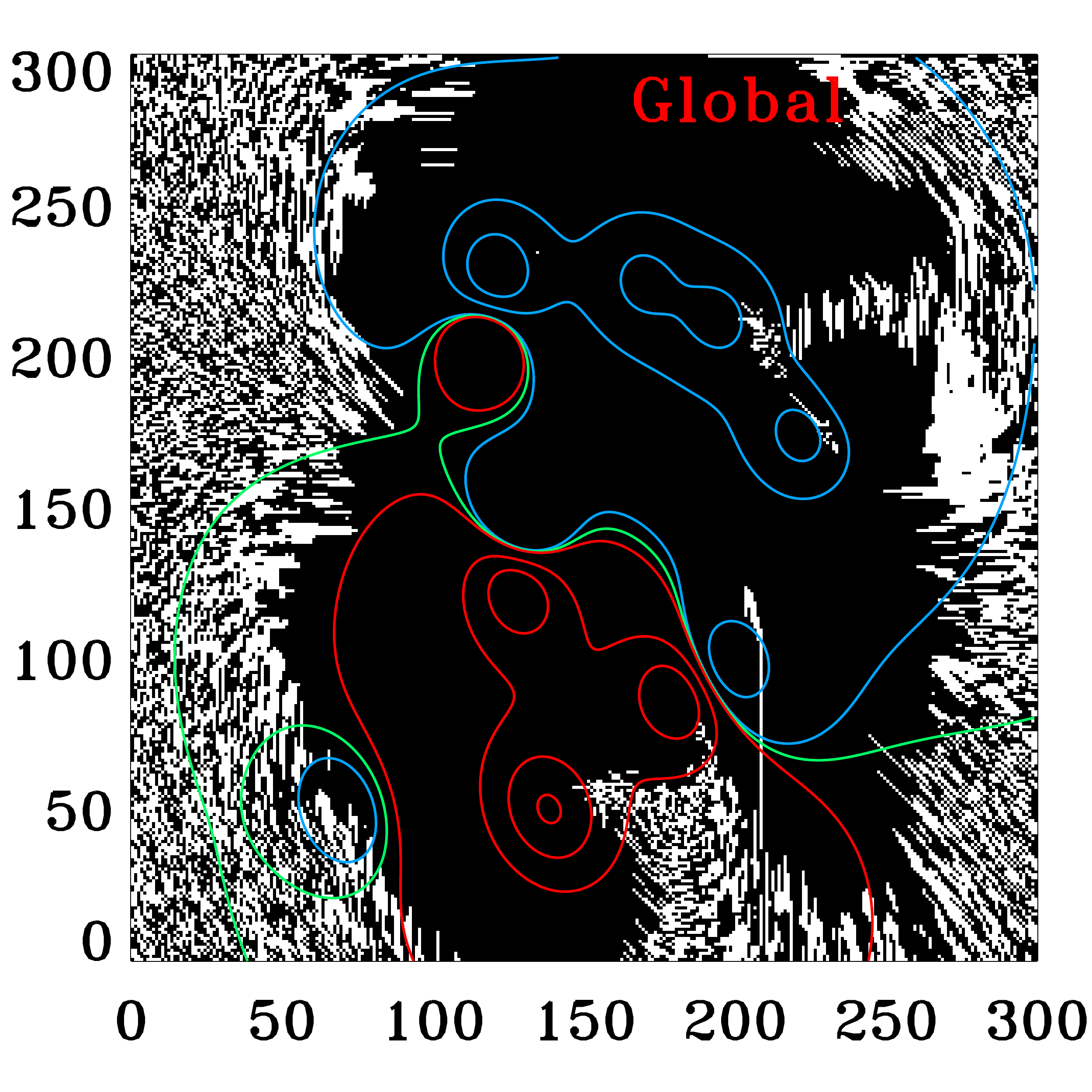}
\end{tabular}
\end{center}
\caption{Disambiguation results at the lower height for the various methods applied to the multipole field with various degrees of noise.
Areas that are correctly/incorrectly resolved are black/white.
Positive/negative vertical magnetic flux for the noise-free case is indicated by red/blue contours at 50, 700, 1500 and 2800~G, and the magnetic neutral line is indicated by the green contour.
(Left to right) No-noise, low-noise and high-noise cases.
(Top row) the Wu and Ai (1990) criterion with initial configuration $\byi > 0$.
(Middle row) The sequential minimisation method.
(Bottom row) The global minimisation method (Equation~(\ref{eglobal})).
}
\label{tpd1}
\end{figure}

\begin{figure}[ht]
\begin{center}
\begin{tabular}{c@{\hspace{0.005\textwidth}}c@{\hspace{0.005\textwidth}}c}
\includegraphics[width=0.3\textwidth]{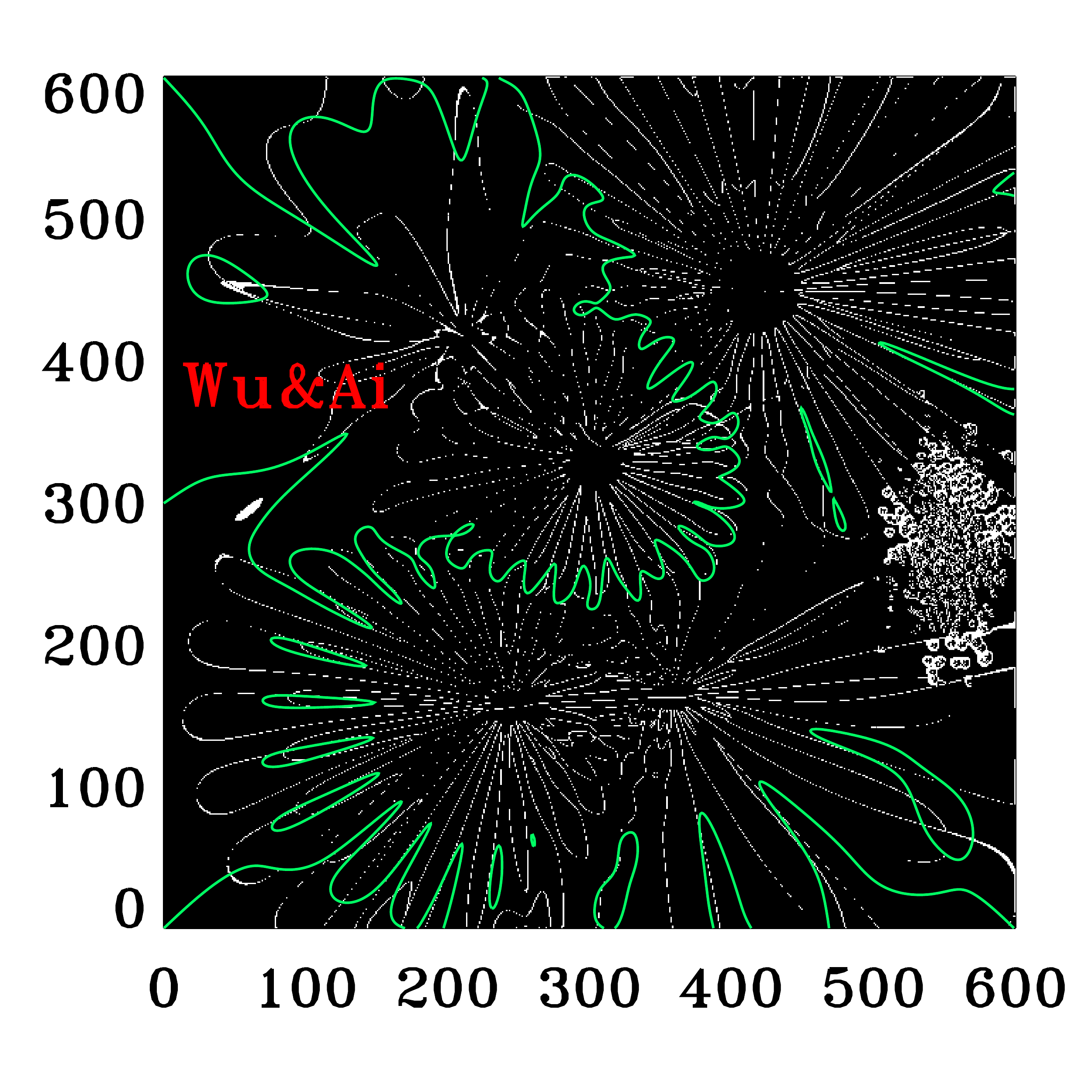} &
\includegraphics[width=0.3\textwidth]{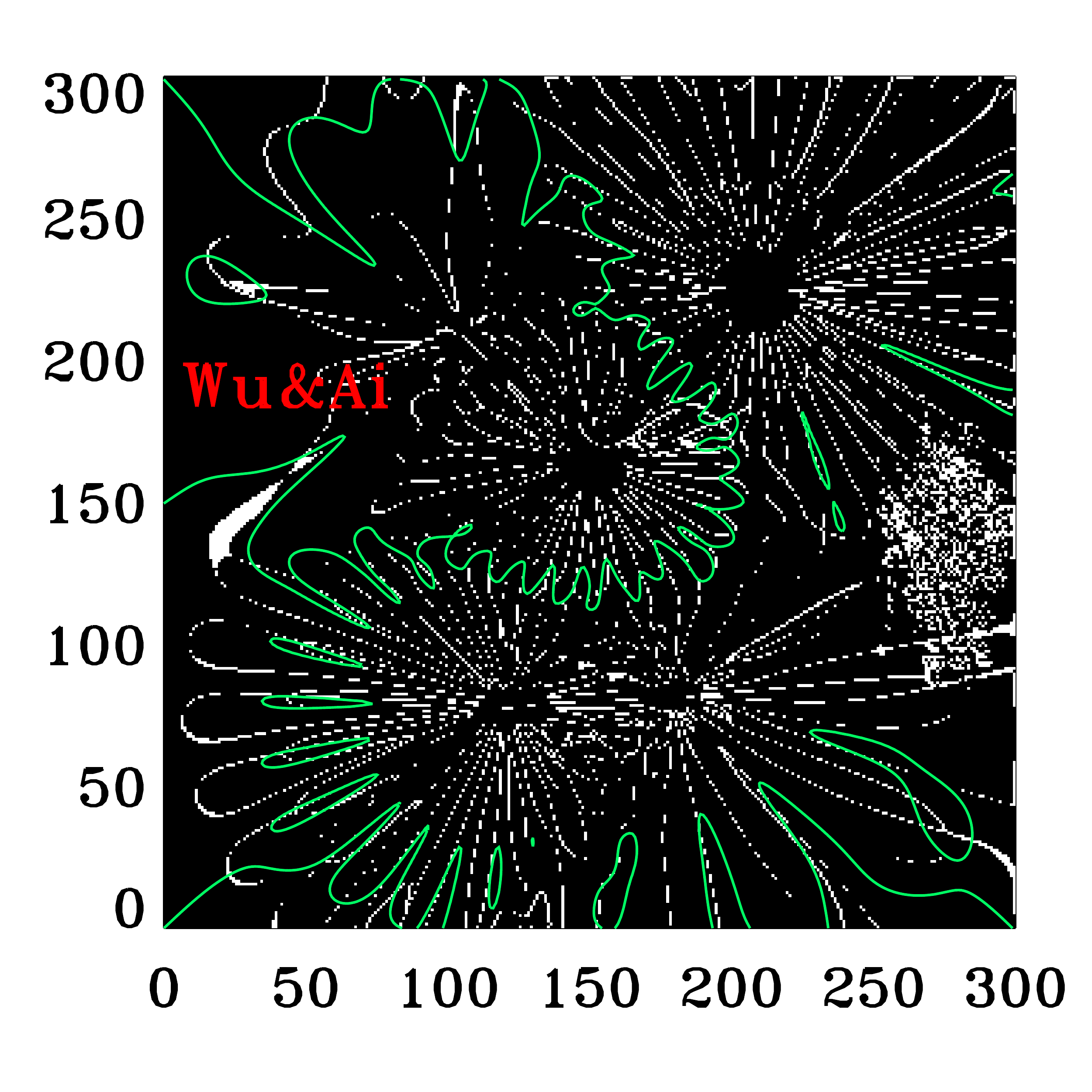} &
\includegraphics[width=0.3\textwidth]{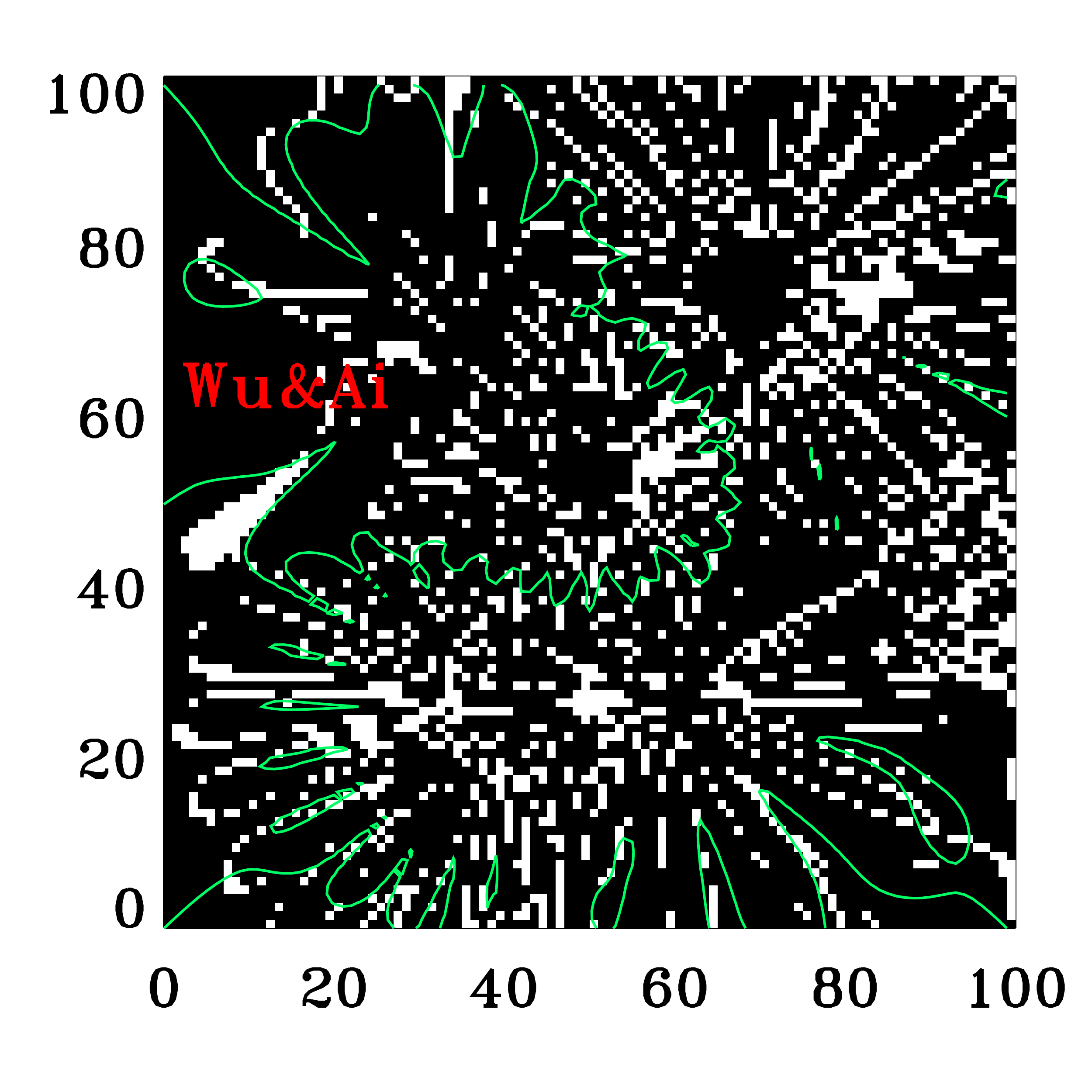} \\[-0.01\textwidth]
\includegraphics[width=0.3\textwidth]{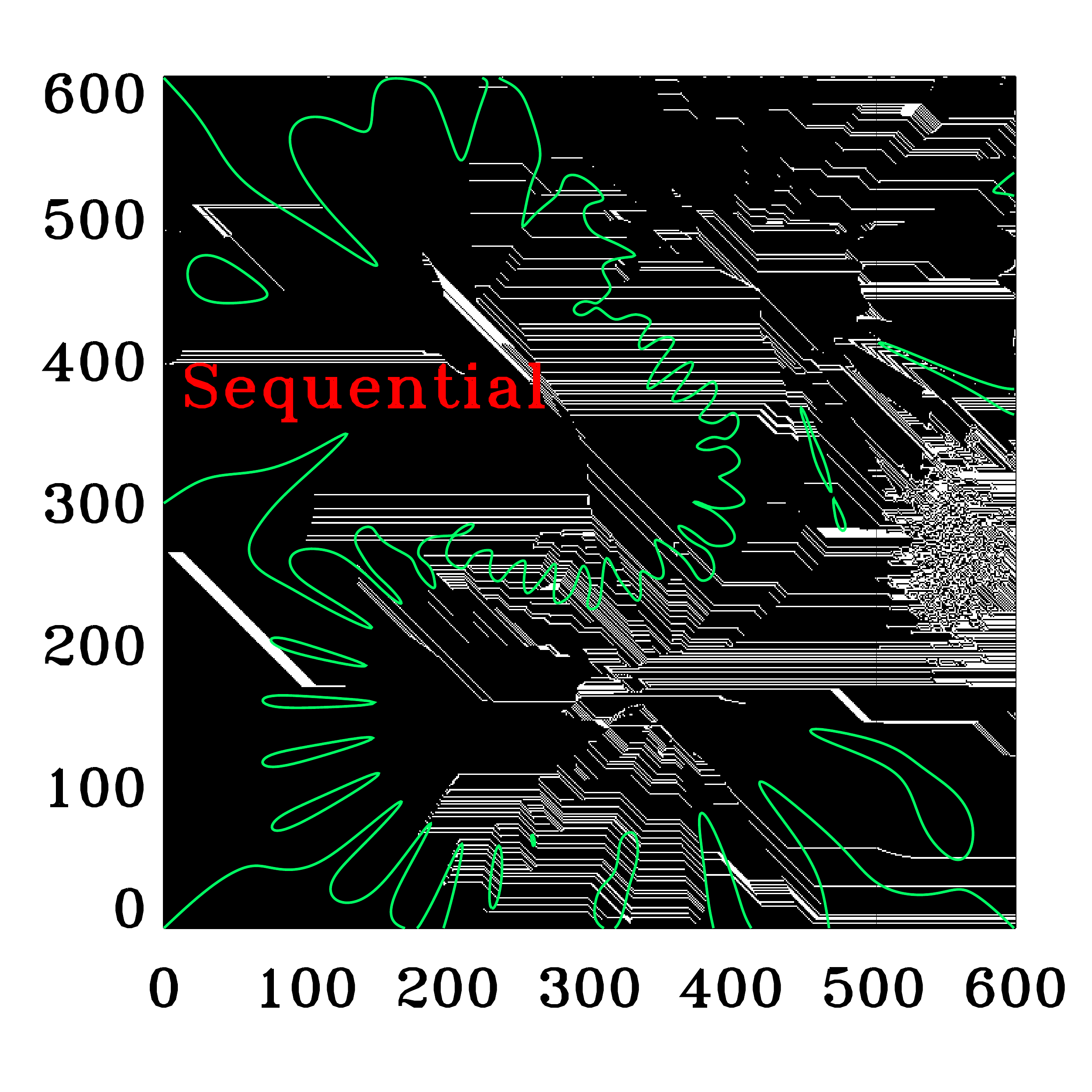} &
\includegraphics[width=0.3\textwidth]{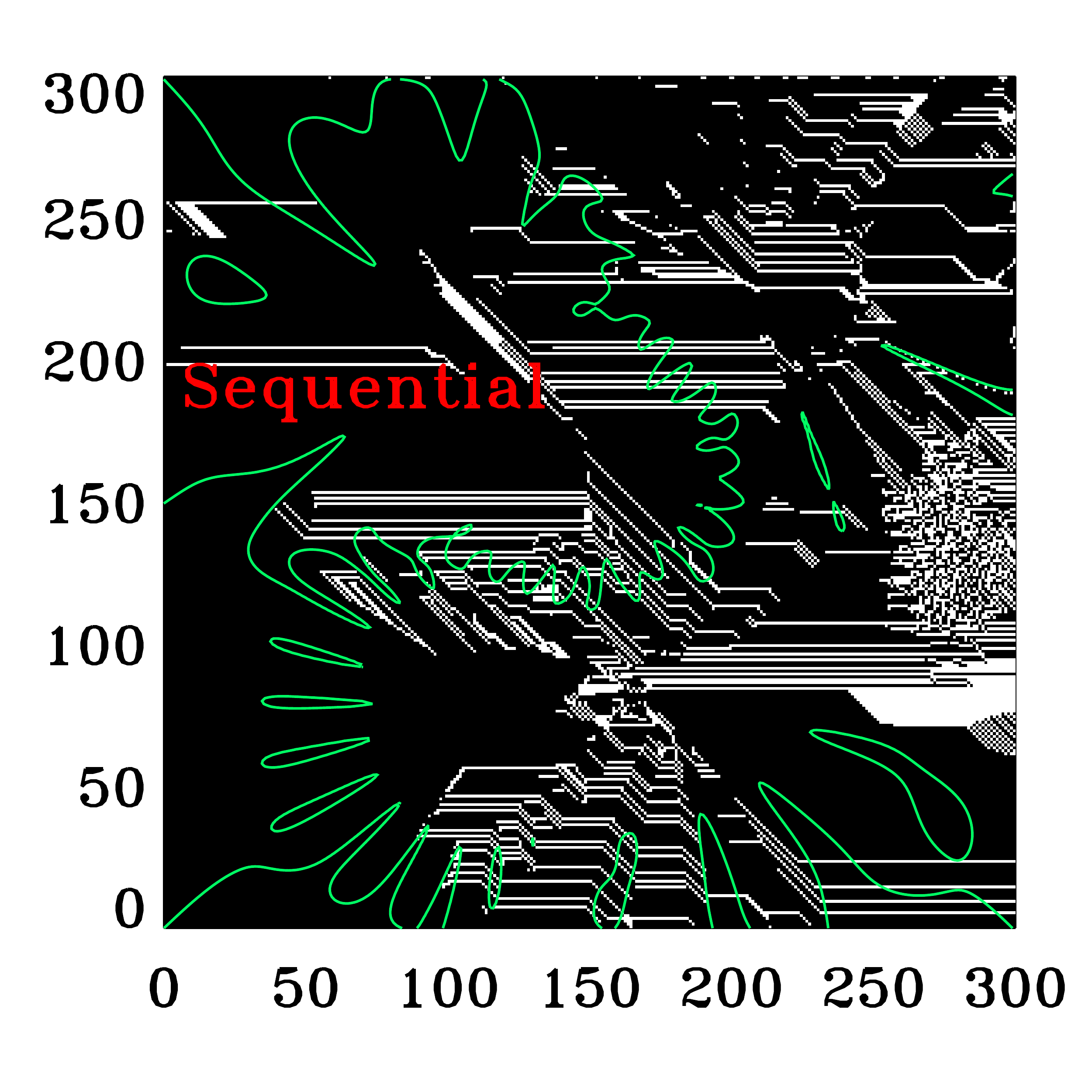} &
\includegraphics[width=0.3\textwidth]{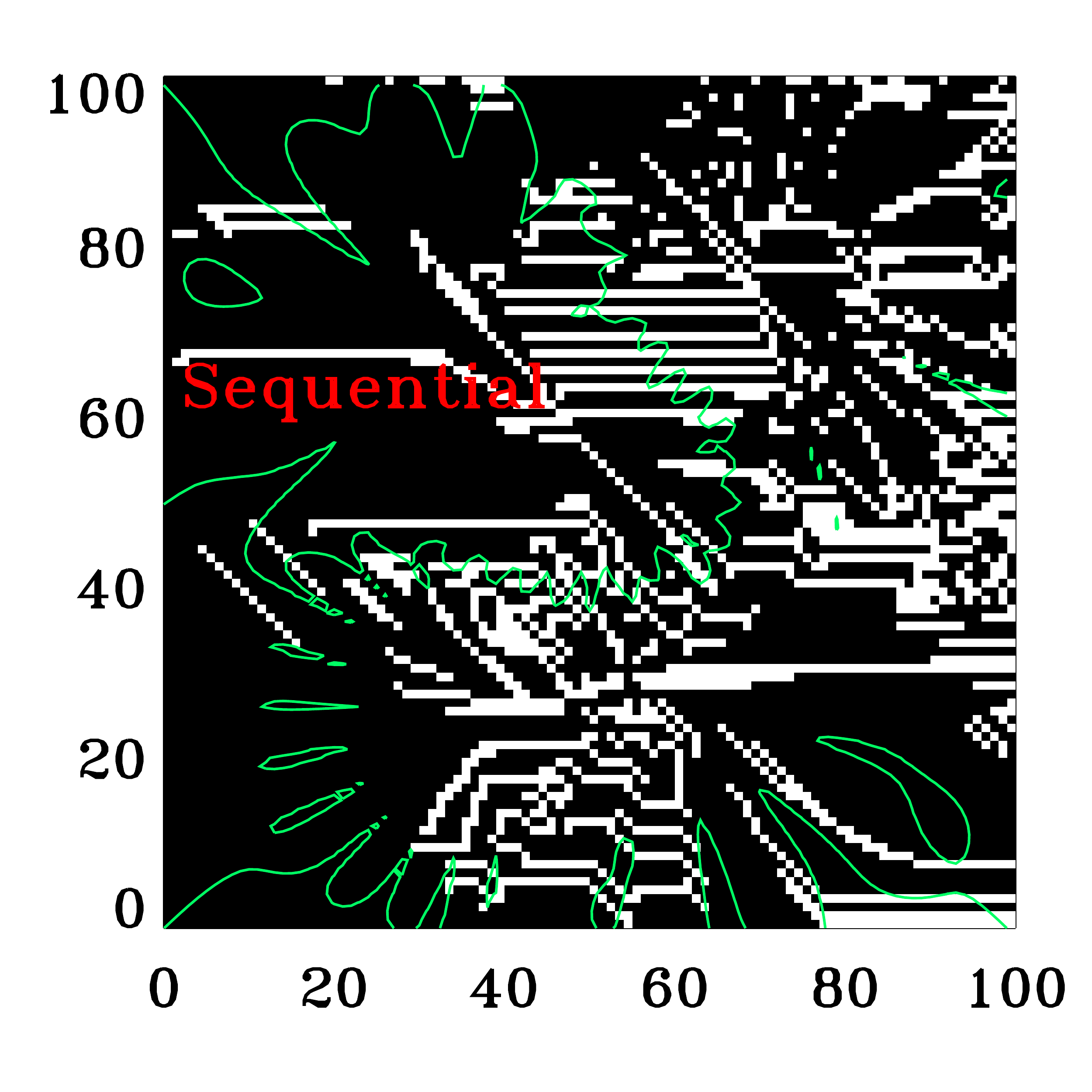} \\[-0.01\textwidth]
\includegraphics[width=0.3\textwidth]{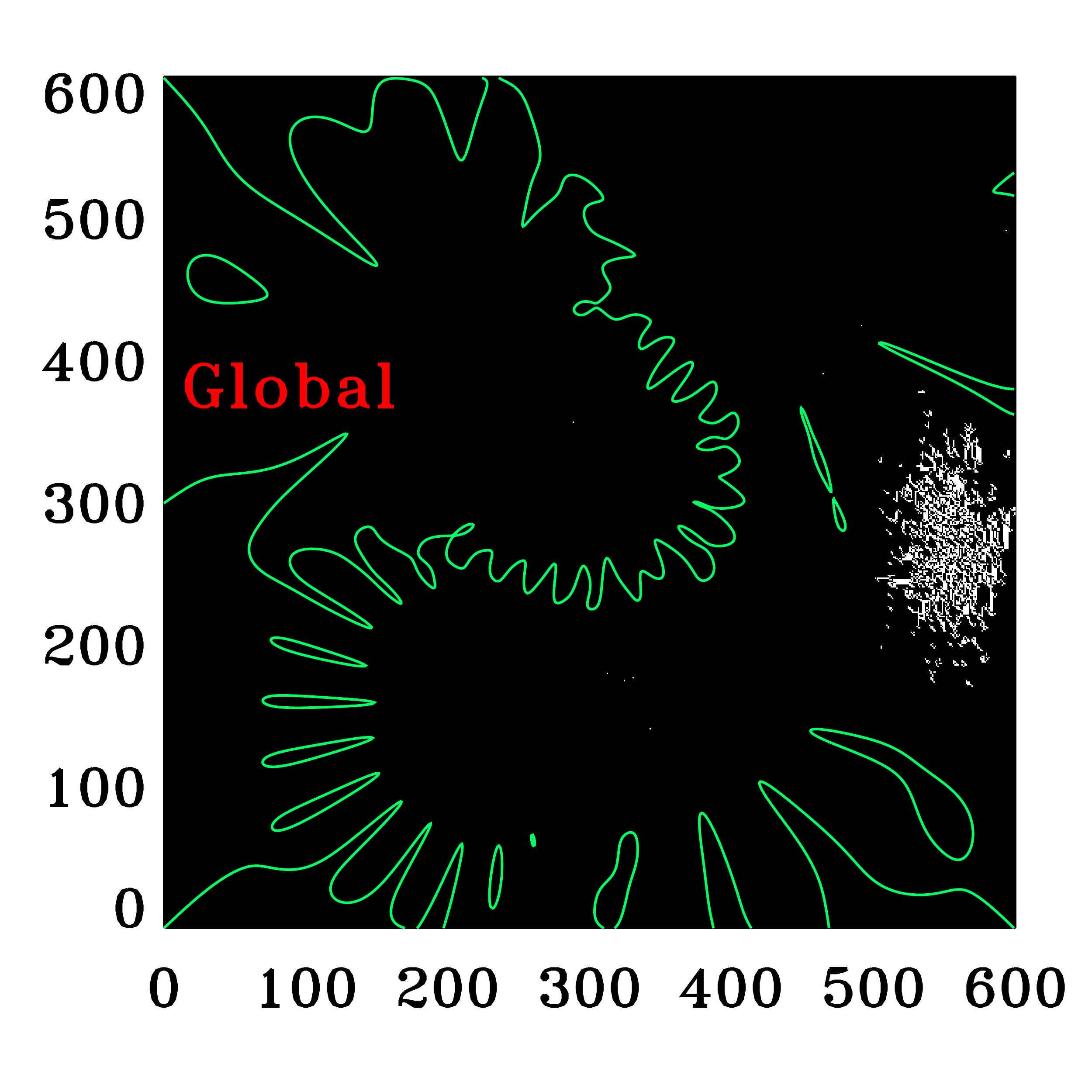} &
\includegraphics[width=0.3\textwidth]{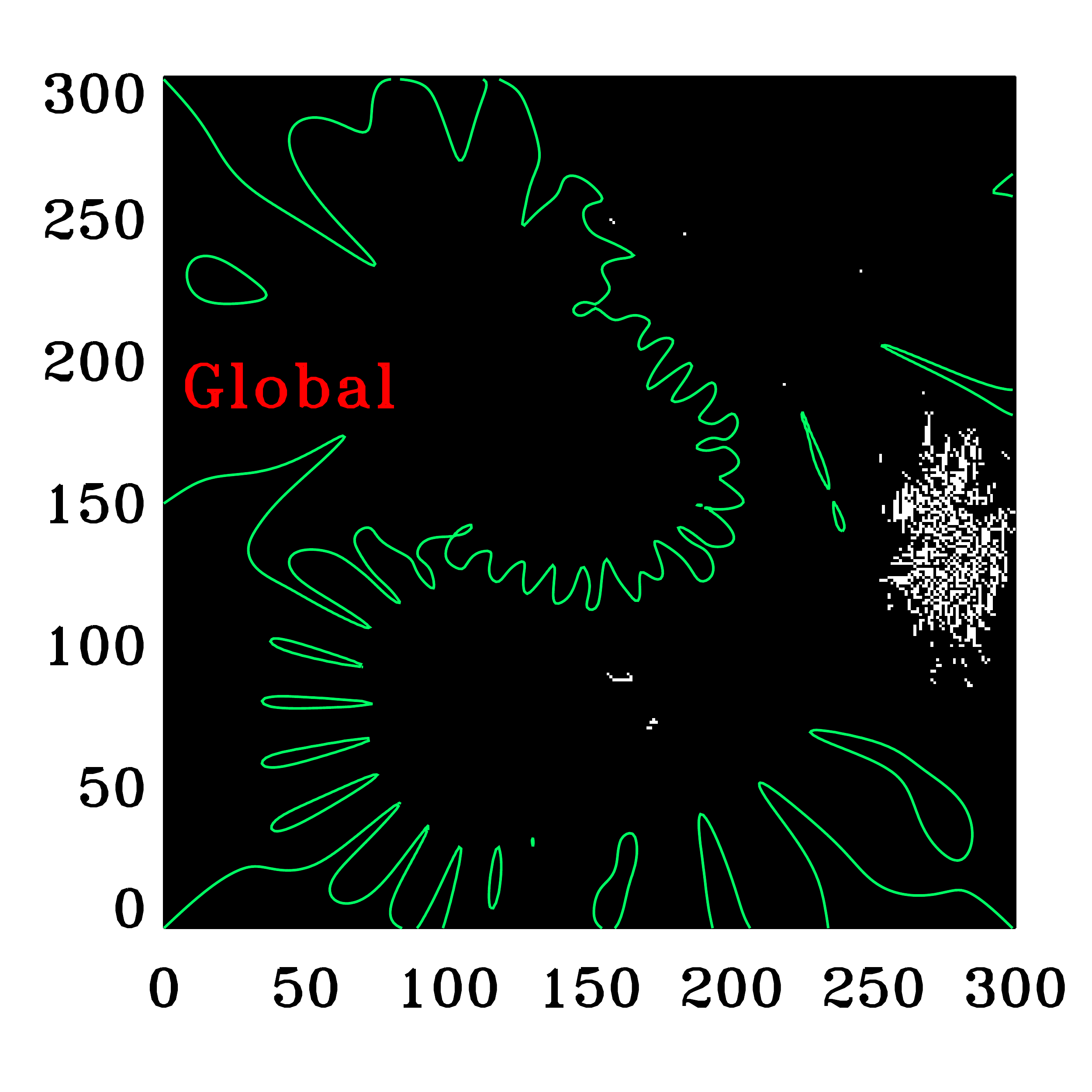} &
\includegraphics[width=0.3\textwidth]{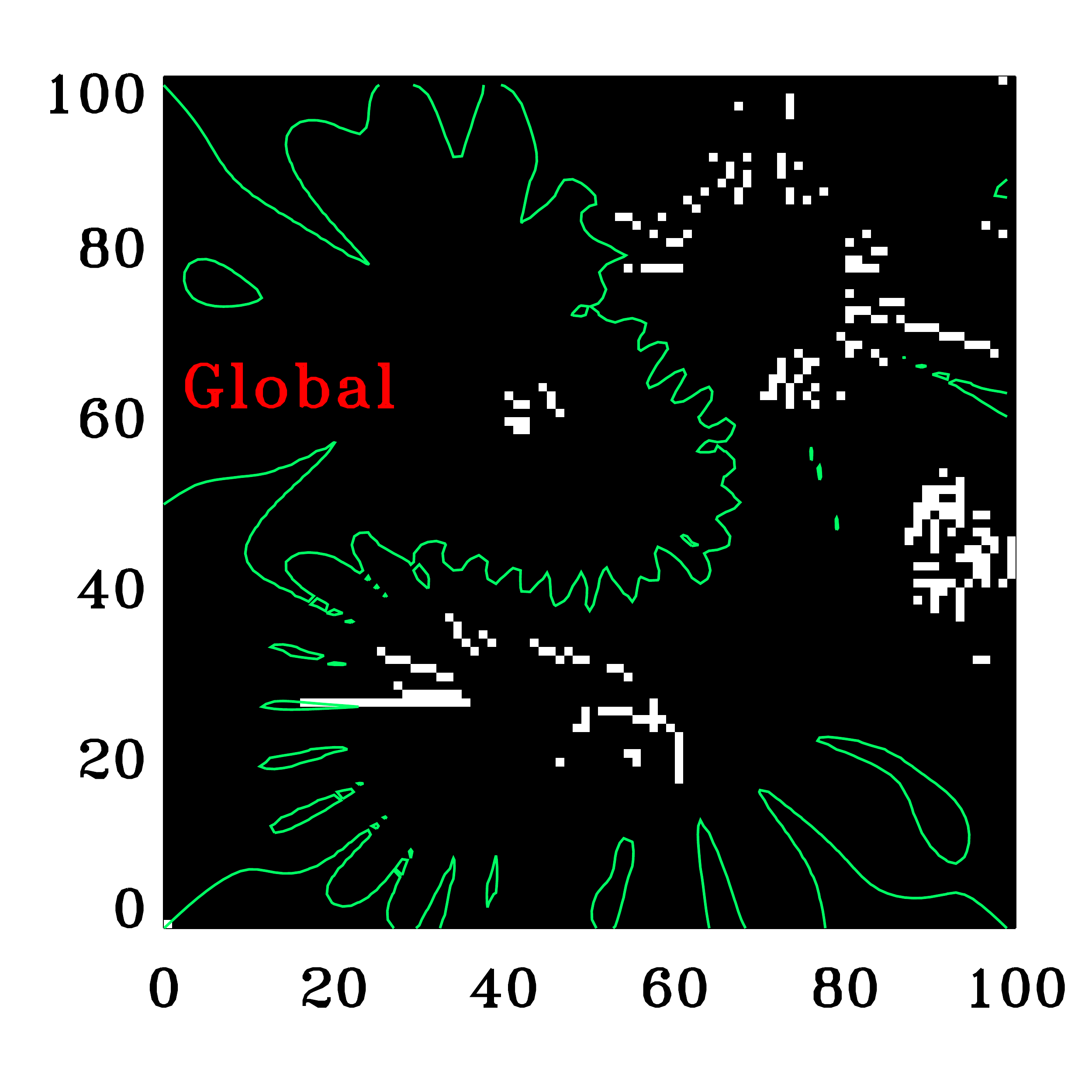}
\end{tabular}
\end{center}
\caption{Disambiguation results for the tests of spatial resolution.
Areas that are correctly/incorrectly resolved are black/white.
The green contour is the magnetic neutral line.
(Left to right)  $0.15''$, $0.3''$ and $0.9''$ cases.
(Top row) The Wu and Ai (1990) criterion in which the initial configuration has $\byi > 0$.
(Middle row) The sequential minimisation method.
(Bottom row) The global minimisation method (Equation~(\ref{eglobal})).
}
\label{flowers1}
\end{figure}

\section{Further Development of the Global Minimisation Method}
\label{sec_extra}

In the previous section we found that all currently available disambiguation methods that are based on the divergence are sensitive to both photon noise and limited spatial resolution, if implemented as described in \cbl{}.
Generally speaking, we found the same trend that was found in \cbl{}, in that the global minimisation method is more robust than both the \citeauthor{1990AcApS..10..371W} criterion and the sequential minimisation method, in the presence of both types of noise.
However, the performance of the global minimisation method is significantly compromised when photon noise or unresolved structure is present in the magnetogram data.
In this section we present results from a series of experiments aimed at improving the performance of the global minimisation method.

\subsection{Approximation of Derivatives for the Divergence}

In this section we discuss the approximations used to compute derivatives for the various terms in the divergence. 
For all three algorithms examined in Section~\ref{sec_now}, derivatives with respect to $\xuh$ and $\yuh$ at $(i,j,k)$ are approximated with three-point finite differences using measurements at $(i,j,k)$, $(i+1,j,k)$ and $(i,j+1,k)$, except at the boundaries $i=n_x$ and $j=n_y$ where the approximation is changed to use only pixels within the field of view.
Derivatives with respect to $\zui$ are approximated with two-point finite differences using measurements from the two heights.
Although they are simple, these finite-difference approximations are evidently sufficient for the global minimisation method to retrieve reasonable results for synthetic data without noise, depending on the nature of the field and the resolution of the observations.
On the other hand, this is generally not true in regions where the influence of noise is significant.

In regions with significant unresolved structure (see Figure~\ref{fbin}), the measurement at a given pixel is a complicated average of the magnetic field underlying  the area covered by the pixel.
Consequently, for the tests of limited spatial resolution the  value of the magnetic field at a given point may not be known exactly and variations of the magnetic field in the horizontal directions may not be resolved. 
Under these conditions finite differences and finite elements are not expected to produce accurate approximations for derivatives.

\begin{figure}[ht]
\begin{center}
\begin{tabular}{c@{\hspace{0.005\textwidth}}c}
\includegraphics[width=0.45\textwidth]{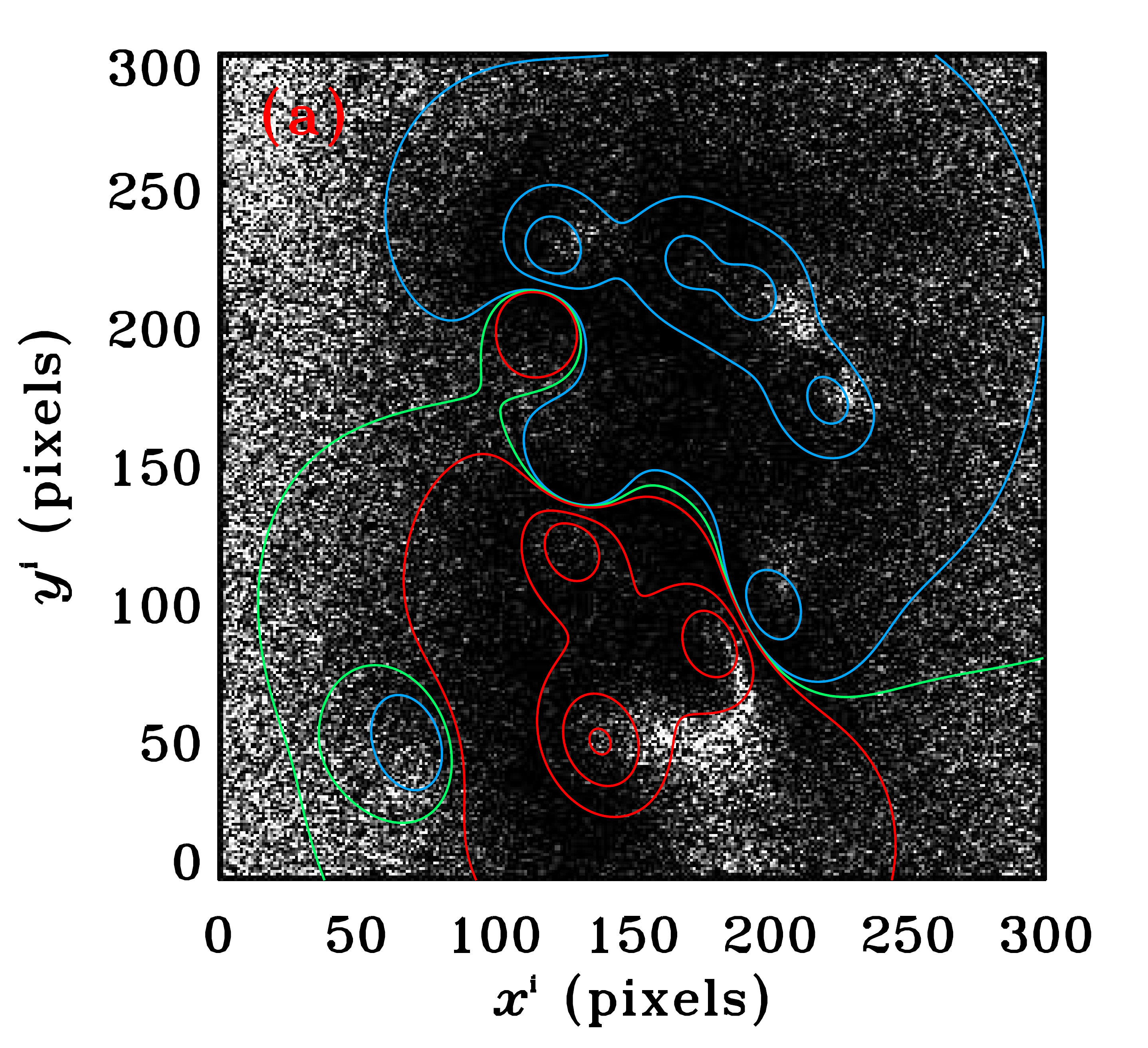} & \includegraphics[width=0.45\textwidth]{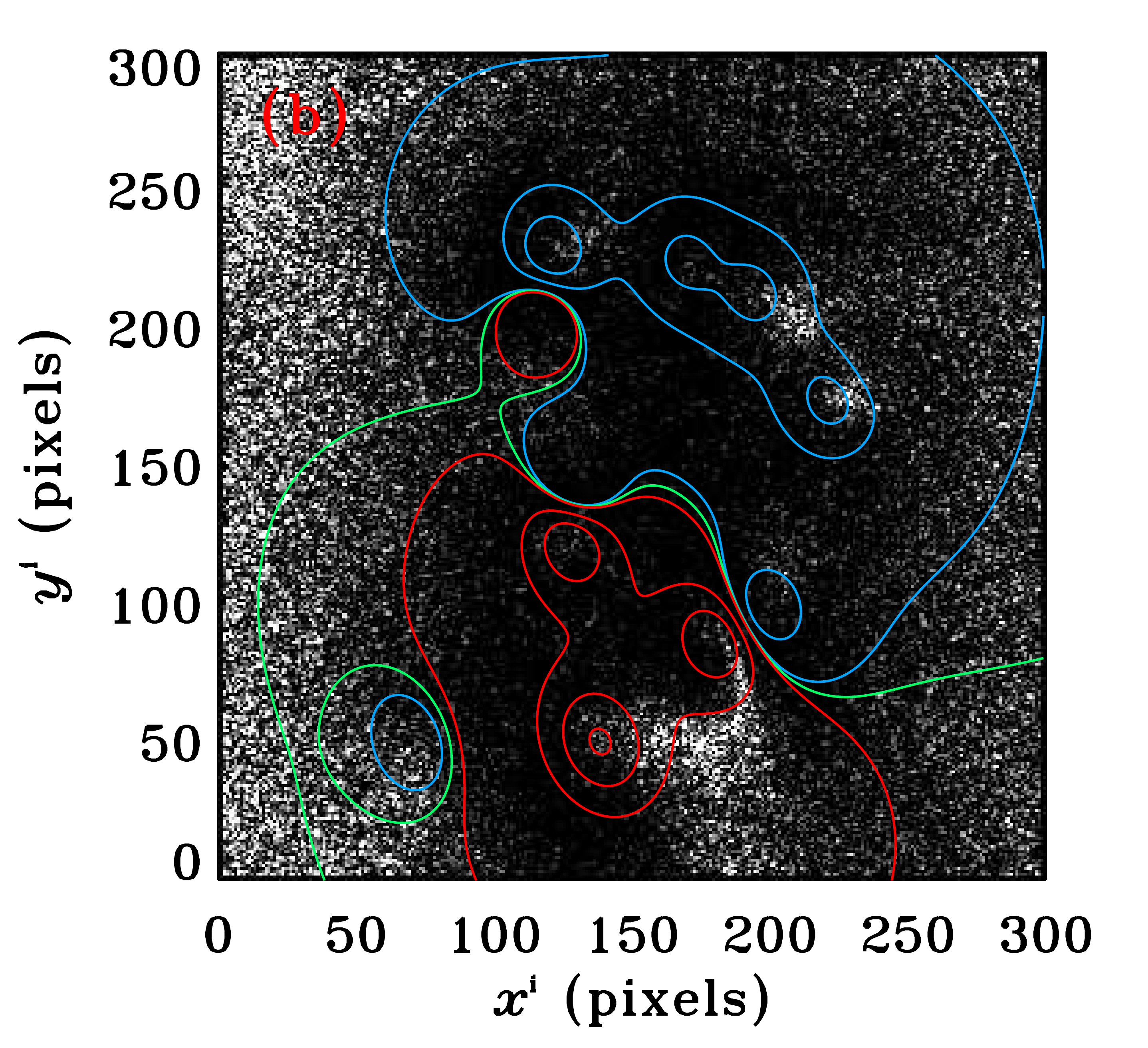} 
\end{tabular}
\begin{tabular}{c@{\hspace{0.005\textwidth}}c@{\hspace{0.005\textwidth}}c}
\includegraphics[width=0.31\textwidth]{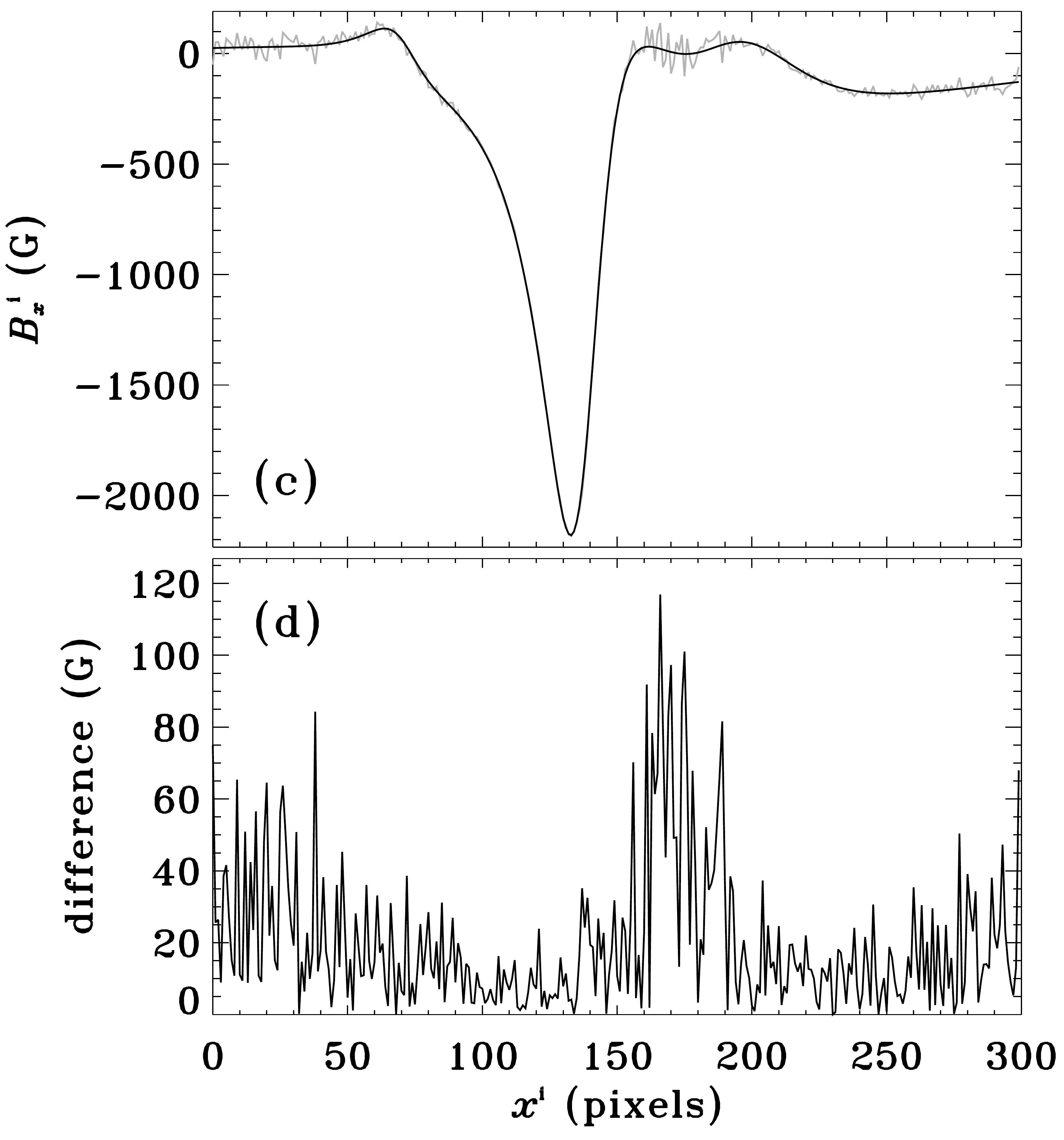} & \includegraphics[width=0.31\textwidth]{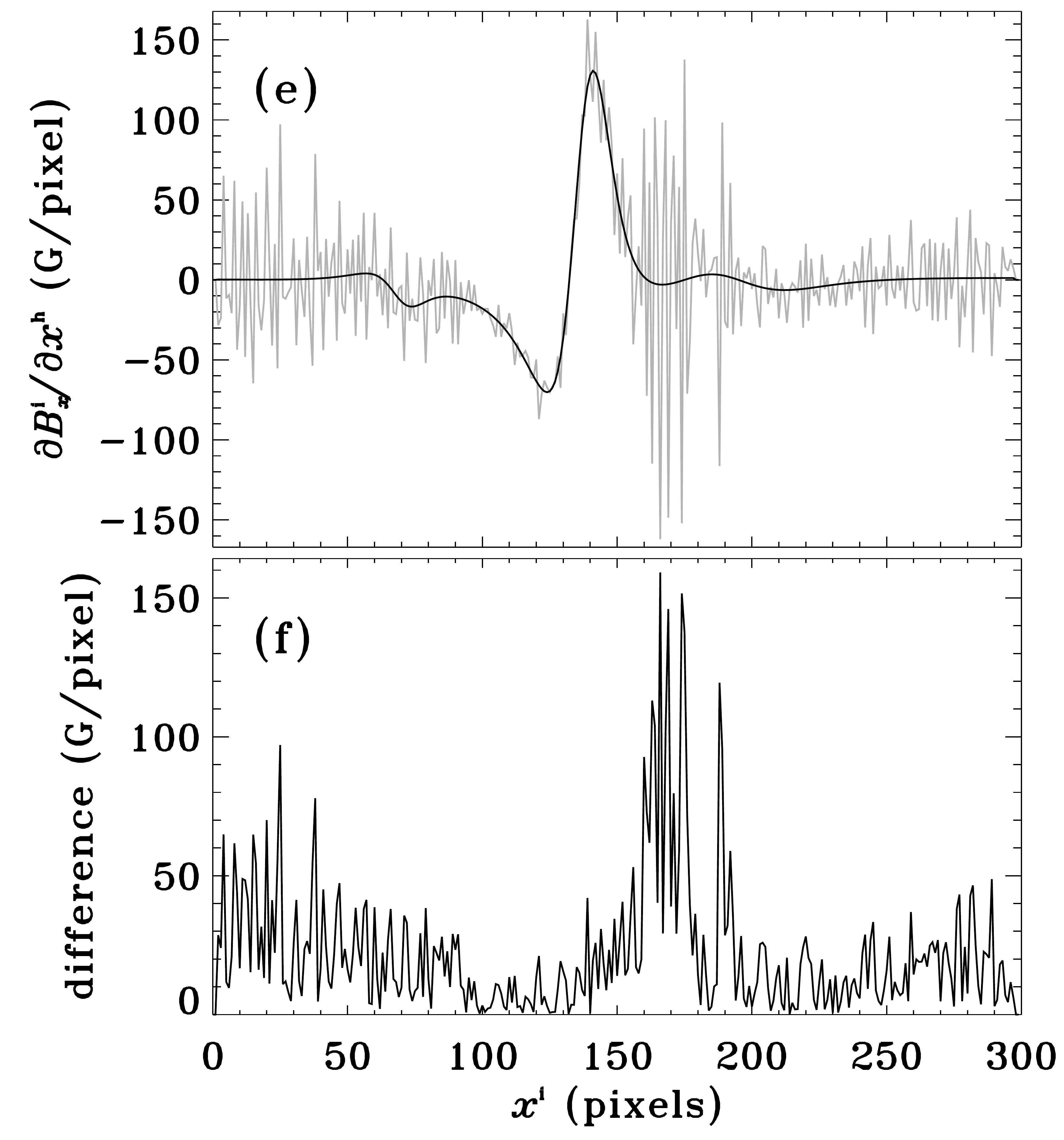} & \includegraphics[width=0.31\textwidth]{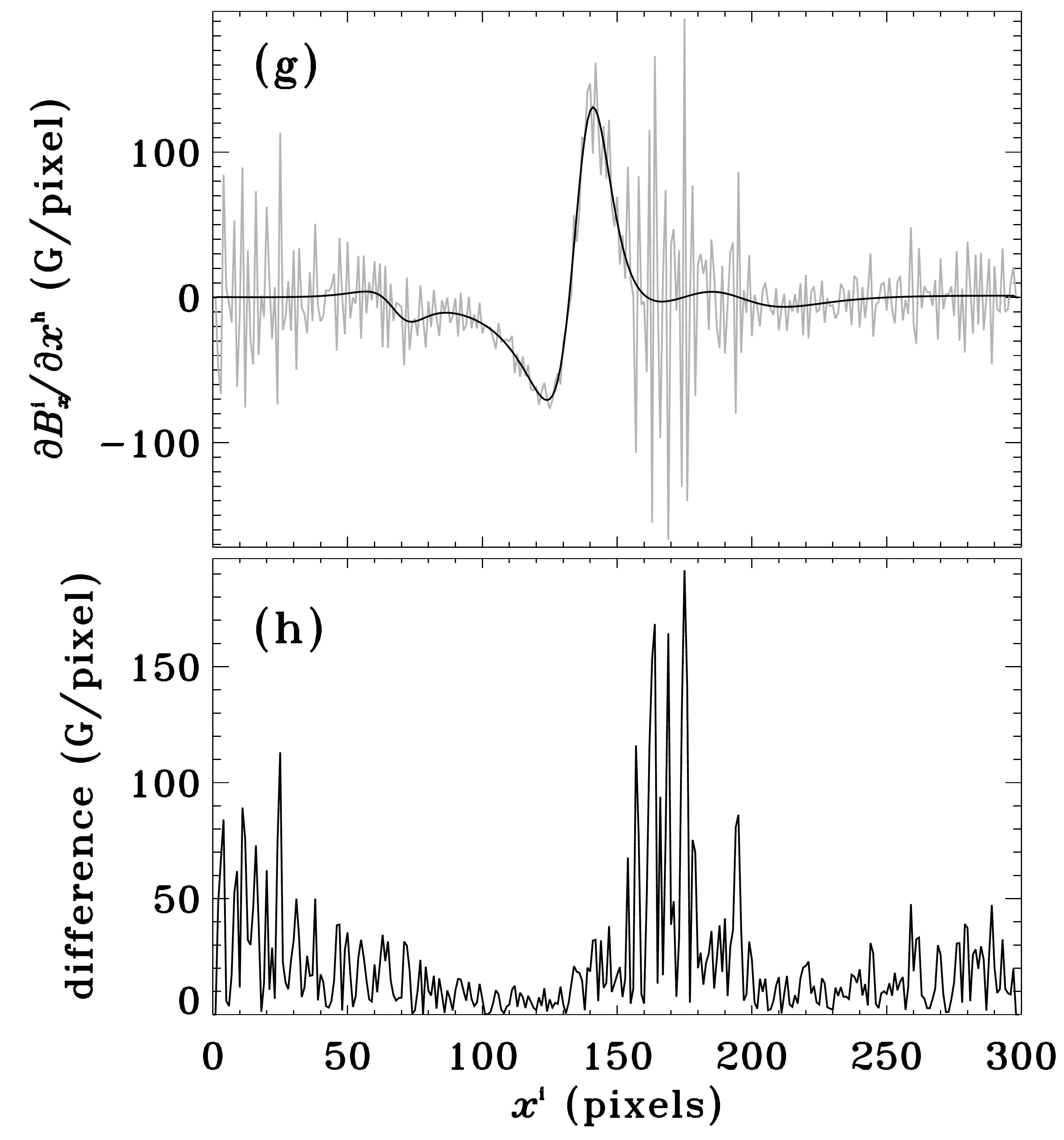}
\end{tabular}
\end{center}
\caption{
(a) The absolute value of the difference between the approximations for the derivative \(\partial \bxi / \partial \xuh \) applied to the no-noise and low-noise  ambiguity resolved answer for the multipole field configuration, as a function of \(\xui\) and \(\yui\) (underlying grey-scale); the range shown is \( [10,60] \)~G/pixel. 
The approximation used for \(\partial \bxi / \partial \xuh \) is a three-point finite-difference approximation (for details see text).
The contours are the same as in Figure~\ref{tpd1}.
(b) Same as (a) except the approximation for \(\partial \bxi / \partial \xuh \) is a 16-point finite-element approximation (for details see text).
(c) \(\bxi\) as a function of \(\xui\) at \(\yui=50\)~pixels. The black/grey curve is for the no-noise/low-noise case; these curves are indistinguishable at many of the points shown.
(d) The absolute value of the difference between the two curves shown in (c).
(e) The three-point finite-difference approximation for \(\partial \bxi / \partial \xuh \) (for details see text) as a function of \(\xui\) at \(\yui=50\)~pixels. The black/grey curve is for the no-noise/low-noise case.
(f) The absolute value of the difference between the two curves shown in (e).
(g) and (h) Same as (e) and (f) except the approximation for \(\partial \bxi / \partial \xuh \) is a 16-point finite-element approximation (for details see text).
} 
\label{dcomp}
\end{figure}

When the influence of photon noise is significant, the measurement of the magnetic field may deviate substantially from the underlying,  no-noise value (see Figure~\ref{adiff}).
Finite differences are therefore not expected to produce accurate approximations for derivatives when the influence of photon noise is significant.
In Figure~\ref{dcomp} we demonstrate how different approximations for the derivative \(\partial \bxi / \partial \xuh \) respond to the effects of photon noise.
In Figures~\ref{dcomp}(a) and (b) we show maps of the difference between the approximation for \(\partial \bxi / \partial \xuh \) for the no-noise and low-noise ambiguity-resolved answers.
It is evident from these maps that areas with large errors in the approximation for this derivative tend to correspond to areas where the global minimisation method retrieves poor results for the ambiguity resolution in the noise-added test cases (for example, compare Figure~\ref{dcomp}(a) with the lower panels of Figure~\ref{tpd1}).

Several additional points are worth noting regarding Figure~\ref{dcomp}: 
(1) Finite-difference and finite-element approximations for derivatives can increase (or amplify) the effect of photon noise.
(2) We find similar results to those shown in Figure~\ref{dcomp} for the approximation for the derivative \(\partial  / \partial \yuh \) and for the other components of the magnetic field (not shown).
(3) We find that the discrepancy between the approximations for the derivatives in the noise-added and noise-free cases generally increases as the level of photon noise increases.
(4) We show results for a 16-point finite-element approximation for \(\partial \bxi / \partial \xuh \) which approximates the derivative at \((i,j)\) using measurements in a four pixel by four pixel region bounded by \(i-1\) and \(i+2\) in the \(\xui\)-direction and \(j-1\) and \(j+2\) in the \(\yui\)-direction.
Figure~\ref{dcomp} shows that the results produced by both this 16-point approximation and the three-point approximation are similar.
When the global minimisation method uses the 16-point finite-element approximation for horizontal heliographic derivatives we find that the ambiguity-resolution results are similar to those retrieved using the three-point approximation, but the 16-point approximation requires more iterations for the global minimisation method to converge.
(5) There are techniques for reliably estimating derivatives from data with noise, which produce encouraging results (\myeg \opencite{1971SJNA....8..254C}; \opencite{2007CoPhC.177..764A}).
However, there are several challenges in applying these techniques to the global minimisation method and we do not attempt to implement these techniques in this article.

\subsection{Smoothing the Data Before Disambiguation}
\label{sec_smooth}

One straightforward approach that may be used to reduce the influence of random noise (like photon noise) is to de-noise or smooth the magnetogram data before disambiguation.
To test the efficacy of smoothing the data, we take the following approach:
(1) Set the initial configuration of azimuthal angles such that \( \byi > 0 \) at each pixel.
This is done to avoid situations where neighbouring pixels get averaged together that initially have very different azimuthal angles.
(2) Apply a two-dimensional spatial smoothing operator to all three components of the magnetic field, \(  \bxi, \byi, \bzi  \), at both heights; this operation does not change the effective pixel size in any direction.
(3) Find the minimum of \(E\) (Equation~(\ref{eglobal})) using the algorithm described in Section~\ref{sec_global}.
(4) Take the realisation of the transverse component of the unsmoothed magnetic field that is closest to the retrieved solution at each pixel.

We have tested several different smoothing operators, including both average (\myie box-car) and triangular smoothing over areas of $3 \times 3$ pixels and $5 \times 5$ pixels.
We find that the results retrieved with these different smoothing operators tend to be quite similar.
For the sake of brevity and to highlight the general trends of smoothing the data before disambiguation, we  show the results for a smoothing operator that takes the two-dimensional box-car average over an area of $3 \times 3$ pixels.

\begin{figure}[ht]
\begin{center}
\begin{tabular}{c@{\hspace{0.005\textwidth}}c@{\hspace{0.005\textwidth}}c}
\includegraphics[width=0.3\textwidth]{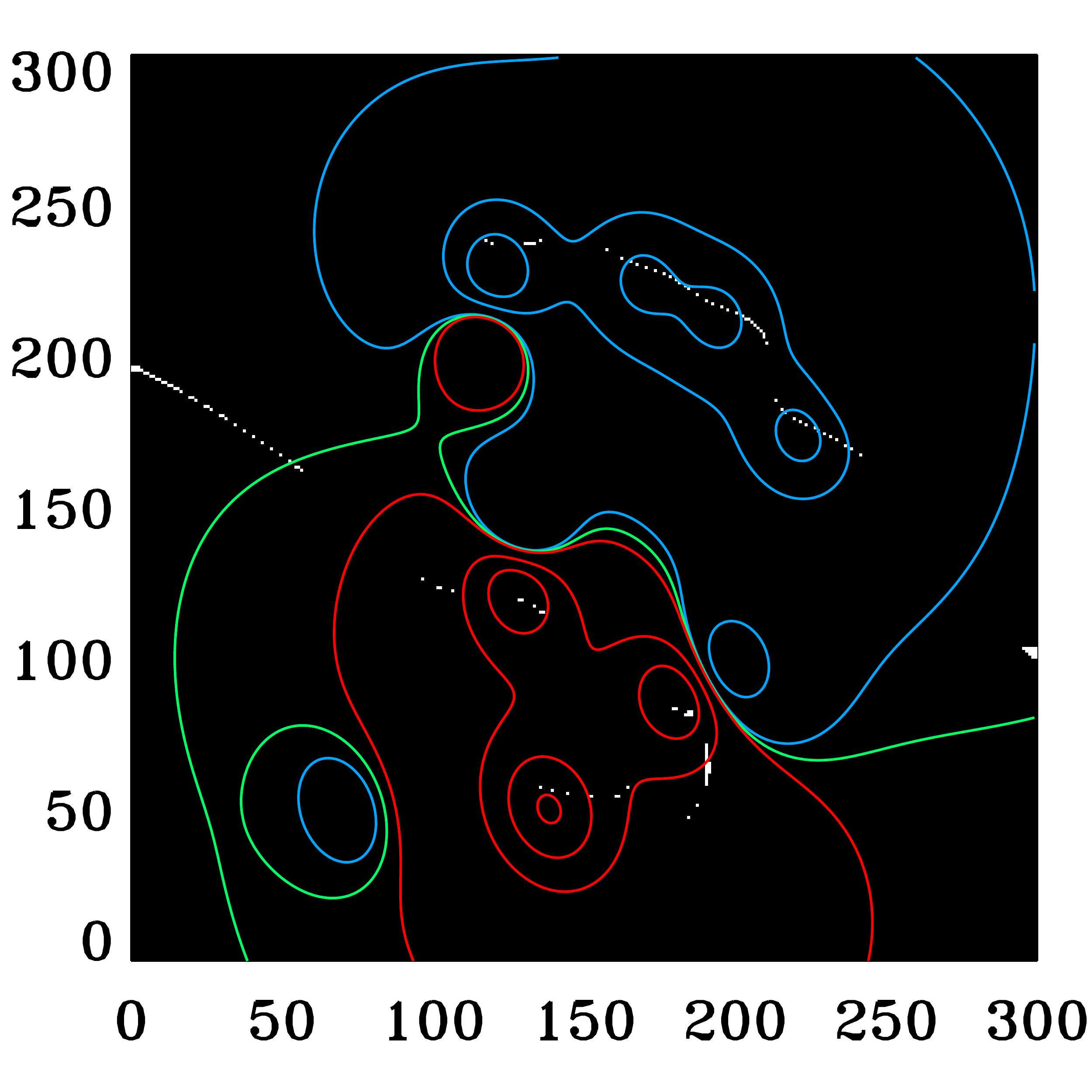} &
\includegraphics[width=0.3\textwidth]{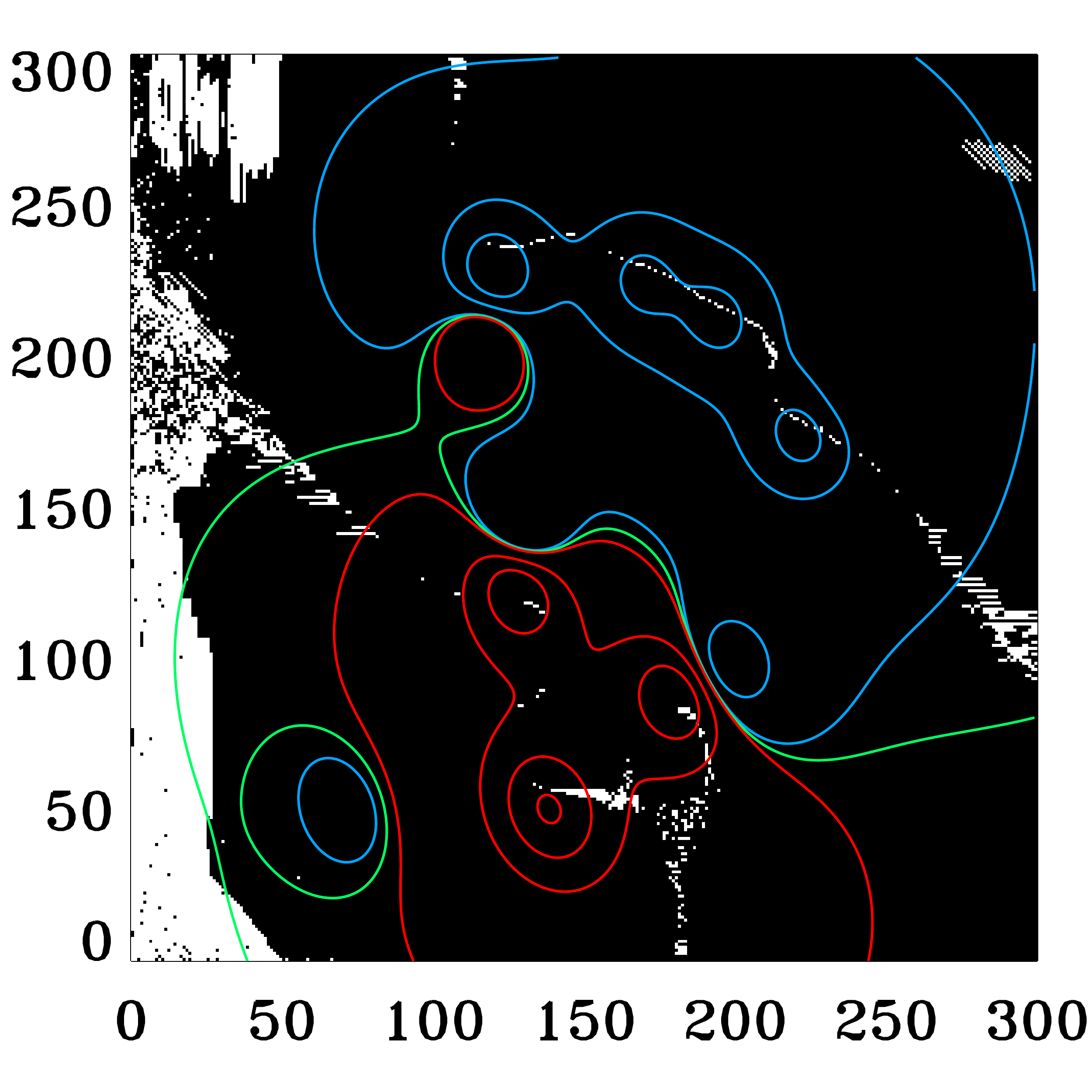} &
\includegraphics[width=0.3\textwidth]{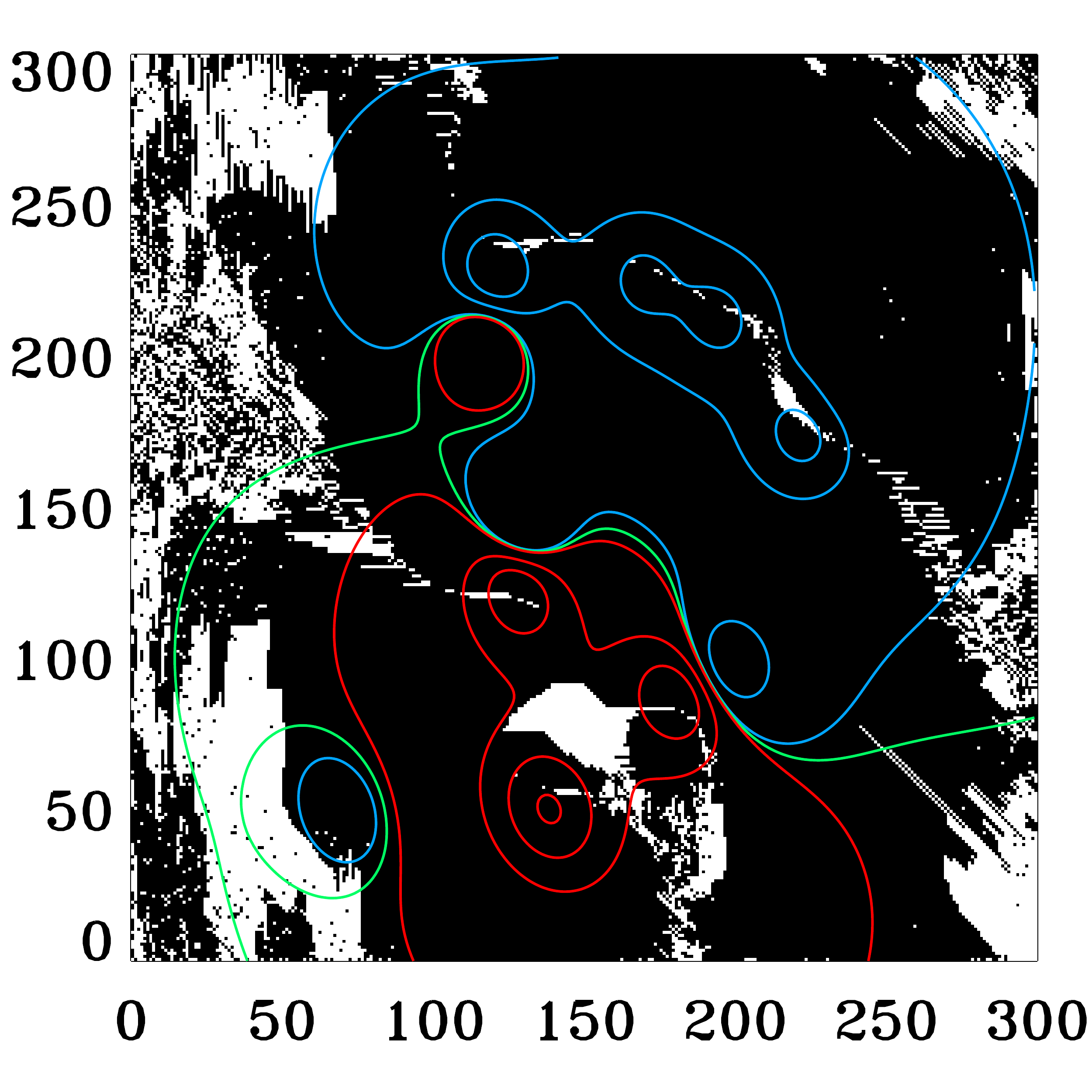}
\end{tabular}
\end{center}
\caption{
Same as the lower panels of Figure~\ref{tpd1}, except the magnetogram data are smoothed before disambiguation with the global minimisation method.
}
\label{tpdsmooth}
\end{figure}

\begin{figure}[ht]
\begin{center}
\begin{tabular}{c@{\hspace{0.005\textwidth}}c@{\hspace{0.005\textwidth}}c}
\includegraphics[width=0.3\textwidth]{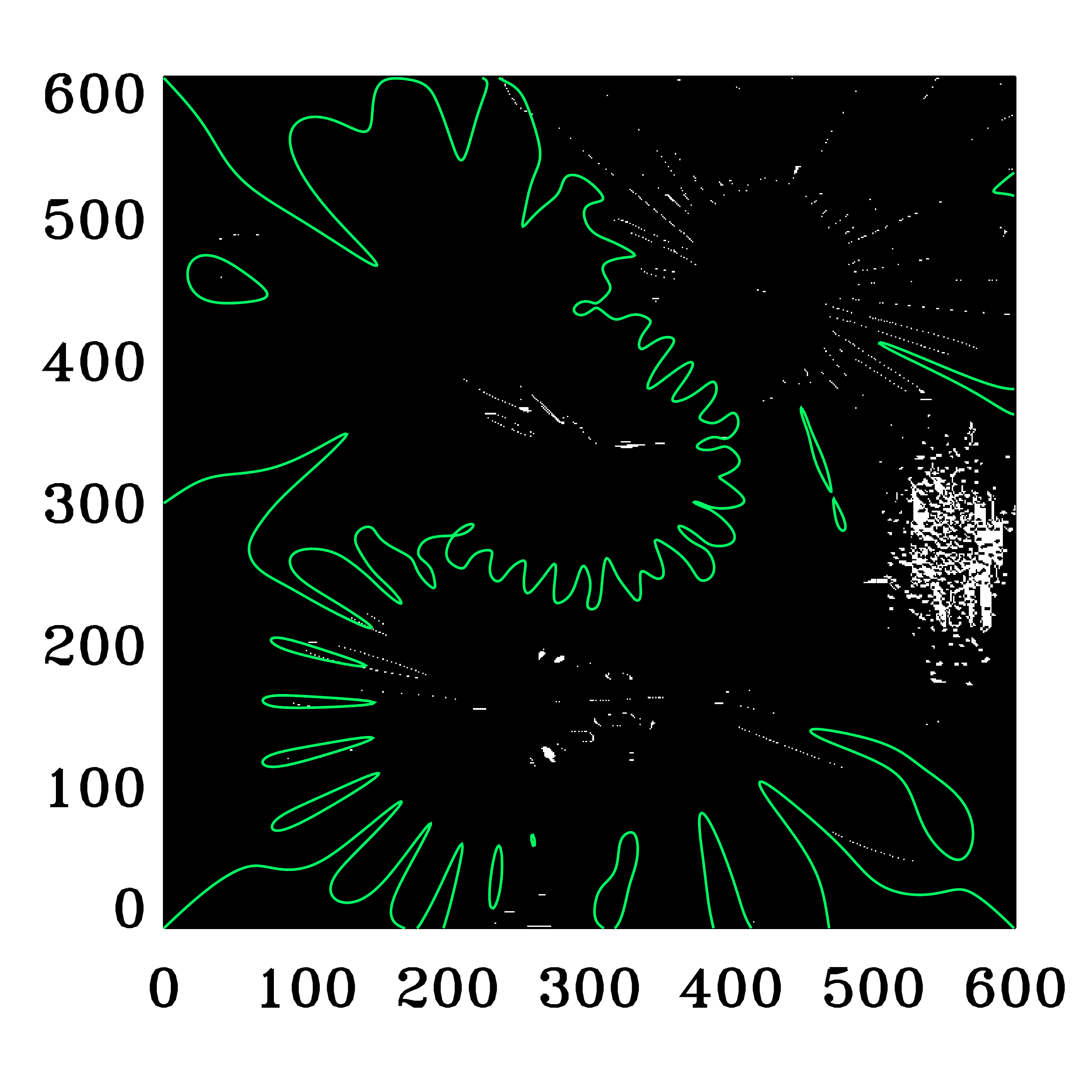} &
\includegraphics[width=0.3\textwidth]{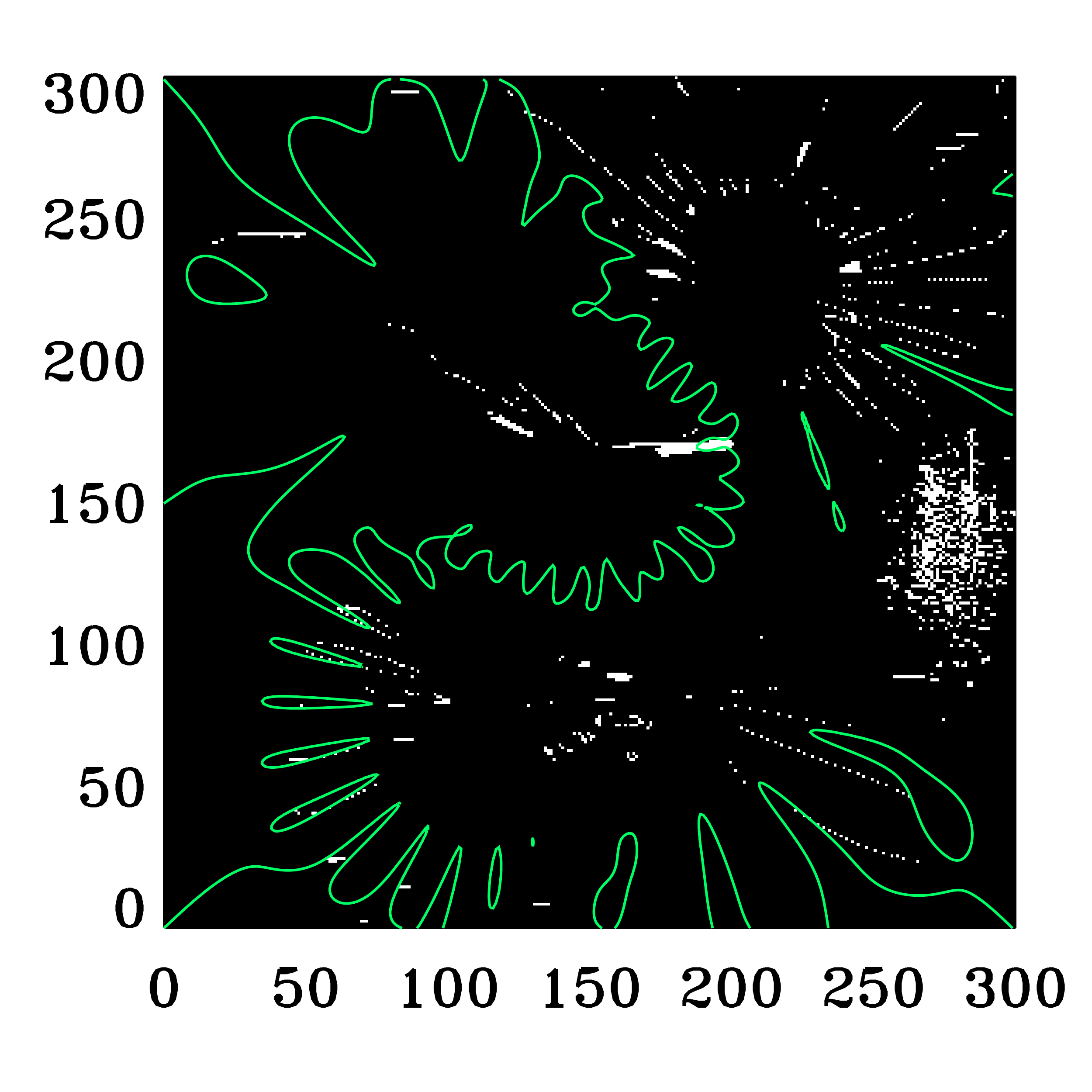} &
\includegraphics[width=0.3\textwidth]{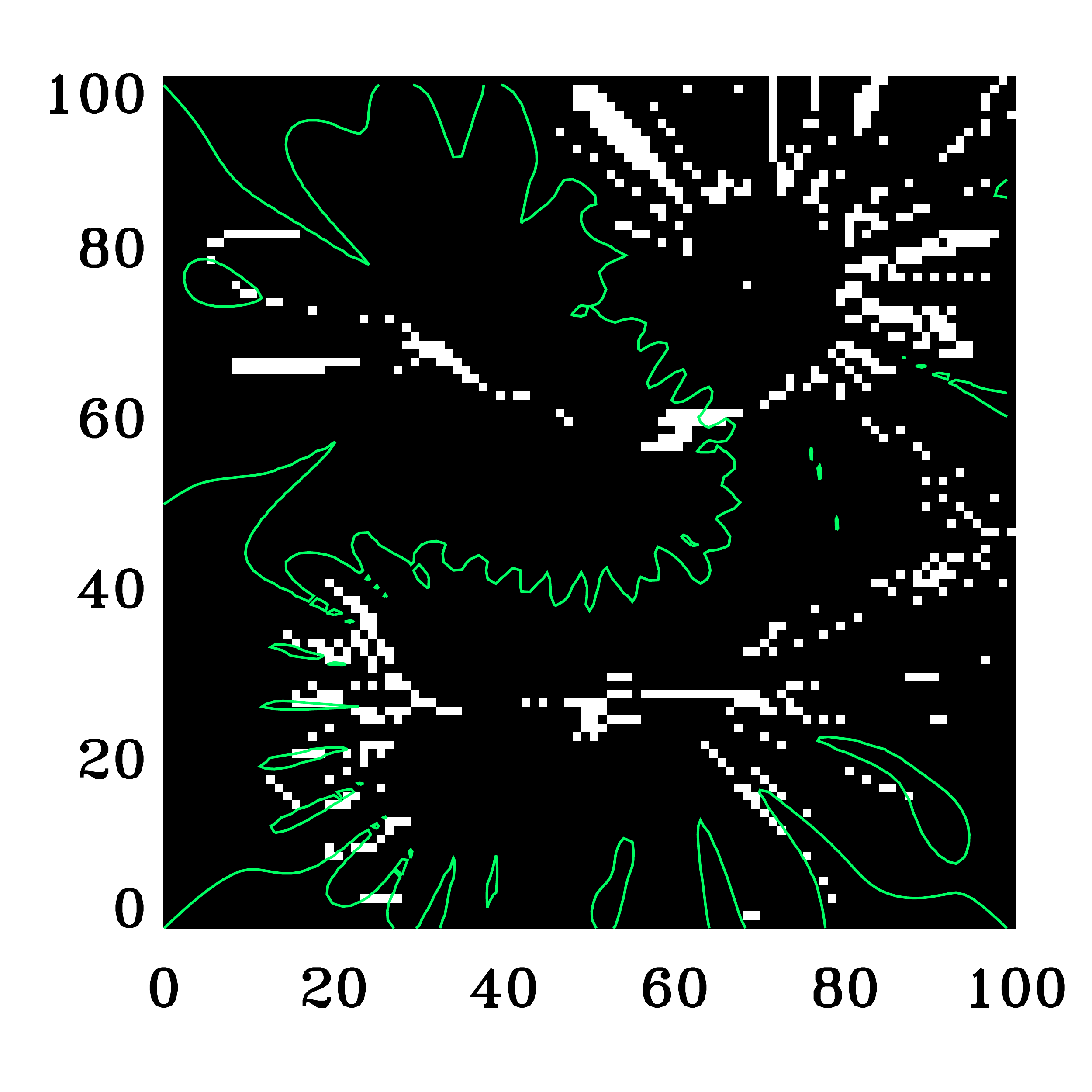}
\end{tabular}
\end{center}
\caption{Same as the lower panels of Figure~\ref{flowers1}, except the magnetogram data are smoothed before disambiguation with the global minimisation method.
}
\label{flowerssmooth}
\end{figure}

Results for the solutions that produce the lowest value of \(E\) are provided in Figures~\ref{tpdsmooth} and \ref{flowerssmooth}  and Tables~\ref{tpd_tab} and \ref{flowers_tab}.
For the multipole field positioned away from disk centre without photon noise this approach does not retrieve the correct solution at every pixel over both heights, indicating that the smoothing procedure introduces some errors when noise is absent. We have confirmed that this is also the case for both of the noise-free test cases examined in \cbl{}.
For the cases with photon noise we find that there is some improvement in the retrieved solutions in some of the regions with a weak transverse component of the magnetic field around the edges of the field of view (especially for the high-noise case).
However, there are regions where the incorrect solution is retrieved for the case with smoothing where the correct solution is retrieved for the case where the data is not smoothed before disambiguation.
For the tests of limited spatial resolution we find that the results retrieved by the method that smoothes the data before disambiguation are generally worse than those from the method that does not smooth the data (this problem tends to get worse as the number of pixels in the smoothing operator gets larger).
Given these mixed results, we conclude that the smoothing the data before disambiguation does not substantially improve the performance of the global minimisation method.

\subsection{The Global Minimisation Method with Additional Constraints}

Several methods that are closely related to the global minimisation method incorporate into the minimisation problem an additional smoothness constraint, which is based on the current density  (\opencite{1994SoPh..155..235M}; \opencite{2006SoPh..237..267M}; \opencite{2009ASPC..415..365L}; \opencite{2009SoPh..260...83L}).
In this section we examine a couple of approaches that incorporate different additional constraints into the global minimisation method.
One of the main reasons for using the divergence-free property of magnetic fields to resolve the azimuthal ambiguity is to avoid unrealistic assumptions.
We acknowledge that incorporating a constraint into the minimisation problem  (in addition to the divergence of the field) may not be not entirely consistent with this philosophy.
However, in this section we will show that the inclusion of an additional constraint can improve the performance of the global minimisation method, especially when photon noise is present in the data.

\subsubsection{Additional Constraints Involving the Current Density}

Here we consider an approach that assumes that the correct configuration of azimuthal angles over the field of view is the one that corresponds to the minimum of

\newcommand{\ljz}{\lambda_{J_z}}

\begin{equation}
E_{J_z} = \sum_{i=1}^{n_x}  \sum_{j=1}^{n_y} \left[ | ( \grad \vdot \B )_{i,j,k} | +  | ( \grad \vdot \B )_{i,j,{k+1}} |  + \mu  \ljz \left(  | ( \jzh )_{i,j,k} | +  | ( \jzh )_{i,j,{k+1}} | \right) \right]
\label{eds2}
\end{equation}

\noindent
where \(\ljz\) is a dimensionless parameter, 
\(\mu\) is the magnetic permeability, and
\( ( \jzh )_{i,j,k} \) is the approximation for the vertical heliographic component of the current density at pixel $(i, j, k)$.
This approach is similar to that taken by the minimum energy method ME0 (\opencite{2009ASPC..415..365L}; \opencite{2009SoPh..260...83L}), where the vertical current density is included in the minimisation problem as an additional constraint.
In terms of observable quantities (\myie derivatives of the image components of the field with respect to $\xuh$, $\yuh$, and $\zui$), the exact expression for  \( \jzh \) is

\[
\mu \jzh  =   a_{21} \frac{\partial \bxi}{\partial \xuh}
          + a_{22} \frac{\partial \byi}{\partial \xuh}
          + a_{23} \frac{\partial \bzi}{\partial \xuh}
    - a_{11} \frac{\partial \bxi}{\partial \yuh}
    - a_{12} \frac{\partial \byi}{\partial \yuh}
    - a_{13} \frac{\partial \bzi}{\partial \yuh} \, ,
\]

\noindent
which does not require the line-of-sight derivative of any of the components of the  magnetic field.
We approximate the horizontal heliographic derivatives using measurements at one height in the same manner as in \( ( \grad \vdot \B )_{i,j,k} \).
Consequently, the approximation \( ( \jzh )_{i,j,k} \) only depends on measurements at the height corresponding to the index \(k\).

\begin{figure}[ht]
\begin{center}
\begin{tabular}{c@{\hspace{0.005\textwidth}}c@{\hspace{0.005\textwidth}}c}
\includegraphics[width=0.3\textwidth]{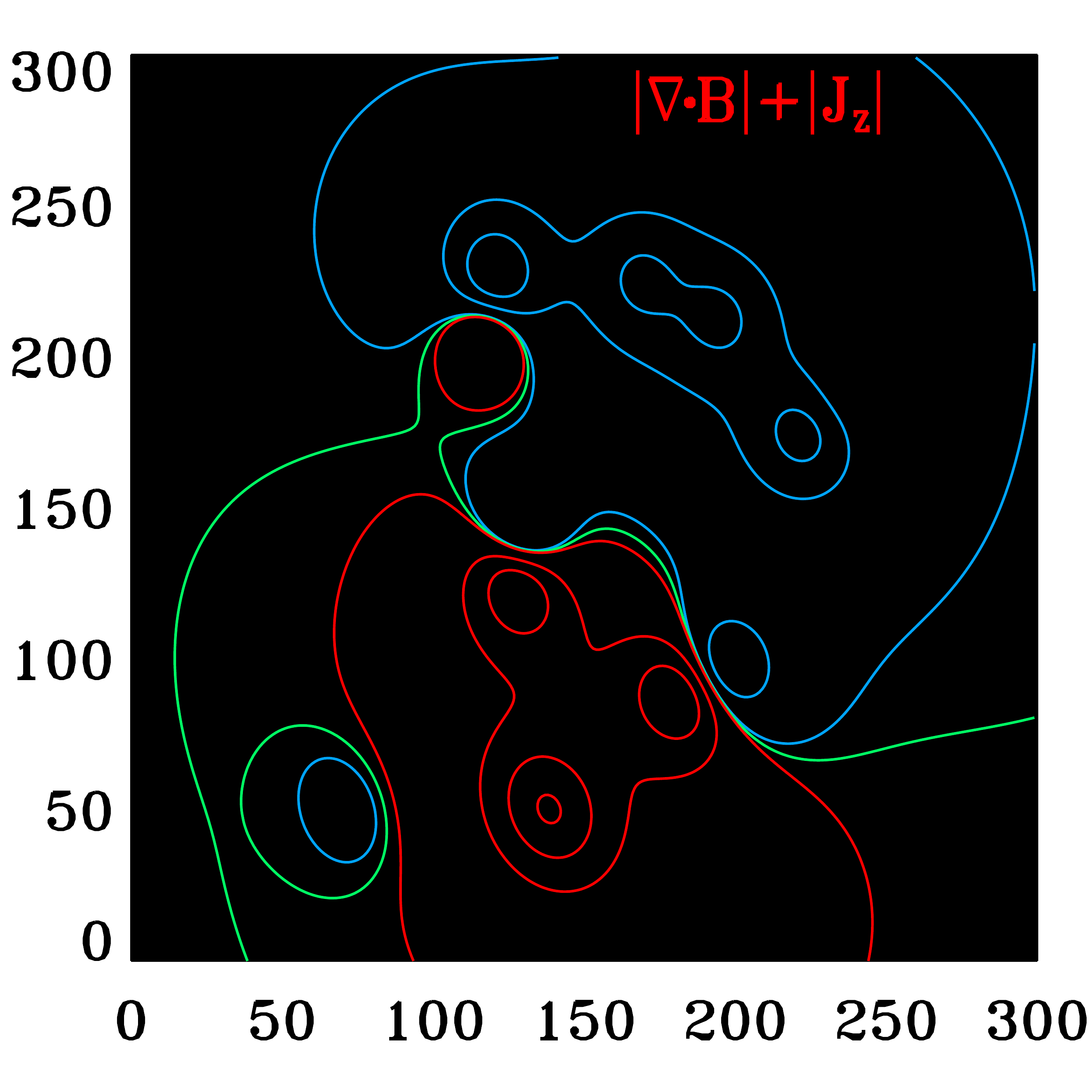} &
\includegraphics[width=0.3\textwidth]{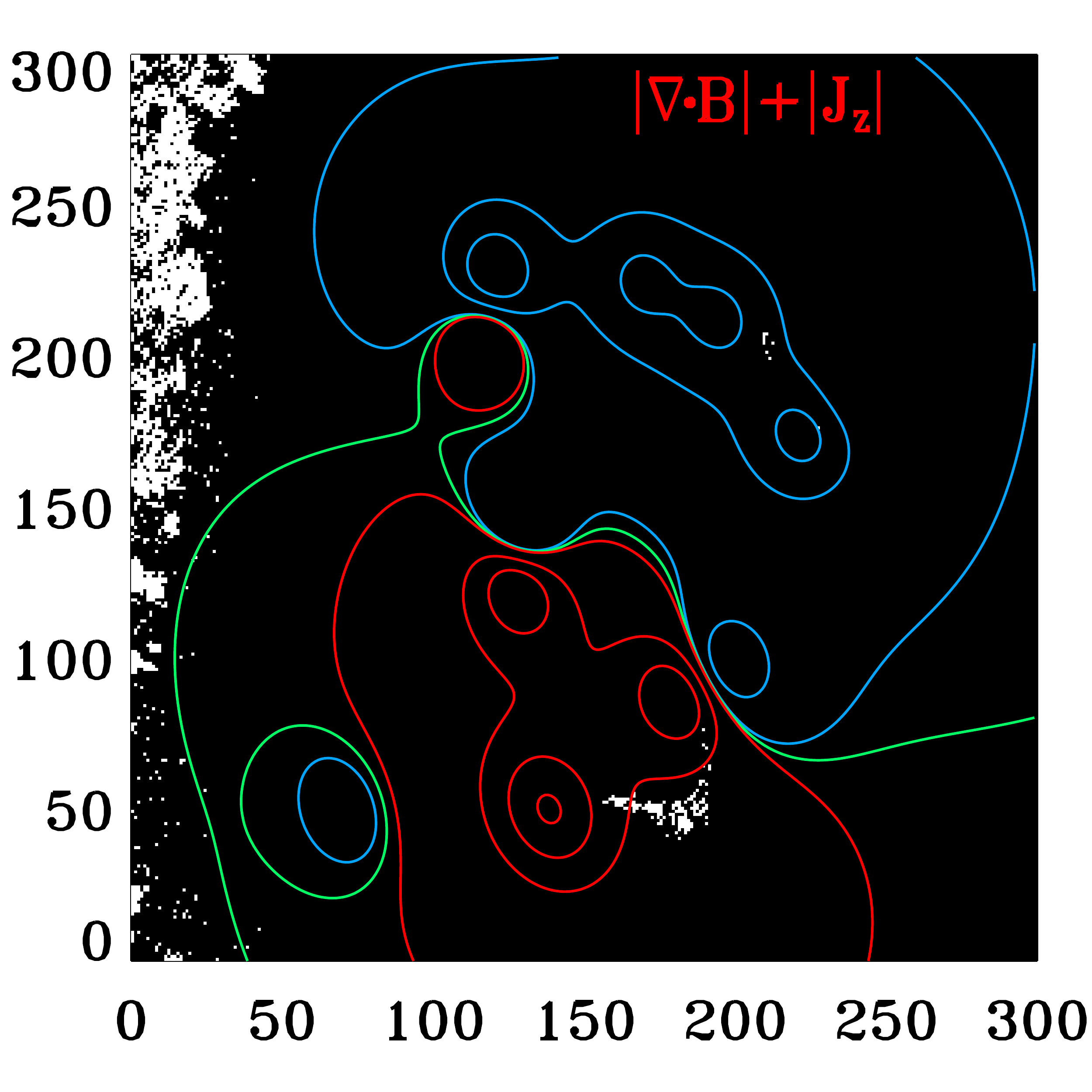} &
\includegraphics[width=0.3\textwidth]{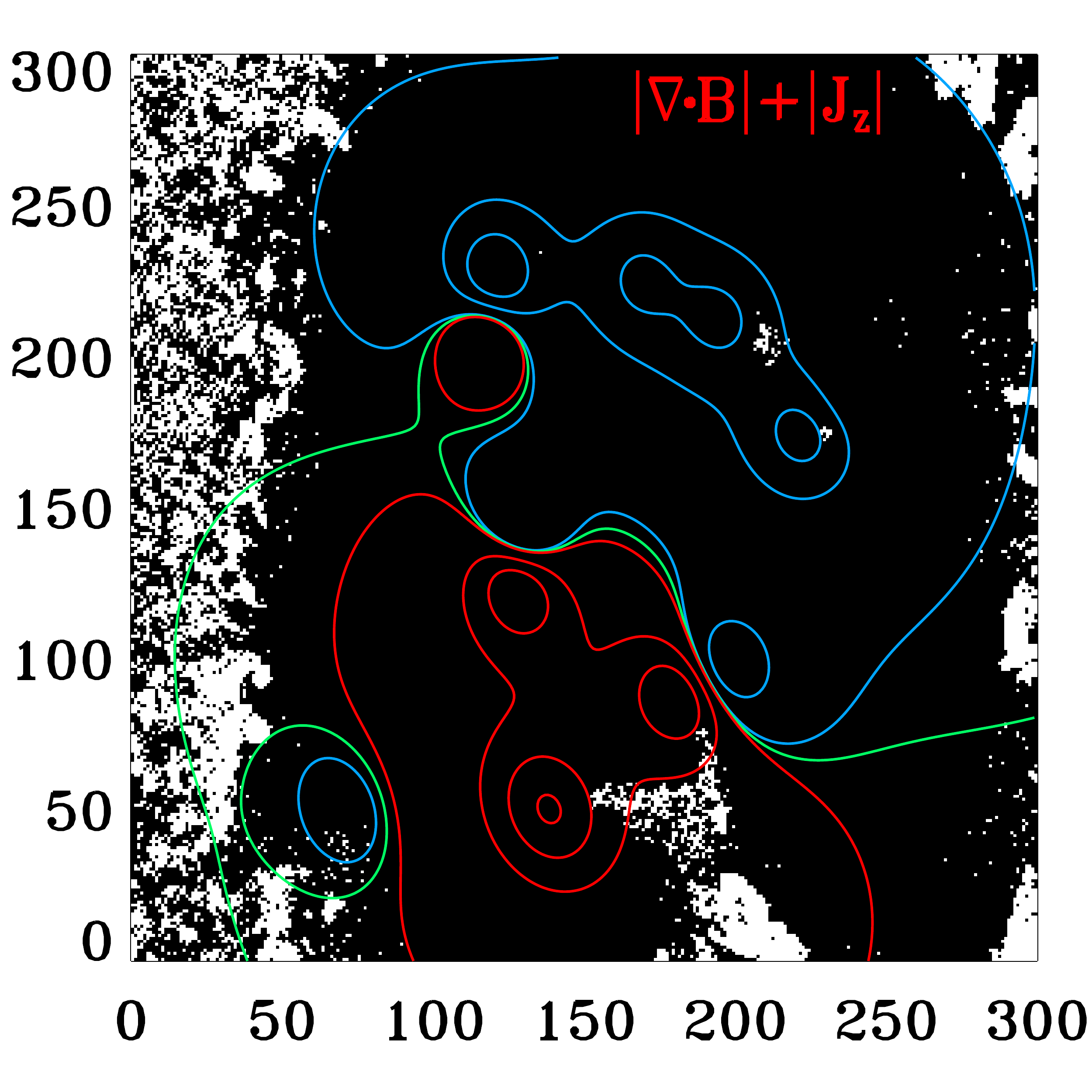}
\end{tabular}
\end{center}
\caption{
Same as the lower panels of Figure~\ref{tpd1} except \( E_{J_z} \) (see Equation~(\ref{eds2})) is minimised.
}
\label{tpd2}
\end{figure}

\begin{figure}[ht]
\begin{center}
\begin{tabular}{c@{\hspace{0.005\textwidth}}c@{\hspace{0.005\textwidth}}c}
\includegraphics[width=0.3\textwidth]{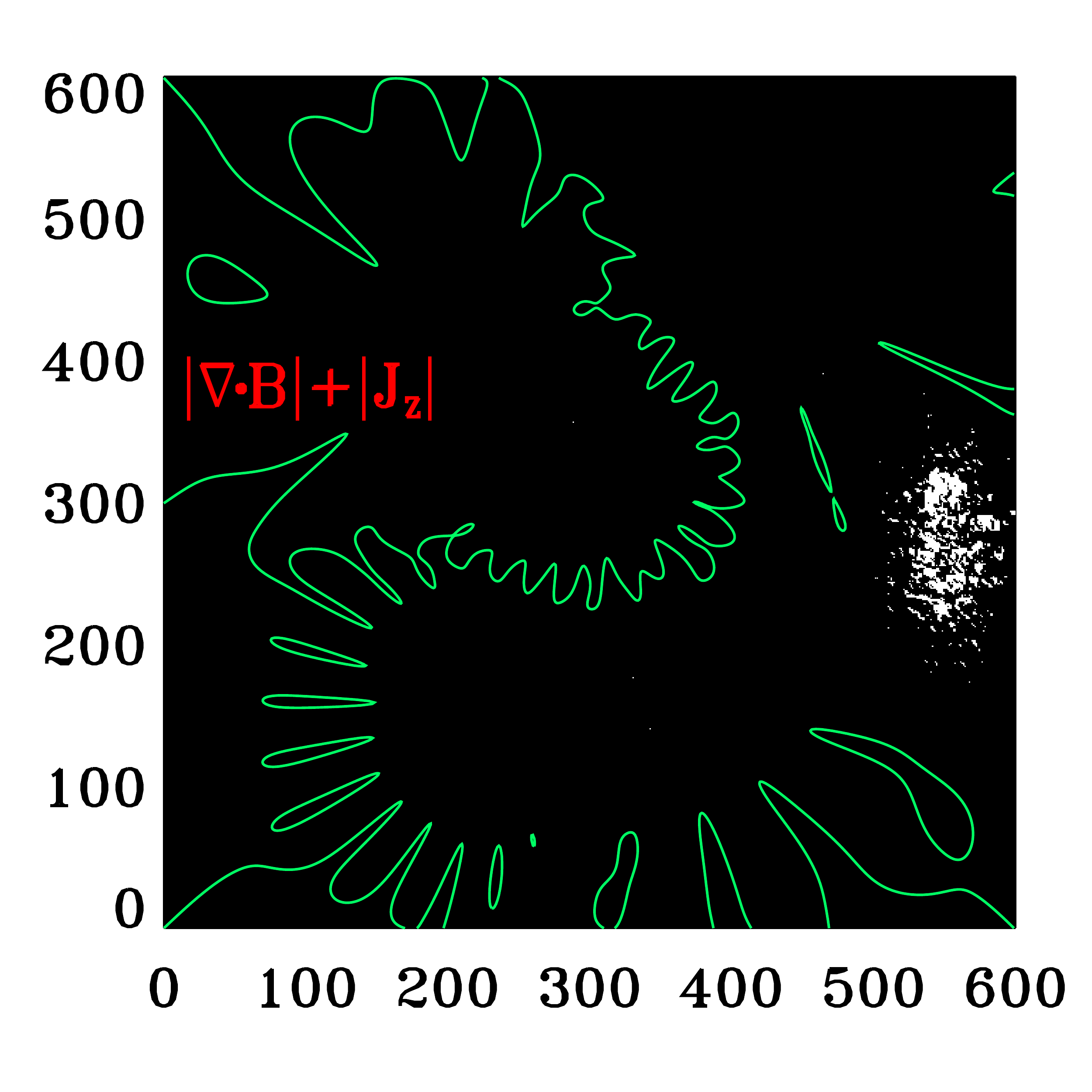} &
\includegraphics[width=0.3\textwidth]{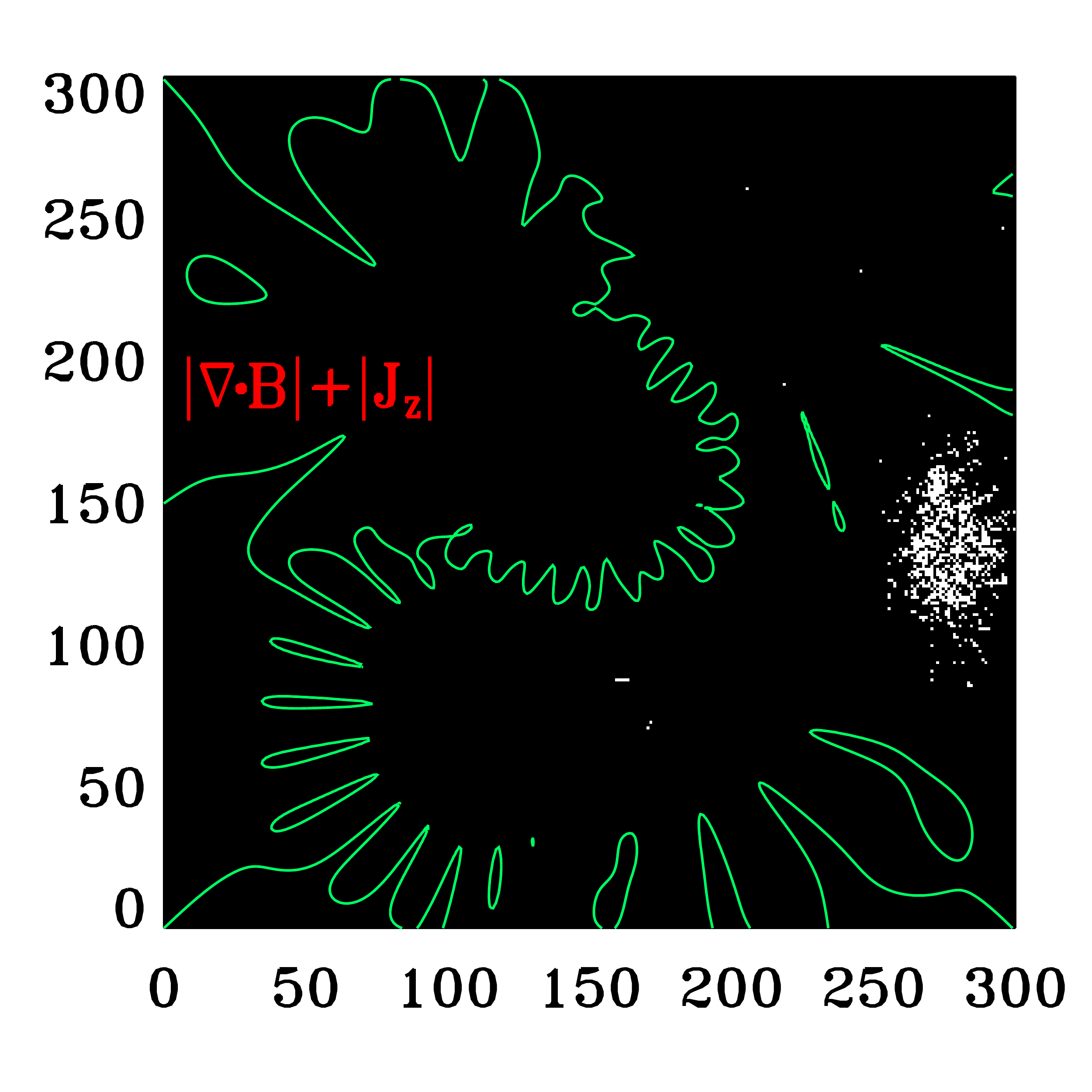} &
\includegraphics[width=0.3\textwidth]{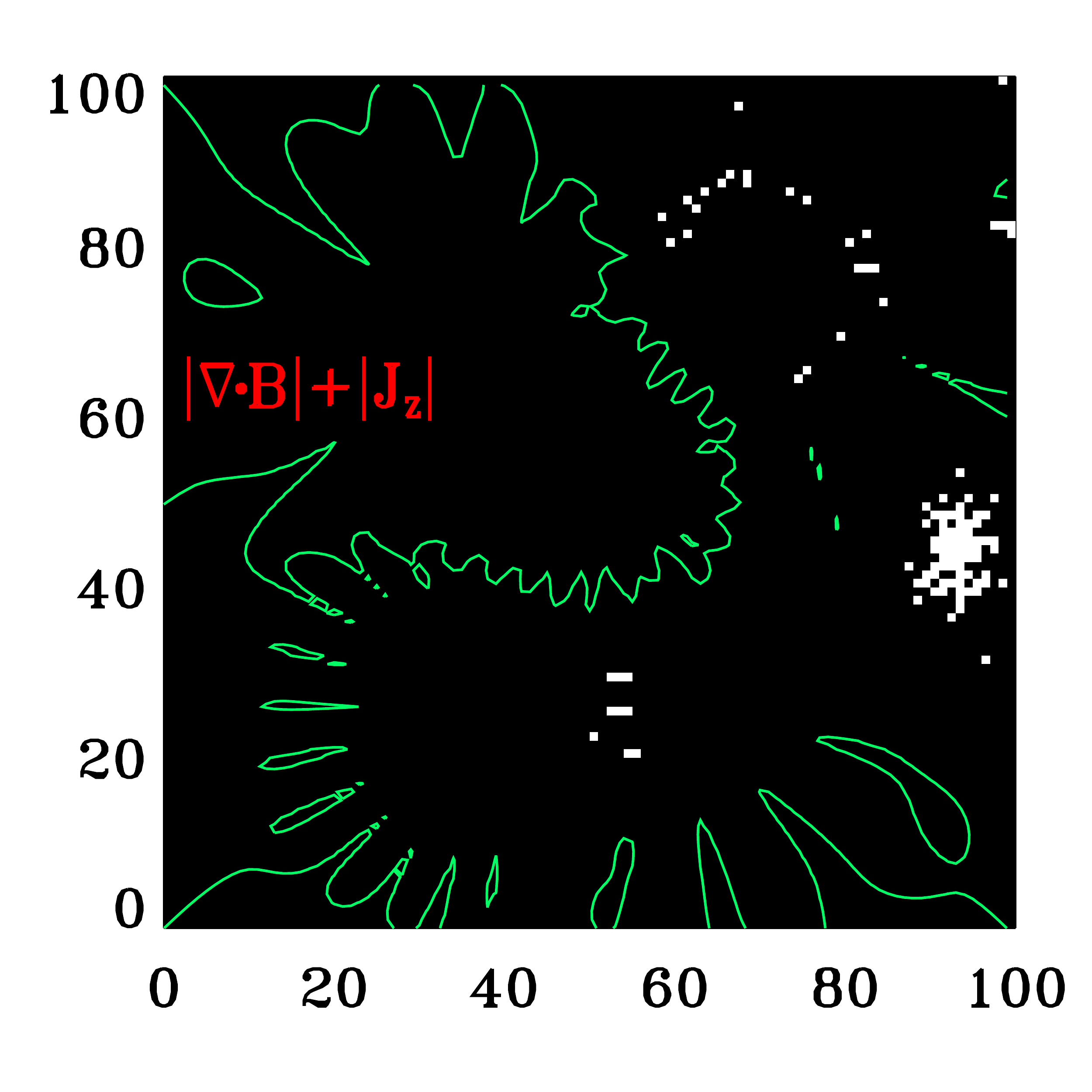}
\end{tabular}
\end{center}
\caption{Same as the lower panels of Figure~\ref{flowers1} except \( E_{J_z} \) (see Equation~(\ref{eds2})) is minimised.
}
\label{flowers2}
\end{figure}

The optimal value of parameter \(\ljz\) may be problem dependent.
We have tested several different values of \(\ljz\) in the range from \(\ljz=0\)  to \(\ljz=10\).
For the set of available test cases we find that values for \(\ljz\)  of approximately unity consistently produce the best results; we use \(\ljz =1\) in what follows.
To find the configuration of azimuthal angles corresponding to the minimum of \(E_{J_z}\) we use the same simulated annealing algorithm that was used for the global minimisation method described in Section~\ref{sec_global}.
Results for the solutions that produce the lowest value of \(E_{J_z}\) are provided in Figures~\ref{tpd2} and \ref{flowers2}  and Tables~\ref{tpd_tab} and \ref{flowers_tab}.

For the multipole field positioned away from disk centre without photon noise this approach retrieves the correct solution at every pixel over both heights; we have confirmed that this is also the case for both of the noise-free test cases examined in \cbl{}.
As expected, as the level of photon noise increases the quality of the results produced by this method decreases (Figure~\ref{tpd2} and Table~\ref{tpd_tab}).
However, for the noise-added cases the results retrieved by this approach  are  better than those retrieved by the global minimisation method that minimises \(|\grad \vdot \B|\) only. This is most pronounced in the regions with a weaker transverse component of the magnetic field (for example, note the trend in \( \mathcal{M}_{B_\perp  > 100~\rm{G}} \) in Table~\ref{tpd_tab}; see also Figure~\ref{mtabplot}).
For the noise-added cases we find that \(E_{J_z}\) retrieved by the minimisation algorithm is less than that for the answer, indicating that the assumption made by this approach is not correct for these test cases over the entire field of view.

For the tests of limited spatial resolution the underlying magnetic field is potential, and, therefore, minimising the current density is somewhat justified in this case.
However, calculation of the vertical component of the current density requires the approximation of horizontal heliographic derivatives.
Consequently, unresolved structure can pose a challenge for this approach (Figure~\ref{flowers2} and  Table~\ref{flowers_tab}).
In the plage region at higher spatial resolution we find that the results are slightly better than those from  the global minimisation method that minimises \(|\grad \vdot \B|\) only. 
Away from the plage region, for the higher resolution test cases, there is not much difference between the results retrieved by the two approaches.
At low resolution (pixel size \(0.9''\))  in and around areas with significant unresolved structure (see Figure~\ref{fbin}) the results retrieved by this approach  are better than those retrieved for the approach that minimises \(|\grad \vdot \B|\) only.
For the test cases with pixel sizes \( 0.15'', 0.3''\) and \( 0.9'' \) we again find that \(E_{J_z}\) retrieved by the minimisation algorithm is less than that for the answer.

We now consider an approach with an additional constraint that involves all three components of the current density vector (\opencite{1994SoPh..155..235M}; \opencite{2006SoPh..237..267M}).
In this case, the correct configuration of azimuthal angles over the field of view is assumed to be the one that corresponds to the minimum of

\newcommand{\lj}{\lambda_J}

\begin{equation}
E_J = \sum_{i=1}^{n_x}  \sum_{j=1}^{n_y} \left[ | ( \grad \vdot \B )_{i,j,k} | +  | ( \grad \vdot \B )_{i,j,{k+1}} |  + \lj \mu \left( J_{i,j,k}  +    J_{i,j,{k+1}}  \right) \right]
\label{eds6}
\end{equation}

\noindent
where 
\(\lj\) is a dimensionless parameter (we use \(\lj=1\) ), 
\( J_{i,j,k}  = \left(  ( \jxh )_{i,j,k}^2  (  +\jyh )_{i,j,k}^2 + ( \jzh )_{i,j,k}^2 \right)^{1/2} \), with \(( \jxh )_{i,j,k}\) and \(( \jyh )_{i,j,k}\) the approximations for the \( \xuh \)- and \( \yuh \)-components of the current density at pixel $(i, j, k)$, respectively.
In terms of observable quantities (\myie derivatives of the image components of the field with respect to $\xuh$, $\yuh$ and $\zui$), the exact expressions for these components of the current density are

\begin{eqnarray}
a_{33} \mu \jxh & = & - a_{21} \frac{\partial \bxi}{\partial \zui} - a_{22} \frac{\partial \byi}{\partial \zui} - a_{23} \frac{\partial \bzi}{\partial \zui} \nonumber \\
& & + a_{13} a_{21} \frac{\partial \bxi}{\partial \xuh} + a_{13} a_{22} \frac{\partial \byi}{\partial \xuh} + a_{13} a_{23} \frac{\partial \bzi}{\partial \xuh} \nonumber \\
& & + \left( a_{33} a_{31} + a_{23} a_{21} \right) \frac{\partial \bxi}{\partial \yuh} 
+ \left( a_{33} a_{32} + a_{23} a_{22} \right) \frac{\partial \byi}{\partial \yuh} 
+ \left( a_{33}^2 + a_{23}^2 \right) \frac{\partial \bzi}{\partial \yuh} \nonumber \, ,
\nonumber
\end{eqnarray}

\noindent
and

\begin{eqnarray}
a_{33} \mu \jyh & = &  a_{11} \frac{\partial \bxi}{\partial \zui} + a_{12} \frac{\partial \byi}{\partial \zui} + a_{13} \frac{\partial \bzi}{\partial \zui} \nonumber \\
& & - \left( a_{11} a_{13} + a_{31} a_{33} \right) \frac{\partial \bxi}{\partial \xuh} 
- \left( a_{13} a_{12} + a_{32} a_{33}\right) \frac{\partial \byi}{\partial \xuh} 
- \left( a_{13}^2 + a_{33}^2 \right) \frac{\partial \bzi}{\partial \xuh} \nonumber \\
& & -  a_{11} a_{23} \frac{\partial \bxi}{\partial \yuh} 
-  a_{12} a_{23} \frac{\partial \byi}{\partial \yuh} 
-  a_{13} a_{23} \frac{\partial \bzi}{\partial \yuh} \, . \nonumber
\nonumber
\end{eqnarray}

\noindent
These equations show that the computation of \(\jxh\) and \(\jyh\) generally requires the line-of-sight derivatives of all three components of the magnetic field, depending on the values of \(a_{11}\), \(a_{12}\), \(a_{21}\) and \(a_{22}\).
As stated previously, the line-of-sight derivatives of the transverse components of the field, $\partial \bxi / \partial \zui$ and $\partial \byi / \partial \zui$, depend on the ambiguity resolution at all heights used to approximate line-of-sight derivatives.
Consequently, for ambiguity-resolution algorithms based on the \(\jxh\) and \(\jyh\), the ambiguity resolution at one height affects the ambiguity resolution at the other heights used to approximate these derivatives.

\begin{figure}[ht]
\begin{center}
\begin{tabular}{c@{\hspace{0.005\textwidth}}c@{\hspace{0.005\textwidth}}c}
\includegraphics[width=0.3\textwidth]{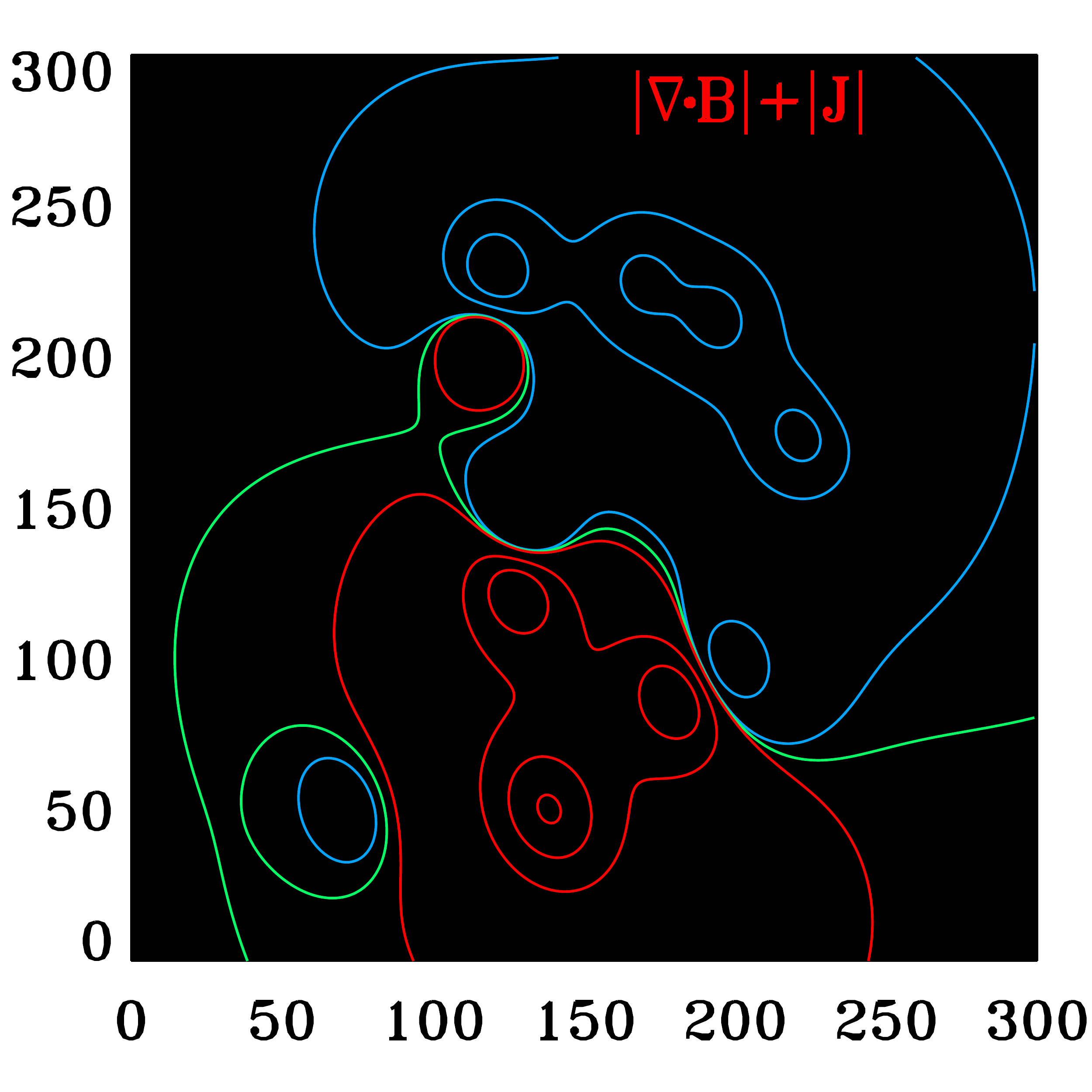} &
\includegraphics[width=0.3\textwidth]{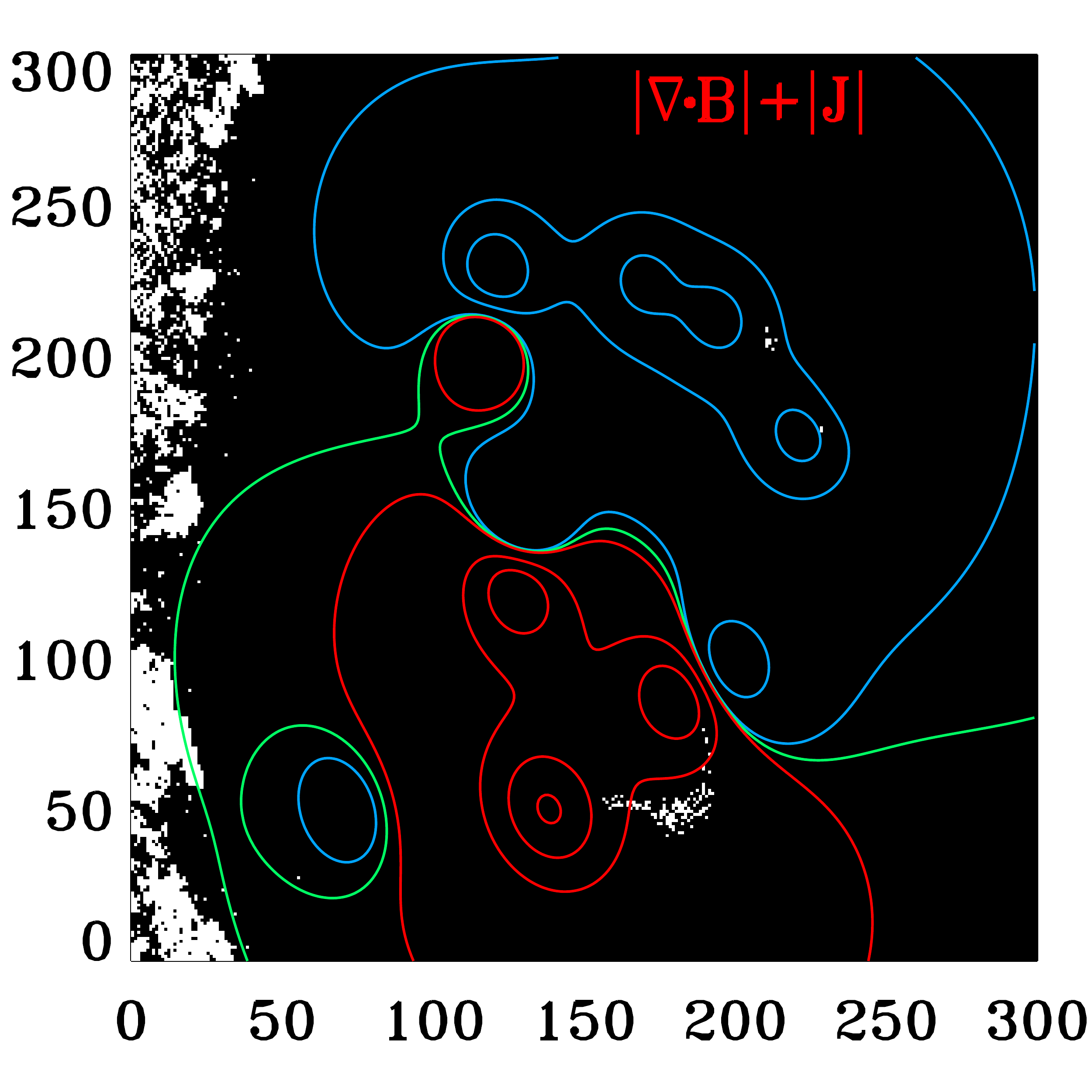} &
\includegraphics[width=0.3\textwidth]{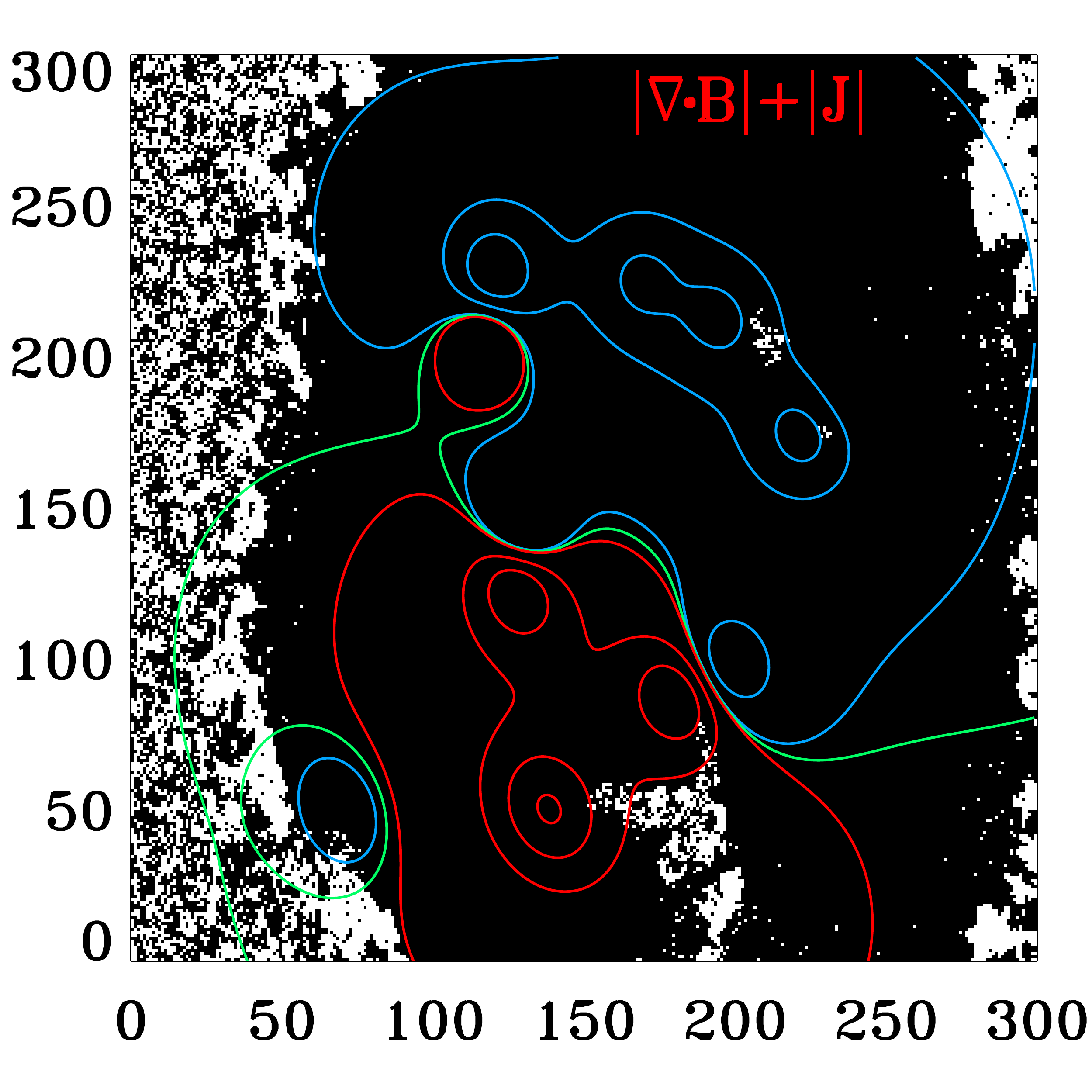}
\end{tabular}
\end{center}
\caption{
Same as Figure~\ref{tpd2} except \( E_{J} \) (see Equation~(\ref{eds6})) is minimised.
}
\label{tpd6}
\end{figure}

\begin{figure}[ht]
\begin{center}
\begin{tabular}{c@{\hspace{0.005\textwidth}}c@{\hspace{0.005\textwidth}}c}
\includegraphics[width=0.3\textwidth]{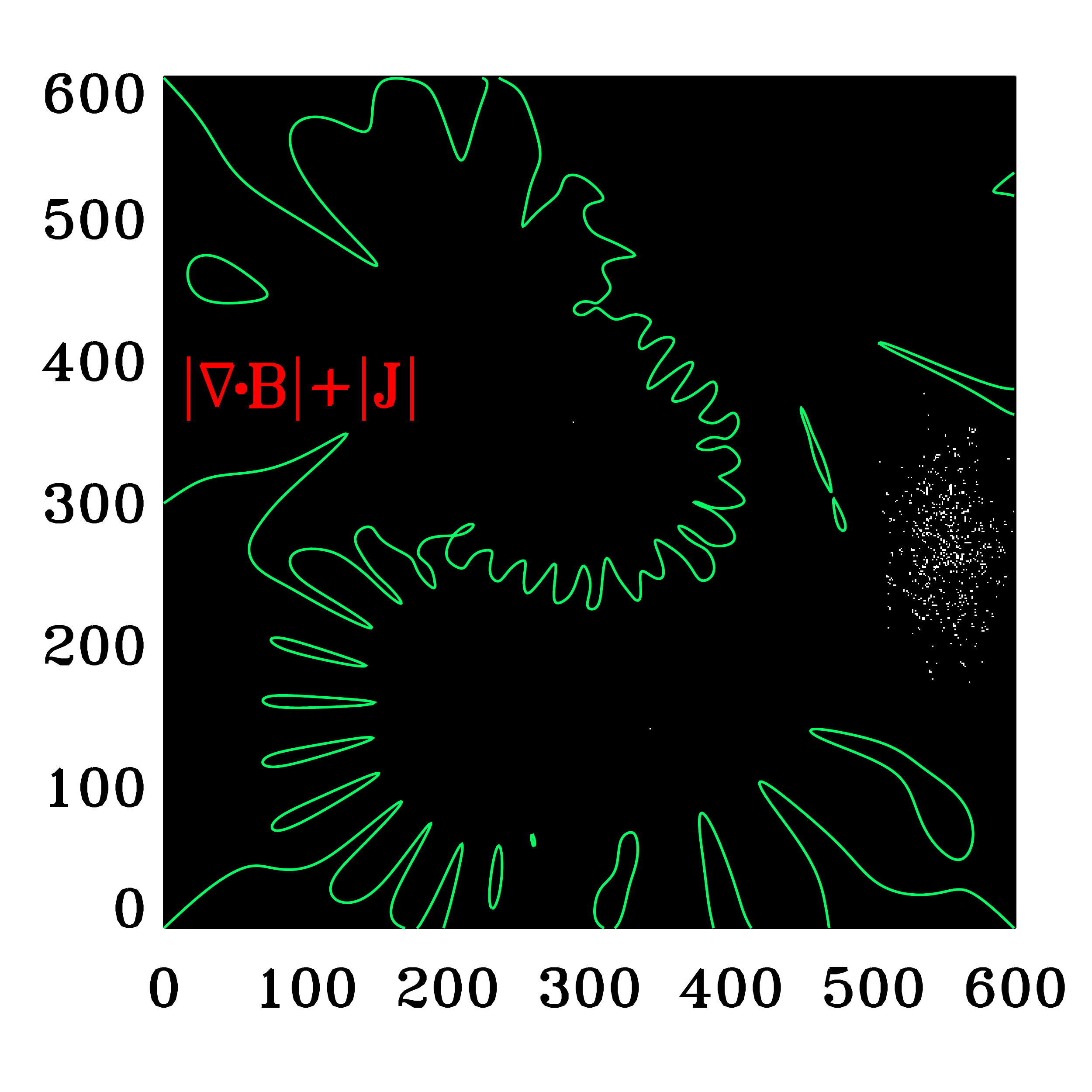} &
\includegraphics[width=0.3\textwidth]{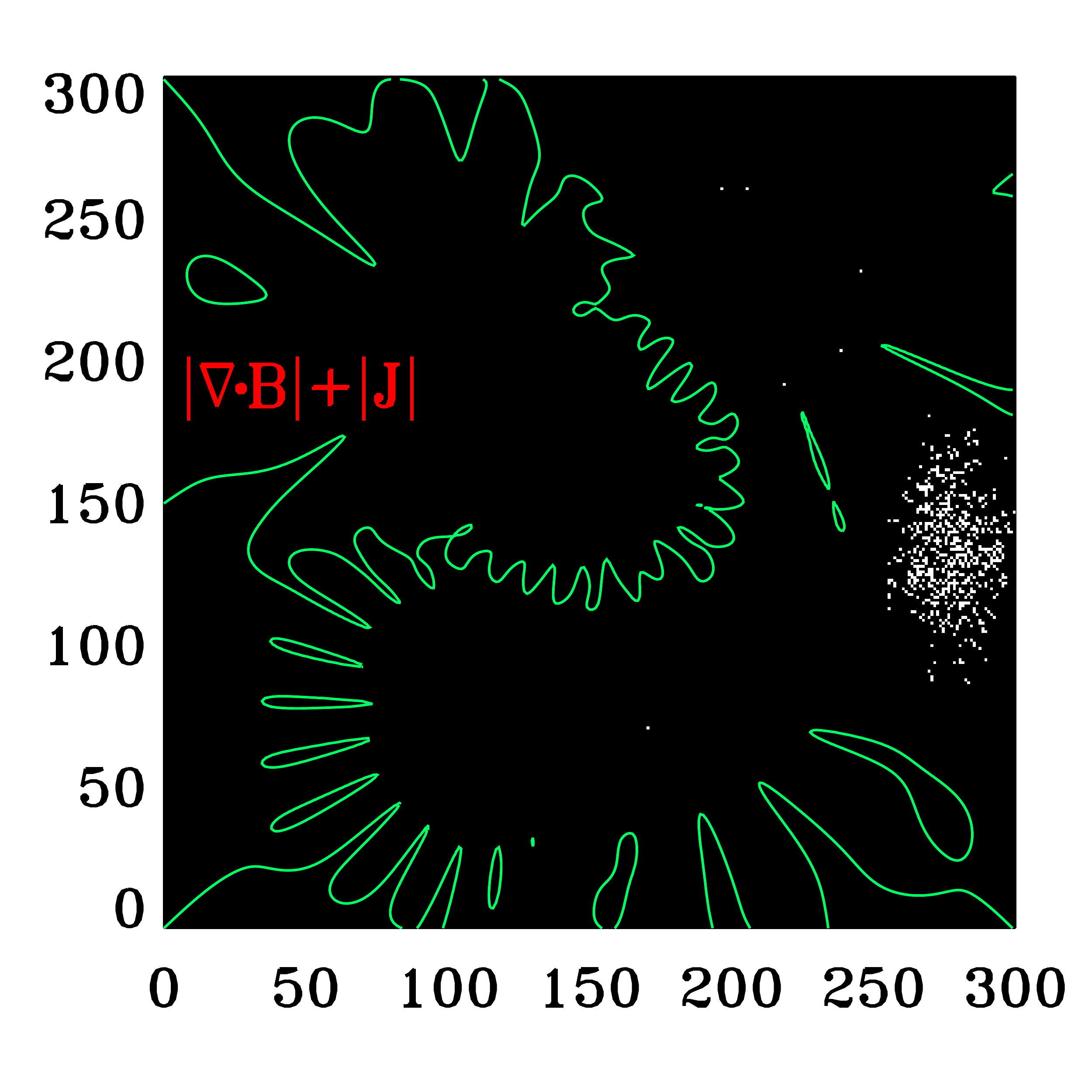} &
\includegraphics[width=0.3\textwidth]{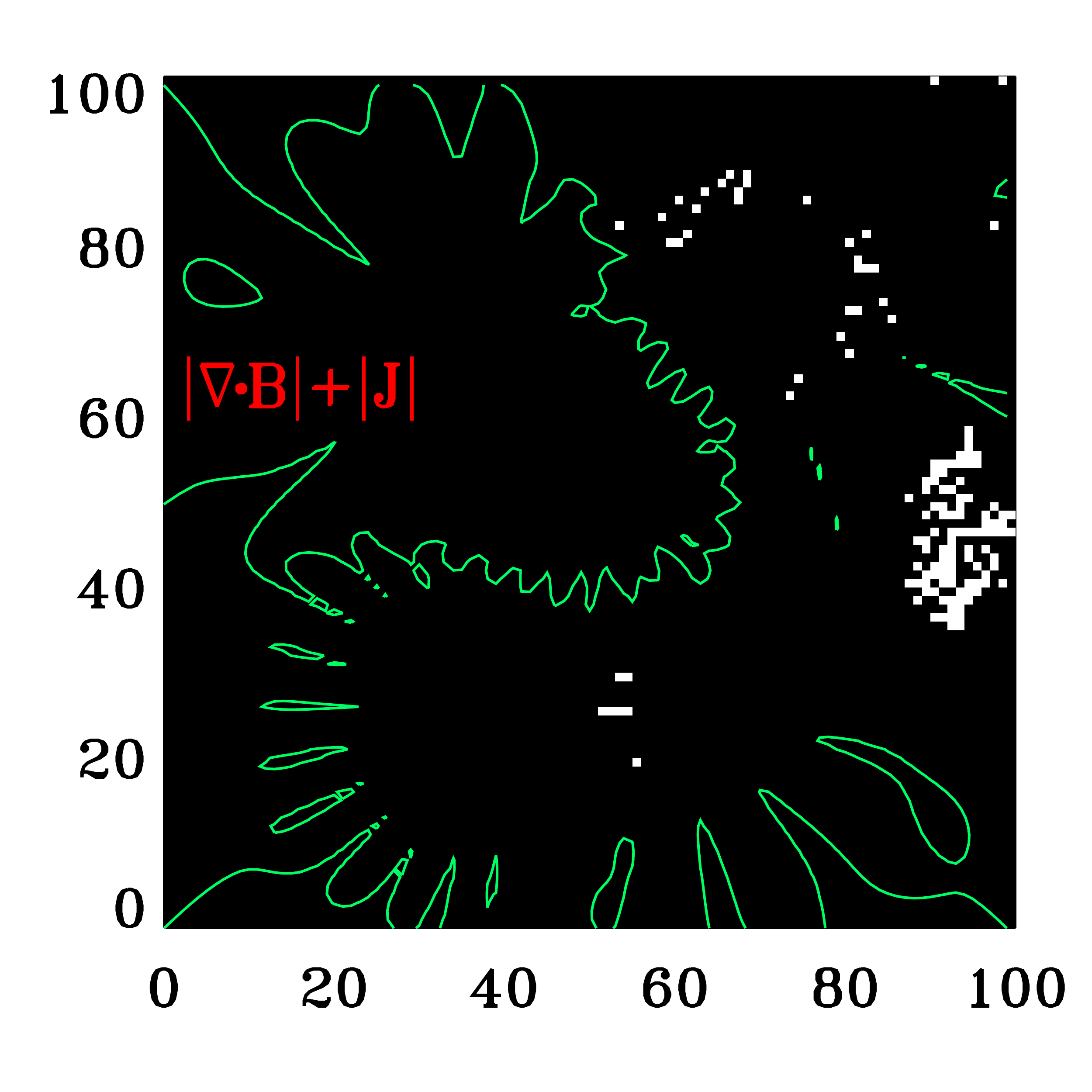}
\end{tabular}
\end{center}
\caption{Same as Figure~\ref{flowers2} except \( E_{J} \) (see Equation~(\ref{eds6})) is minimised.
}
\label{flowers6}
\end{figure}

Results for the solutions that produce the lowest value of \(E_J\) are provided in Figures~\ref{tpd6} and \ref{flowers6} and Tables~\ref{tpd_tab} and \ref{flowers_tab}.
Broadly speaking, results produced by minimising \(E_J\) are very similar to those produced by minimising \(E_{J_z}\).
We again find that \(E_{J}\) retrieved by the minimisation algorithm is less than that for the answer.
For the test cases with photon noise the results for the approach that minimises \(E_J\) are slightly worse than those for \(E_{J_z}\) (see  Figure~\ref{mtabplot}).
For the tests of limited spatial resolution, the results for the approach that minimises \(E_J\) are slightly better than those for \(E_{J_z}\) for the higher resolution cases, predominantly in the plage region.
For the lowest resolution case (with pixel size \(0.9''\)) the results for the approach that minimises \(E_J\) are slightly worse than those for \(E_{J_z}\), again, predominantly in the plage region.

\subsubsection{An Additional Constraint Based on Smoothness of the Magnetic Field}
\label{sec_eds7}

In this section we consider an approach where the additional constraint included in the minimisation problem is designed to suppress pixel-to-pixel variations in the magnetic field.
In this case the correct configuration of azimuthal angles over the field of view is assumed to be the one that corresponds to the minimum of

\newcommand{\ls}{\lambda_{\rm s}}

\begin{equation}
E_{\rm s} = \sum_{i=1}^{n_x}  \sum_{j=1}^{n_y} \left( | ( \grad \vdot \B )_{i,j,k} | +  | ( \grad \vdot \B )_{i,j,{k+1}} |  + \ls \st_{i,j,k} \right) \, ,
\label{eds3}
\end{equation}

\noindent
where \(\ls\) is a dimensionless parameter and \(\st_{i,j,k}\) is defined as

\begin{eqnarray}
\st_{i,j,k} & = & \frac{1}{d_1} \left\{ \left[ \bxi \left( i, j, k \right) - \bxi \left( i+1, j, k \right) \right]^2 + \left[ \byi \left( i, j, k \right) - \byi \left( i+1, j, k \right) \right]^2 \right\}^{1/2} \nonumber\\
& + & \frac{1}{d_2}  \left\{ \left[ \bxi \left( i, j, k \right) - \bxi \left( i, j+1, k \right) \right]^2 + \left[ \byi \left( i, j, k \right) - \byi \left( i, j+1, k \right) \right]^2 \right\}^{1/2} \nonumber\\
& + & \frac{1}{d_1} \left\{ \left[ \bxi \left( i, j, k+1 \right) - \bxi \left( i+1, j, k+1 \right) \right]^2 + \left[ \byi \left( i, j, k+1 \right) - \byi \left( i+1, j, k+1 \right) \right]^2 \right\}^{1/2} \nonumber\\
& + & \frac{1}{d_2} \left\{ \left[ \bxi \left( i, j, k+1 \right) - \bxi \left( i, j+1, k+1 \right) \right]^2 + \left[ \byi \left( i, j, k+1 \right) - \byi \left( i, j+1, k+1 \right) \right]^2 \right\}^{1/2} \nonumber\\
& + & \frac{1}{\Delta \zui} \left\{ \left[ \bxi \left( i, j, k \right) - \bxi \left( i, j, k+1 \right) \right]^2 + \left[ \byi \left( i, j, k \right) - \byi \left( i, j, k+1 \right) \right]^2 \right\}^{1/2}\, ,
\end{eqnarray}

\noindent
where \(d_1\) is the distance between the centres of the pixels at \( \left( i, j \right) \) and \( \left(  i+1, j \right) \) at fixed \(k\), and 
\(d_2\) is the distance between the centres of the pixels at \( \left( i, j \right) \) and \( \left(  i, j+1 \right) \) at fixed \(k\).
At the boundaries of the field of view $i=n_x$ and $j=n_y$ the definition of \( \st_{i,j,k}\) is modified to use only pixels within the field of view, in the same manner as the finite differencing stencil used to approximate horizontal heliographic derivatives.
The purpose of \(\st_{i,j,k}\)  is to minimise the difference between the magnetic field in neighbouring pixels, with the difference normalised by the physical distance between measurements.
It is worth noting that \( \st_{i,j,k}\) includes terms that depend on the ambiguity resolution at both of the heights \(k\) and \(k+1\) and, therefore, the ambiguity resolution produced by this approach at one height depends on the ambiguity resolution at the other height used in the comparison.

\begin{figure}[ht]
\begin{center}
\begin{tabular}{c@{\hspace{0.005\textwidth}}c@{\hspace{0.005\textwidth}}c}
\includegraphics[width=0.3\textwidth]{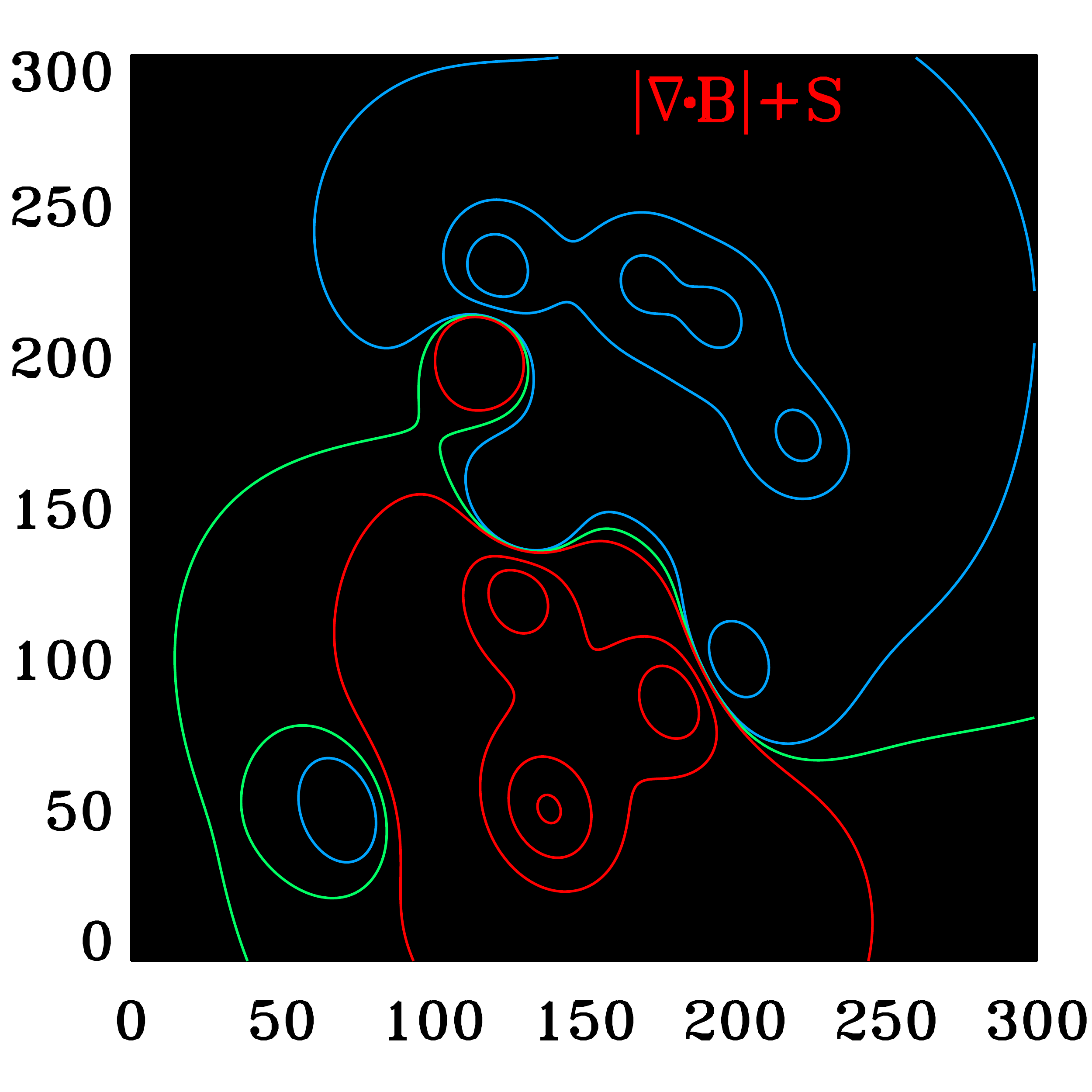} &
\includegraphics[width=0.3\textwidth]{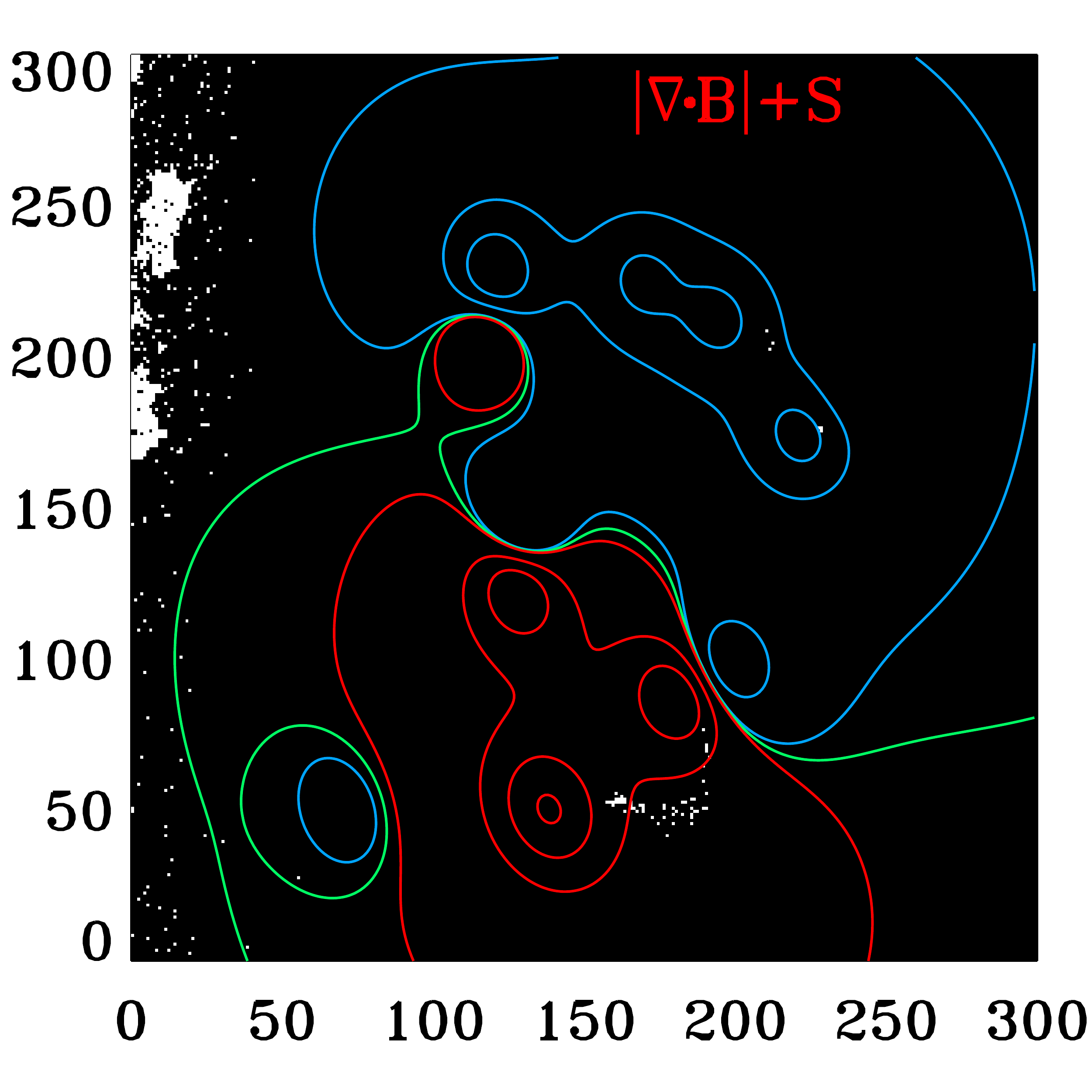} &
\includegraphics[width=0.3\textwidth]{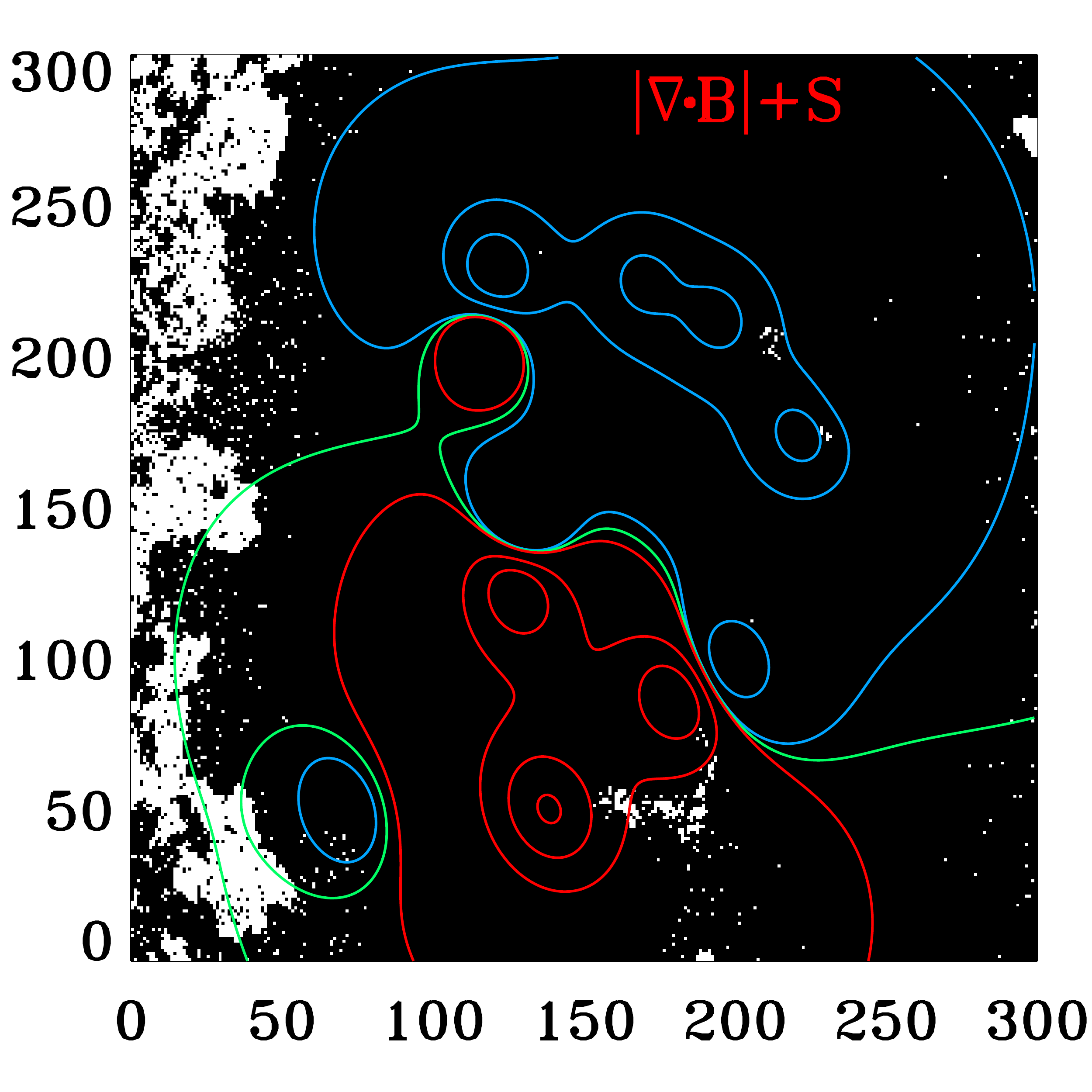}
\end{tabular}
\end{center}
\caption{Same as Figure~\ref{tpd2} except \( E_{\rm s} \) (see Equation~(\ref{eds3})) is minimised.}
\label{tpd3}
\end{figure}

\begin{figure}[ht]
\begin{center}
\begin{tabular}{c@{\hspace{0.005\textwidth}}c@{\hspace{0.005\textwidth}}c}
\includegraphics[width=0.3\textwidth]{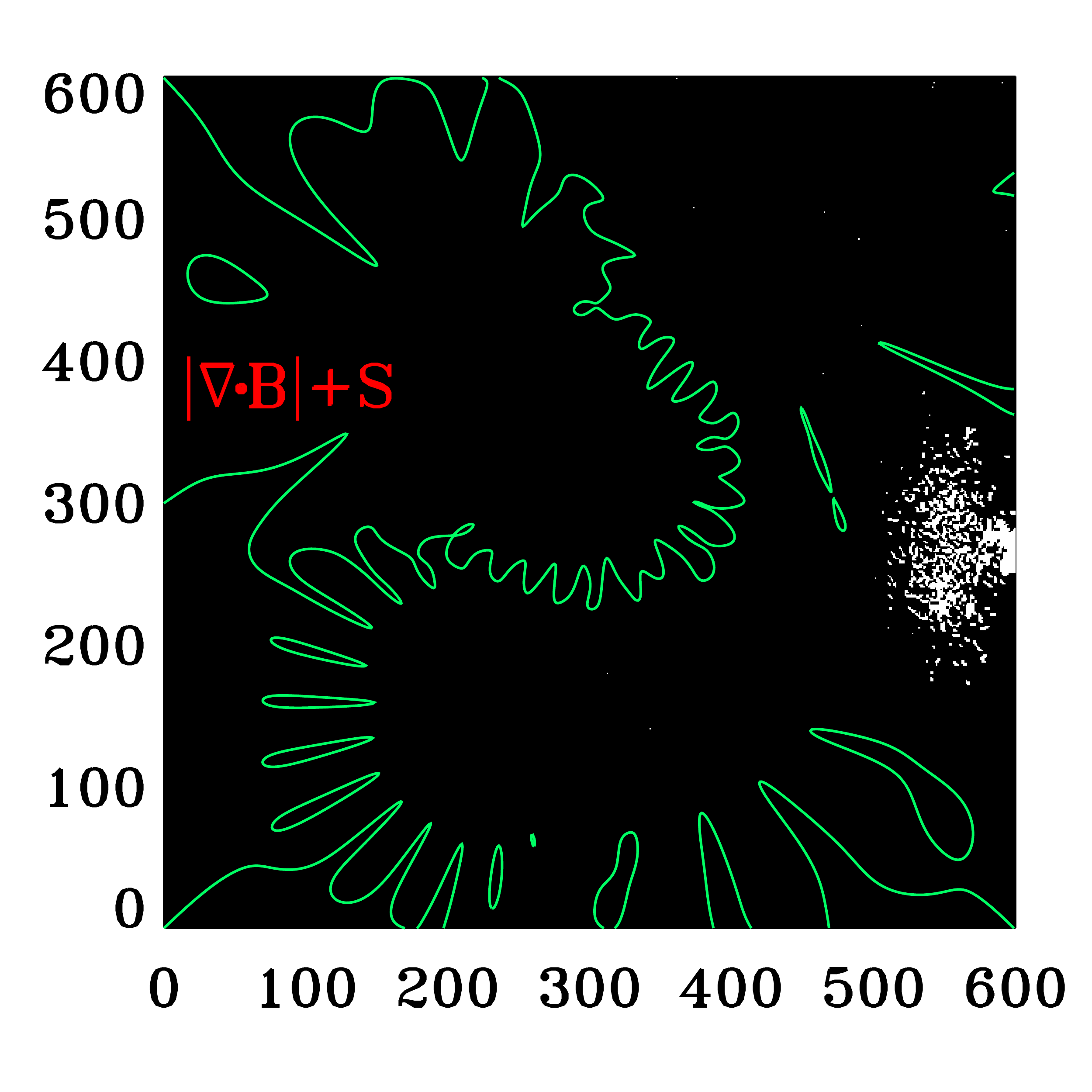} &
\includegraphics[width=0.3\textwidth]{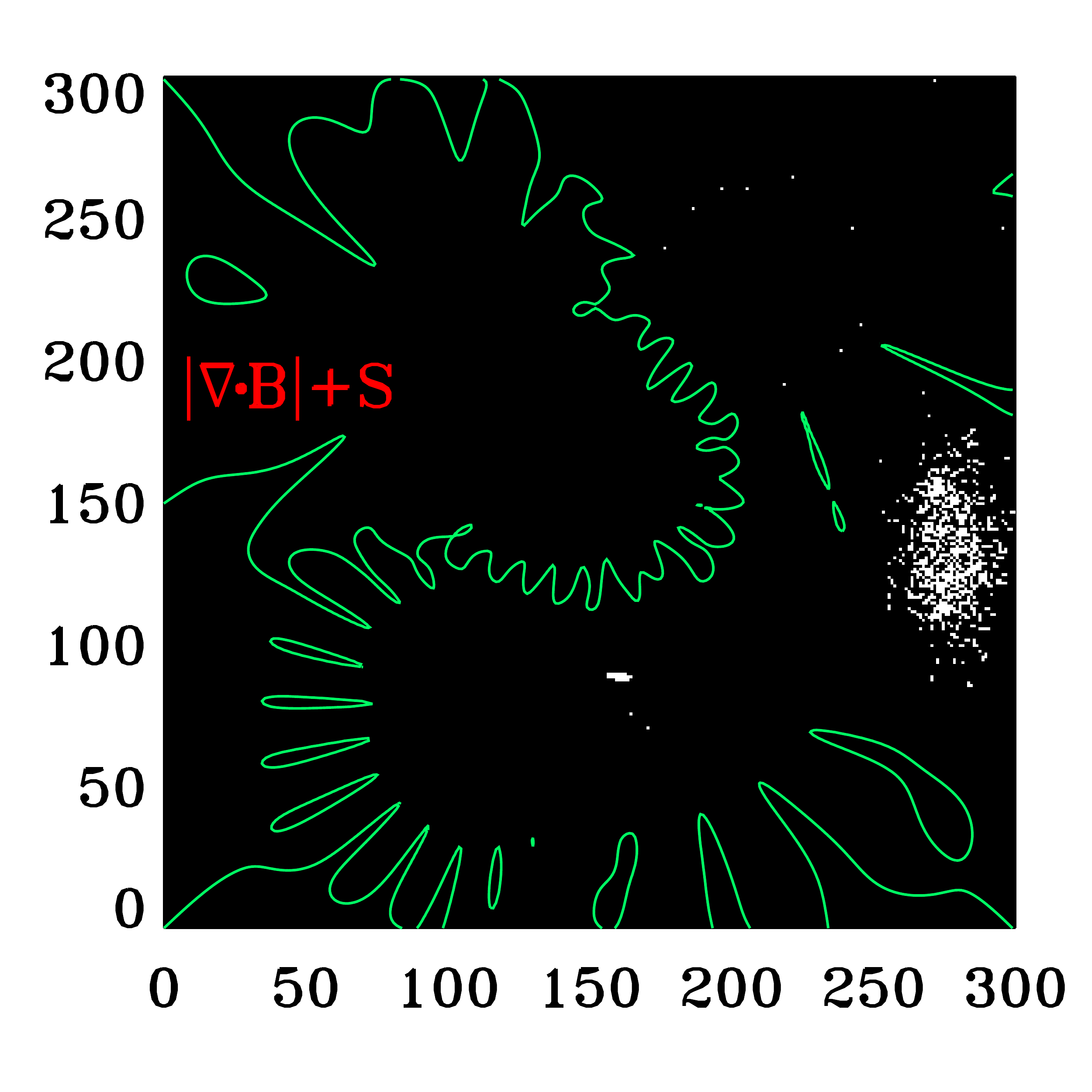} &
\includegraphics[width=0.3\textwidth]{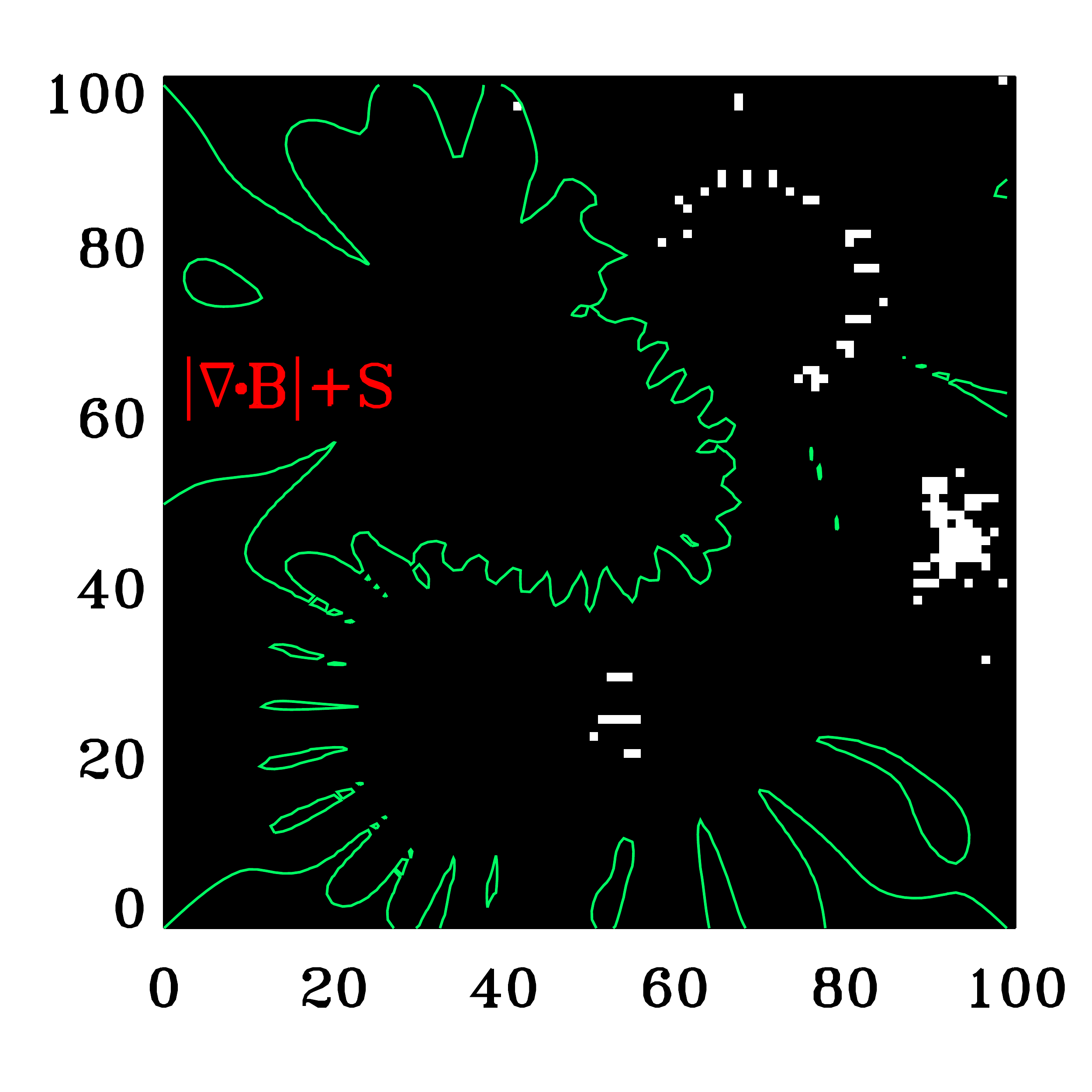}
\end{tabular}
\end{center}
\caption{Same as Figure~\ref{flowers2} except \( E_{\rm s} \) (see Equation~(\ref{eds3})) is minimised.}
\label{flowers3}
\end{figure}

We find that the optimal value of parameter \(\ls\) is problem dependent.
We have tested several values for \(\ls\) in the range \(\ls=0\) to \(\ls=20\).
We find that values of approximately \(\ls=1\) consistently produce reasonable results for the various test cases (notable exceptions are discussed below);  we set \(\ls=1\) in what follows.
The simulated annealing algorithm described in Section~\ref{sec_global} is used to search for the configuration of azimuthal angles that corresponds to the minimum of \( E_{\rm s} \).
Results for the solutions that produce the smallest value of \( E_{\rm s} \) found are given in Figures~\ref{tpd3} and \ref{flowers3} and Tables~\ref{tpd_tab} and \ref{flowers_tab}.

For the multipole field configuration without noise this approach retrieves the correct solution at each pixel over both heights: This is also the case for both of the noise-free test cases examined in \cbl{}. 
For both noise-added test cases this approach produces  better results than both the global minimisation method that minimises \(|\grad \vdot \B|\) only and that which minimises a combination of \(|\grad \vdot \B|\) and the current density (\myie \( E_{J_z} \) or \(E_{J}\)).
This is especially noticeable in regions with  \(\xui \approx 0 \), \(\xui \approx 300 \), and at \(\xui \approx 170 \) and \(\yui \approx 50 \), where the transverse component of the magnetic field is relatively weak and the signal tends to be dominated by noise (see Figure~\ref{adiff} and Figure~\ref{mtabplot}).
Again, for the noise-added test cases we find that \( E_{\rm s} \) retrieved by the minimisation algorithm is less than that for the answer.
For the noise-added test cases we also find that slightly better results than those shown in Figure~\ref{tpd3} can be achieved with this approach  in the regions with a weaker transverse component of the field with values of \(\ls\) slightly larger than unity.

Broadly speaking, for the tests of limited spatial resolution the results produced by this approach are not much different to those retrieved by the global minimisation method that minimises \(|\grad \vdot \B|\) only and that which minimises a combination of \(|\grad \vdot \B|\) and the current density.
There are some subtle differences that are worth noting for the case with pixel size \( 0.9'' \). The results produced by minimising \( E_{\rm s} \) tend to be a little better in the plage region than those produced by minimising \(E\) or \(E_{J_z}\),  but a little worse  around the flux concentration at the top, right of the field of view.
Nevertheless, it is encouraging that this approach  performs relatively well for the tests of limited spatial resolution, as the assumption that the magnetic field is smooth made by this approach may not be correct in regions with significant unresolved structure.
Again, for the test cases with pixel sizes \( 0.15'', 0.3''\) and \( 0.9'' \) we find that \( E_{\rm s} \) retrieved by the minimisation algorithm is less than that for the answer, indicating that the assumption made by this approach is indeed not correct.
For the tests of limited spatial resolution we find that the results produced by minimising  \( E_{\rm s} \) tend to get slightly worse for values of  \(\ls\)  larger than unity, especially for the case with pixel size \( 0.9'' \).

We have also tried an approach where the additional smoothness constraint in the minimisation problem is based on the second horizontal heliographic derivatives of the magnetic field (results not shown).
For the tests of Poisson noise the results produced by this approach are similar to those produced by minimising \( E_{J_z} \) but not as good as the results produced by minimising \( E_{\rm s} \).
For the tests of limited spatial resolution the results produced by the approach based on second derivatives are similar to those produced by minimising \( E_{J_z} \) or \( E_{\rm s} \).

\subsection{A Hybrid Method: the Global Minimisation Method Combined with a Smoothing Algorithm}
\label{sec_hybrid}

\begin{figure}[ht]
\begin{center}
\begin{tabular}{c@{\hspace{0.005\textwidth}}c}
\includegraphics[width=0.45\textwidth]{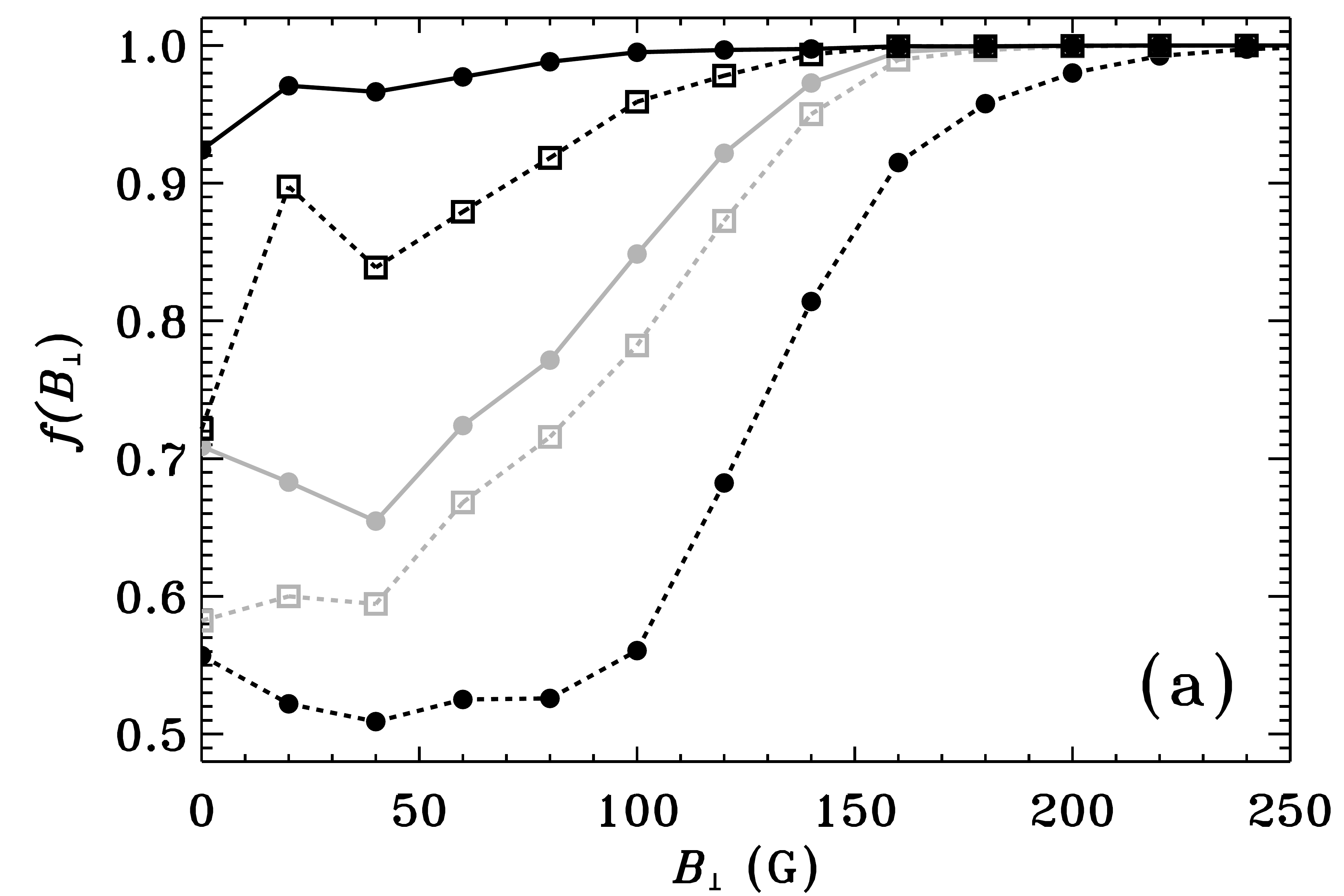} &
\includegraphics[width=0.45\textwidth]{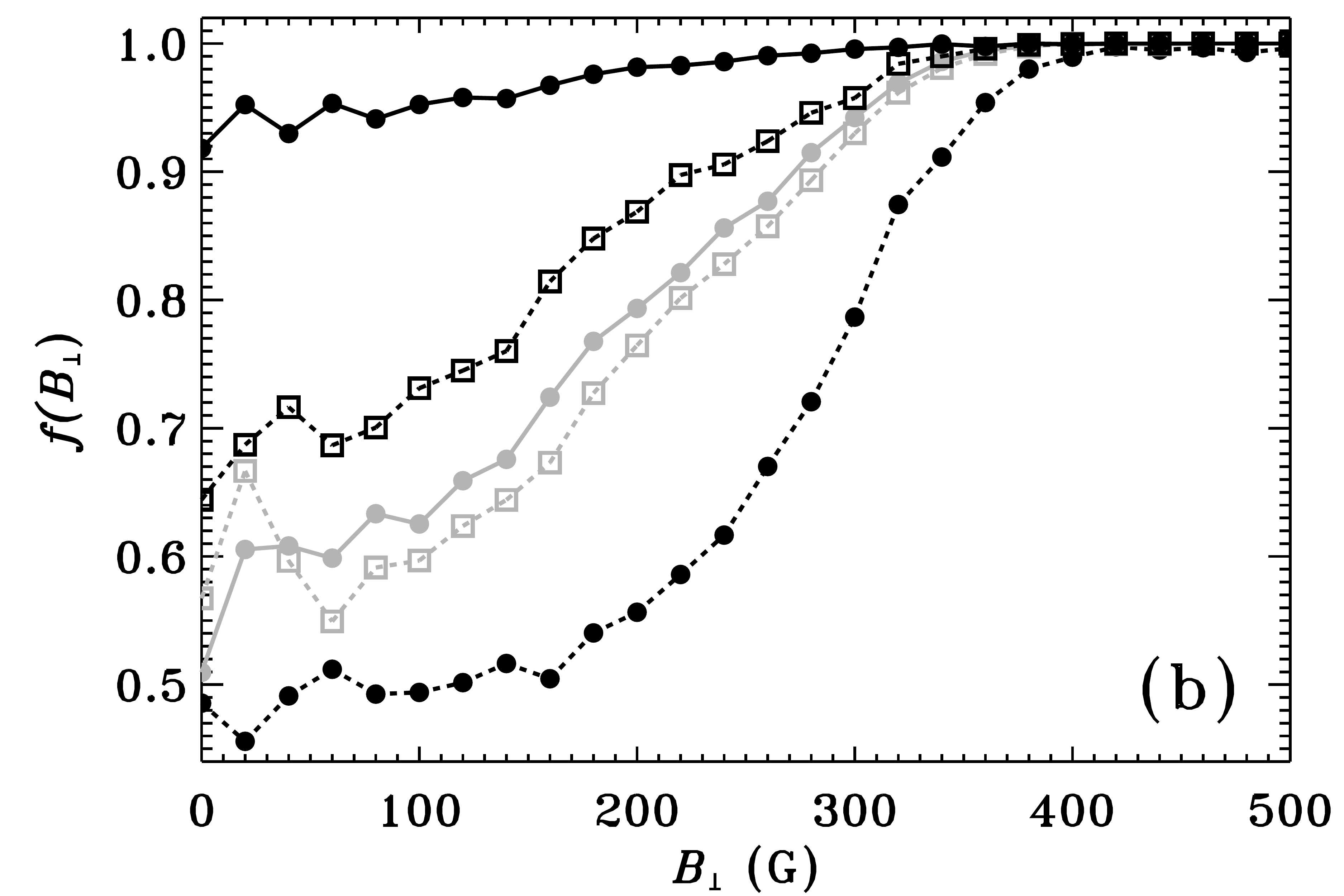}
\end{tabular}
\end{center}
\caption{
(a) \( f \left( B_\perp \right) \), the fraction of pixels correctly resolved in the interval  \( \left[ B_\perp , B_\perp + \delta B_\perp \right) \) (where \( \delta B_\perp =20\)G) as a function of the magnitude of the transverse component of the magnetic field, \( B_\perp \), for the test case with a low level of photon noise.
The black circles joined by a dashed line are for the global minimisation method that minimises \(|\grad \vdot \B|\) only.
The grey circles joined by a full line are for the global minimisation method that minimises \( E_{J_z} \) (see Equation~(\ref{eds2})).
The grey squares joined by a dashed line are for the global minimisation method that minimises \( E_J \) (see Equation~(\ref{eds6})).
The black squares joined by a dashed line are for the global minimisation method that minimises  \( E_{\rm s} \) (see Equation~(\ref{eds3})).
The black circles joined by a full line are for the hybrid method (see Section~\ref{sec_hybrid}).
(b) Same as (a) except for the high-noise test case.
Note that, in both panels, the curves joining the points are only included as a guide.
}
\label{mtabplot}
\end{figure}

Evidently, the global minimisation method that seeks to minimise a combination of the divergence and the difference between the magnetic field in neighbouring pixels can produce better results than the global minimisation method based only on the divergence.
However, none of the various global minimisation methods implemented in this article produce desirable results in regions where the transverse component of the magnetic field is strongly affected by photon noise.
Figure~\ref{mtabplot} shows the fraction of pixels correctly resolved as a function of the magnitude of the transverse component of the magnetic field for the two test cases with photon noise. 
This figure clarifies how the various methods perform in regions with a weaker transverse component of the magnetic field where the influence of photon noise is most pronounced.
Figure~\ref{mtabplot} shows that for each method as \( B_\perp \) decreases the fraction of pixels correctly resolved generally decreases, for low values of the transverse component of the magnetic field.
Figure~\ref{mtabplot} also shows that, of the global minimisation methods, the one that minimises \(|\grad \vdot \B|\) only performs worst and the one that minimises \( E_{\rm s} \) performs best. 
The global minimisation methods with an additional constraint involving the current density produce intermediate results for these test cases.
We note that in practice the choice for the additional constraint used by the global minimisation method may need to be adjusted depending on the expected nature of the field.

Given the results for the global minimisation methods shown in Figure~\ref{mtabplot} we present a hybrid method that first minimises \( E_{\rm s} \) (Equation~(\ref{eds3})), as described in Section~\ref{sec_eds7}, then revisits pixels in regions where the transverse component of the magnetic field is weak (determined by a prescribed ``threshold'') with a smoothing algorithm that does not use the divergence-free condition.
The smoothing algorithm implemented here is similar to that used by ME0 (\opencite{2009ASPC..415..365L}; \opencite{2009SoPh..260...83L}) and aims to minimise the difference between the magnetic field at a target pixel with that in a small neighbourhood of surrounding pixels.

The smoothing algorithm proceeds as follows.
The azimuthal angles in pixels considered above the prescribed threshold are fixed.
Pixels below threshold are visited according to the following criteria:
(1) Pixels with more neighbours considered above threshold are visited first.
(2) For pixels with the same number of neighbours considered above threshold, those with a stronger transverse component of the magnetic field are visited first.
Once a pixel has been examined it is treated as if it is above threshold.
After each pixel is visited the neighbourhood information for each below-threshold pixel surrounding it is updated to ensure that pixels are visited in the  order described above.
At each target pixel we select the configuration of azimuthal angles that corresponds to the smallest average difference between the magnetic field in the target pixel and that in surrounding pixels.
In doing so, the azimuthal angles in any of the surrounding pixels considered below threshold are allowed to change independently. Consequently, the results produced by this smoothing algorithm do not depend on the initial configuration of azimuthal angles for below-threshold pixels.

\begin{figure}[ht]
\begin{center}
\begin{tabular}{c@{\hspace{0.005\textwidth}}c@{\hspace{0.005\textwidth}}c}
\includegraphics[width=0.3\textwidth]{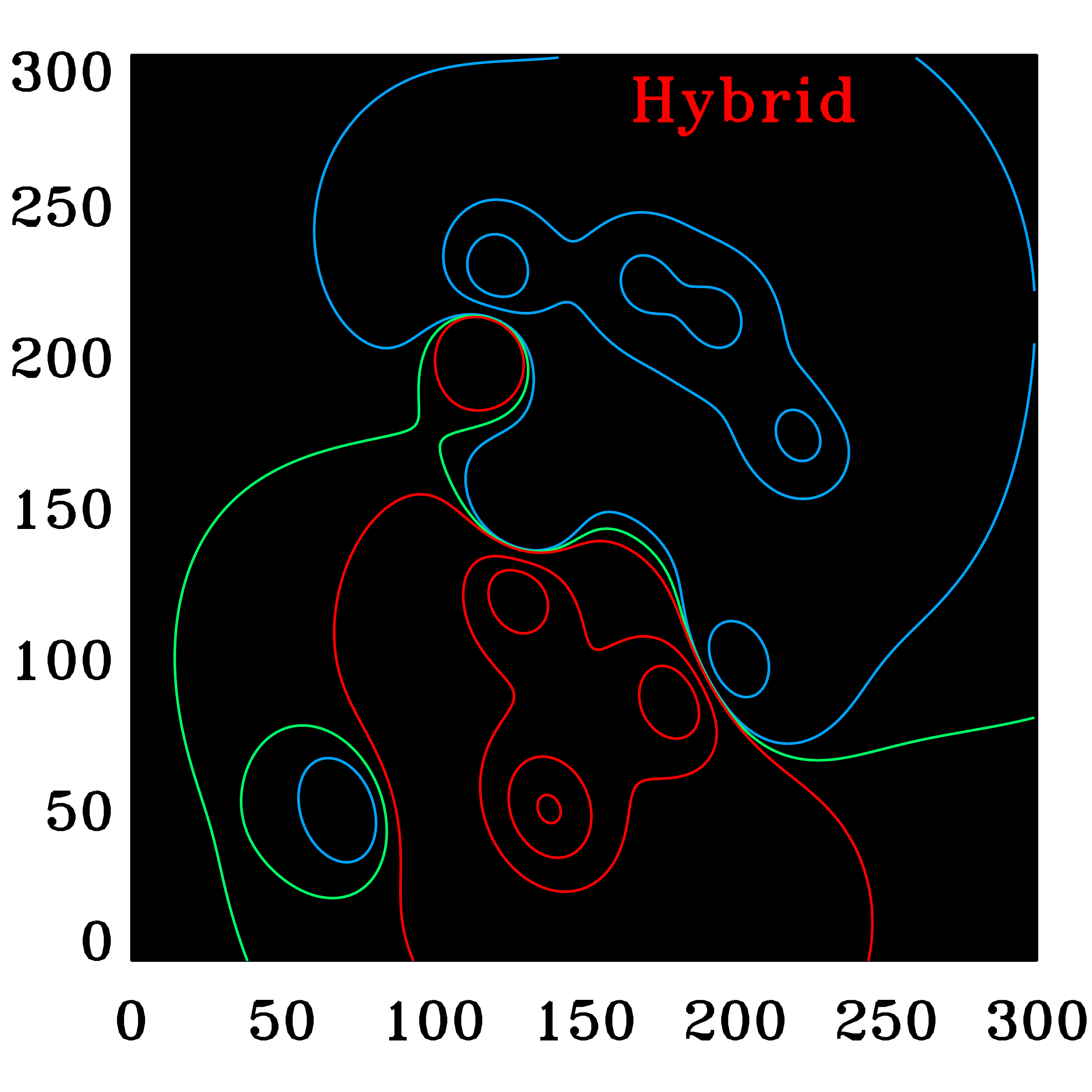} &
\includegraphics[width=0.3\textwidth]{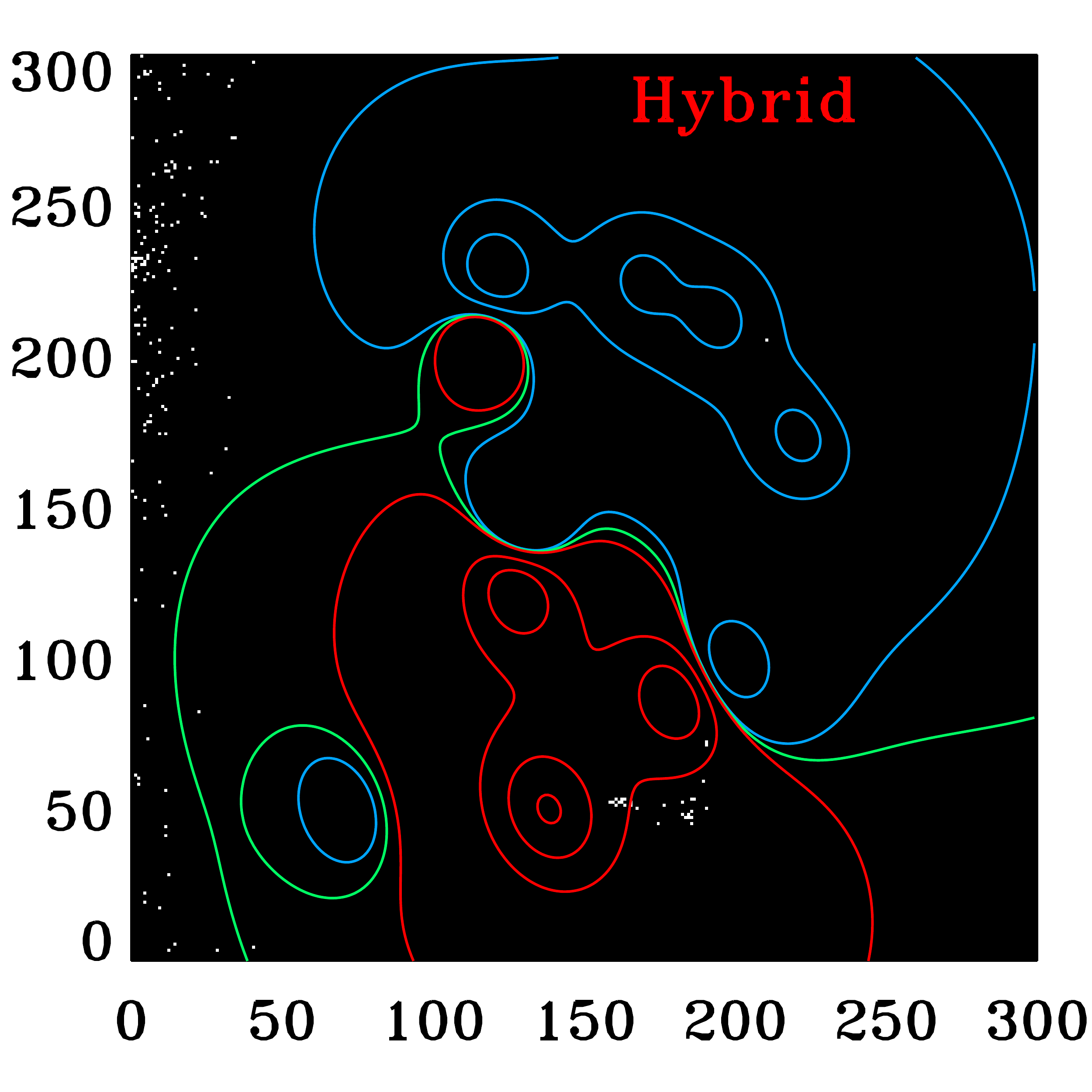} &
\includegraphics[width=0.3\textwidth]{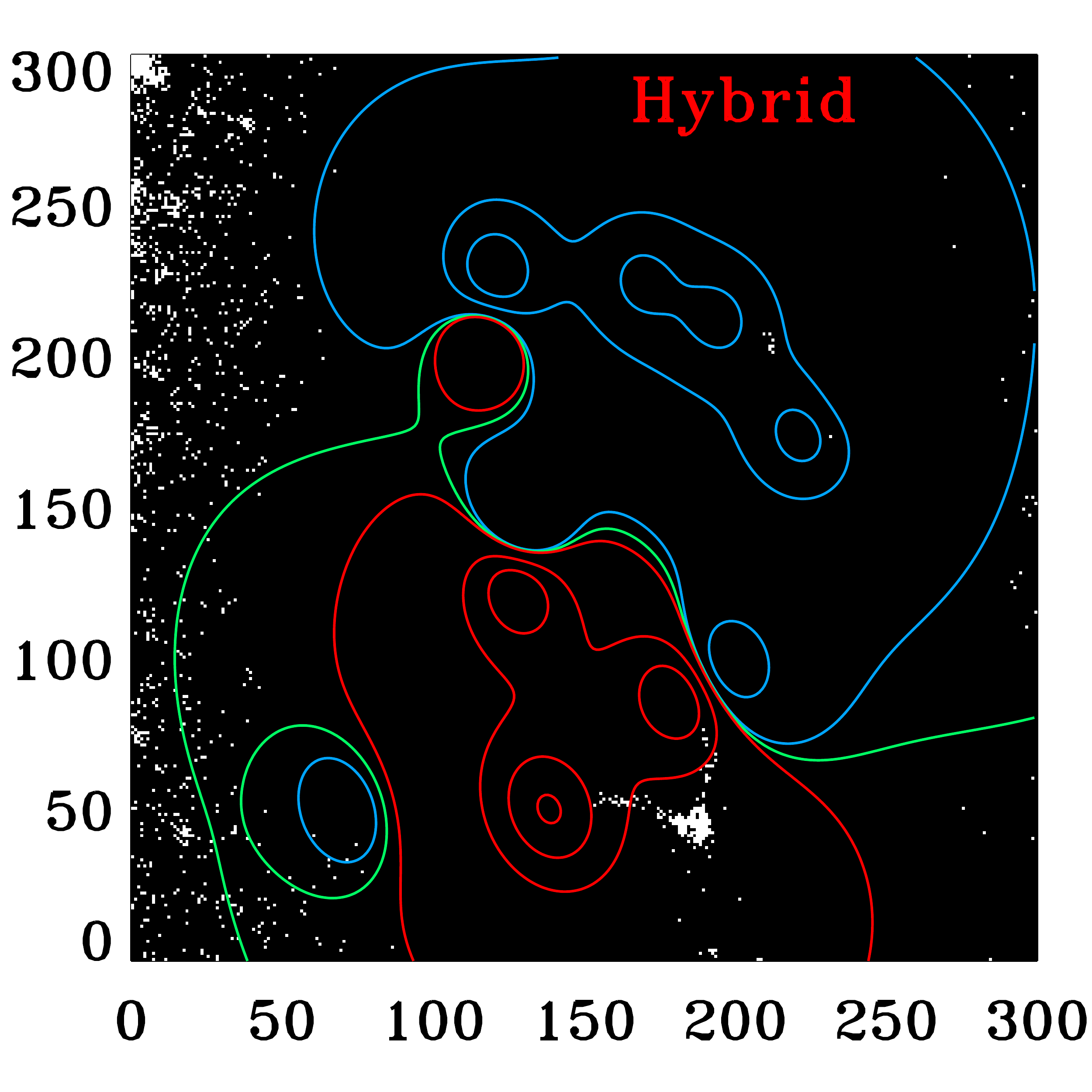}
\end{tabular}
\end{center}
\caption{
Same as Figure~\ref{tpd2} except the hybrid method is employed, as described in Section~\ref{sec_hybrid} (\myie, first \( E_{\rm s} \) (see Equation~(\ref{eds3})) is minimised, then a smoothing algorithm is applied to pixels below a threshold transverse magnetic field strength).
}
\label{tpdh}
\end{figure}

\begin{figure}[ht]
\begin{center}
\begin{tabular}{c@{\hspace{0.005\textwidth}}c@{\hspace{0.005\textwidth}}c}
\includegraphics[width=0.3\textwidth]{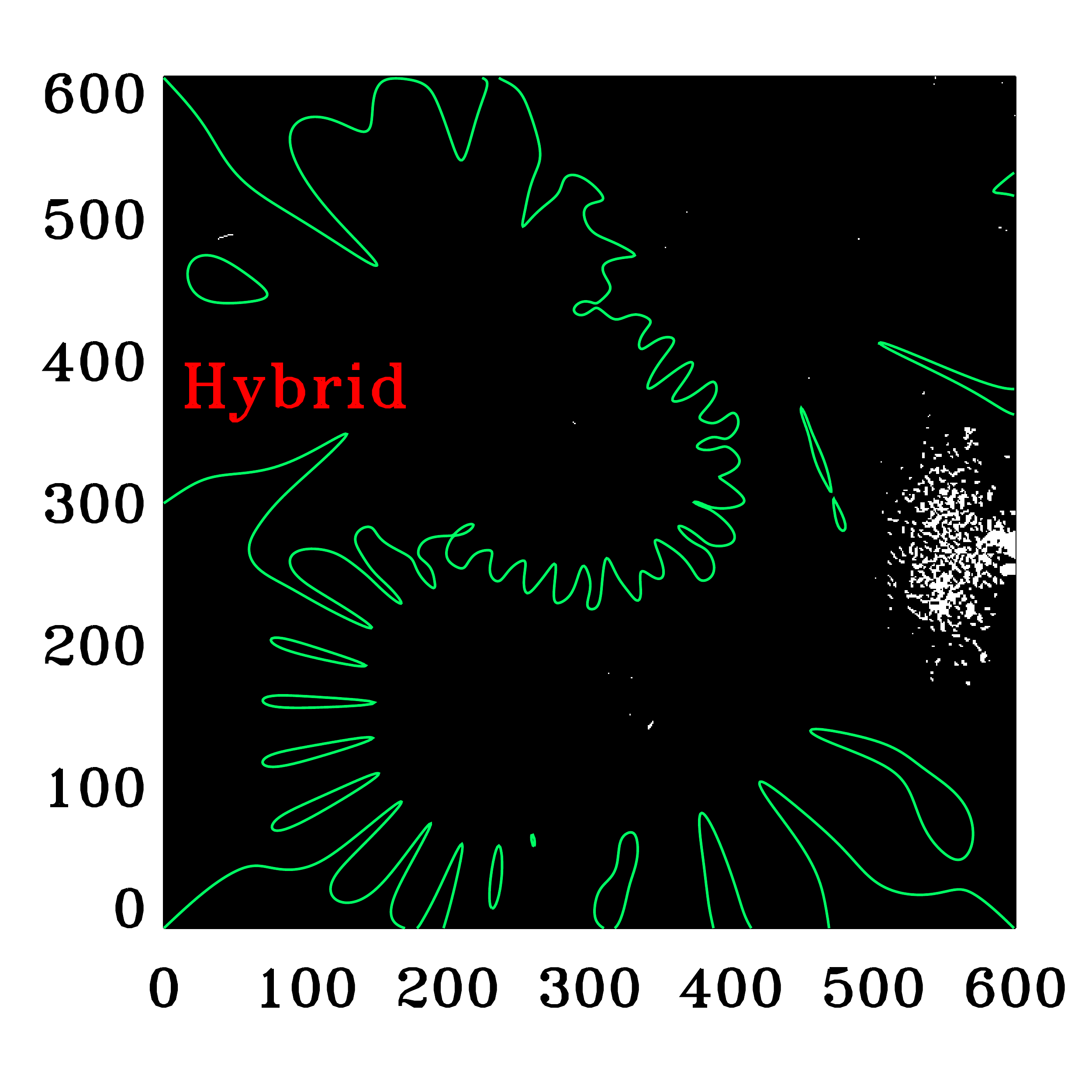} &
\includegraphics[width=0.3\textwidth]{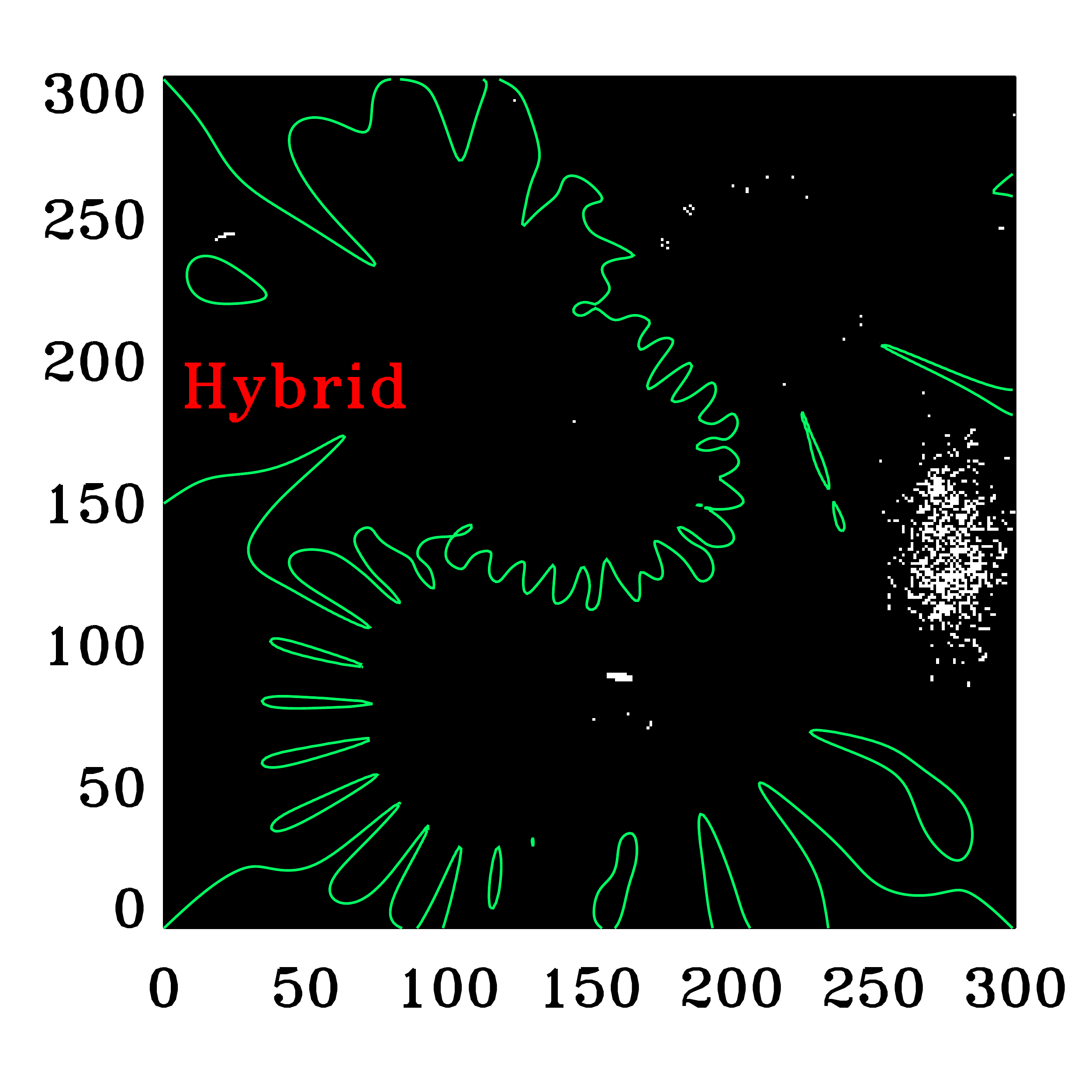} &
\includegraphics[width=0.3\textwidth]{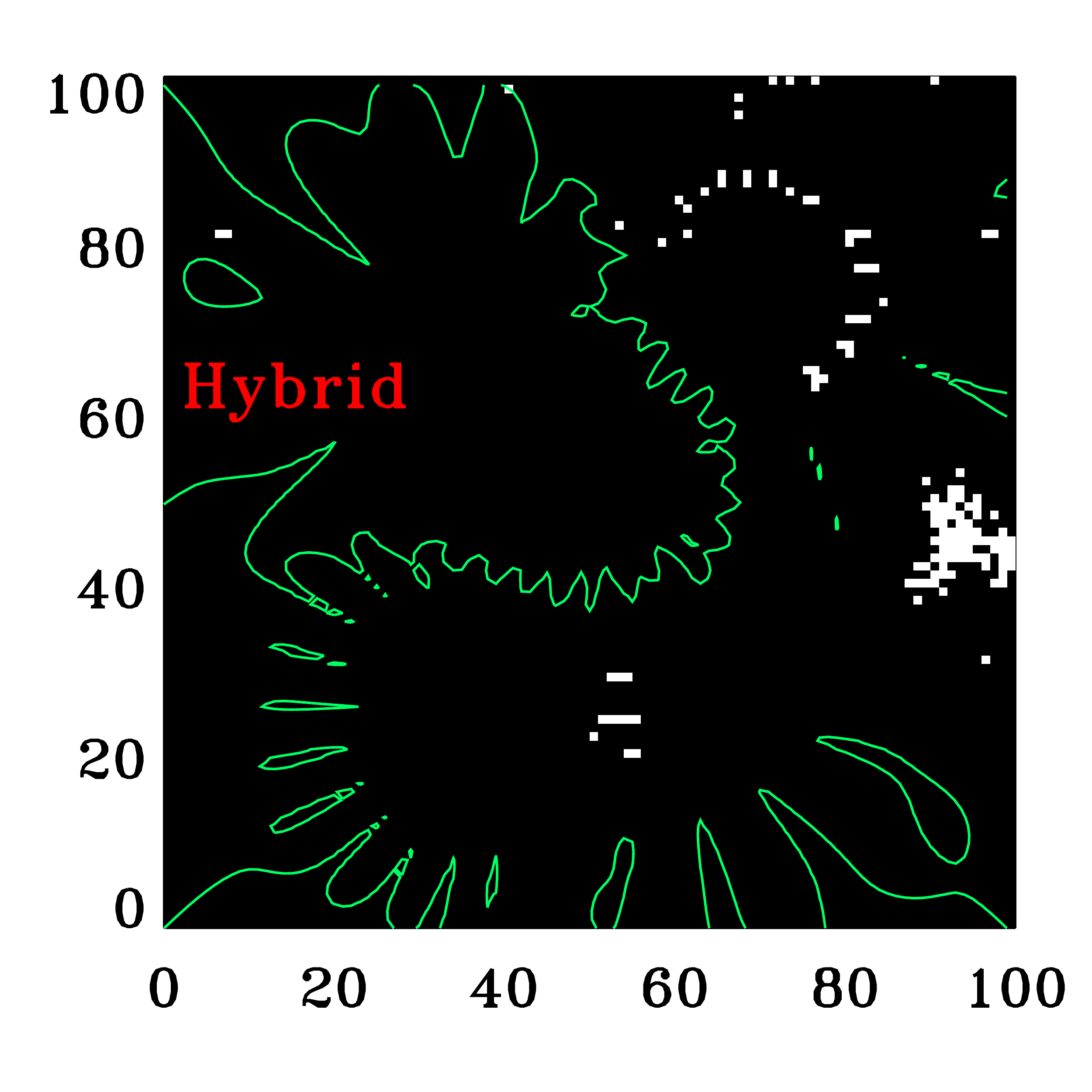}
\end{tabular}
\end{center}
\caption{
Same as Figure~\ref{flowers2} except the hybrid method is employed, as described in Section~\ref{sec_hybrid} (\myie, first \( E_{\rm s} \) (see Equation~(\ref{eds3})) is minimised, then a smoothing algorithm is applied to pixels below a threshold transverse magnetic field strength).
}
\label{flowersh}
\end{figure}

We have experimented with several different definitions for the neighbourhood used by the smoothing algorithm.
Broadly speaking, we find that neighbourhoods involving more pixels perform better than those with fewer pixels, and those that involve pixels from two heights perform better than those that use only pixels at one height.
In Figures~\ref{tpdh} and \ref{flowersh} and Tables~\ref{tpd_tab} and \ref{flowers_tab} we show results for this hybrid approach where the smoothing algorithm uses a neighbourhood of five pixels by five pixels at the same height as the target pixel plus the single pixel directly above (or below) the target pixel.

We find that the ambiguity-resolution results depend on the prescribed threshold.
For the test cases with photon noise, the ambiguity-resolution results generally improve as the threshold increases from zero to a point whereafter the results gradually get worse.
This indicates that in practice the threshold may need to be adjusted according to the level of photon noise present in the data.
For demonstration purposes we use a threshold of 150~G for the low-noise case and 400~G for the high-noise case: These thresholds are chosen because they correspond roughly to the value where the smoothing algorithm produces optimal results in experiments.
In these cases, the smoothing algorithm clearly produces results that are greatly improved in the regions below the prescribed threshold, in comparison to the results produced by minimising \( E_{\rm s} \) (see Figure~\ref{mtabplot}, Figure~\ref{tpdh} and Table~\ref{tpd_tab}).

For the sake of comparison, for the test cases without photon noise we use a threshold of 100~G. This value is somewhat arbitrary since there is no photon noise in these cases.
For cases without photon noise we find that the results produced by the hybrid approach are very similar to those produced by minimising \( E_{\rm s} \).
However, a small number of pixels below threshold do have their azimuthal angles altered by the smoothing algorithm.
This indicates that the smoothing algorithm may not be necessary, and may be counterproductive, for regions where the transverse component of the magnetic is only weakly affected by photon noise.

Figure~\ref{mtabplot} shows that a reasonable hybrid method may also be constructed by combining one of the other global minimisation methods with the smoothing algorithm, although this may require a higher threshold (ideally, the threshold for the smoothing algorithm is chosen such that all pixels above the threshold are correctly resolved).
For example, in other experiments we apply the smoothing algorithm to the results produced by the global minimisation method that minimises \(|\grad \vdot \B|\) only.
In this instance, to obtain optimal solutions for the low-noise test case we find that the threshold needs to be increased to approximately 250~G; the results produced are comparable to those shown in Figure~\ref{tpdh}.
We acknowledge that it may not be desirable in practice to use the smoothing algorithm over large areas in general because it does not use the divergence-free condition.
If this is the case, the results in Figure~\ref{mtabplot} suggest that the most reliable hybrid method is the one that first uses the global minimisation method that minimises \( E_{\rm s} \), as this allows the smoothing algorithm to start from a lower threshold.

\section{Conclusions}
\label{sec_conc}

We investigate how the azimuthal ambiguity, present in solar vector magnetogram data, can be resolved by using the divergence-free property of magnetic fields, along with information regarding the variation of the field in the line-of-sight and horizontal heliographic directions.
In \cbl{} several methods were tested that each make different assumptions about how to use the divergence to resolve the ambiguity, using error-free synthetic data that did not account for noise that is typical of solar observational data.
In this article we use more realistic synthetic data to test how the various methods examined in \cbl{} respond to the effects of two kinds of noise: noise to simulate Poisson photon noise in the observed polarization spectra, and a spatial binning to simulate the effects of limited instrumental spatial resolution in the directions perpendicular to the line-of-sight.
We find the same general trend that was found in \cbl{} in that the global minimisation method is more robust than both the \citeauthor{1990AcApS..10..371W} criterion and the sequential minimisation method \cite{1999A+A...347.1005B}.
However, we find that all methods  based on the divergence-free property examined in \cbl{} can produce undesirable results when photon noise or unresolved structure are present in the data.

We conduct a series of experiments aimed at improving the performance of the global minimisation method and reducing the effects of noise on the disambiguation results.
Based on the results from these experiments we present a two-step, hybrid method that produces reasonable results in tests with synthetic data.
The first step of the hybrid method is similar to the global minimisation method of \cbl{} and involves the minimisation of a combination of the approximation for the divergence and a smoothness constraint that seeks to minimise the difference between the magnetic field in neighbouring pixels.
We test several versions of this modified global minimisation method that have different additional constraints and find that all of the methods tested produce undesirable results in regions where the transverse component of the magnetic field is strongly affected by photon noise.
Consequently, in the second step of the hybrid approach pixels with measurements of the transverse component of the magnetic field below a prescribed threshold are revisited with a smoothing algorithm that also seeks to minimise the difference between the magnetic field in neighbouring pixels.
For synthetic data with photon noise the hybrid approach produces results that are significantly better than those produced by the global minimisation method that minimises only the divergence.
For synthetic data with unresolved structure the hybrid approach produces results that are slightly better than those produced by the global minimisation method that minimises only the divergence.
The hybrid method uses an assumption about the smoothness of the magnetic field that is not explicitly related to the divergence of the field. 
We acknowledge that this assumption may not be appropriate for solar magnetic fields in general.

For disambiguation methods based on the divergence-free condition there are some sources of uncertainty that are not modelled by the synthetic data employed in this investigation.
First, limited spatial resolution in the line-of-sight direction; the synthetic data employed here assume that the locations of the observation heights are known exactly, but this may not be accurate for solar observational data.
Second, the synthetic data is constructed assuming that the magnetic field is measured at constant geometrical height over the field of view, but in practice this may not be a reasonable approximation as the solar atmosphere may vary significantly in the directions perpendicular to the line-of-sight.
Third, the line-of-sight distance between the two observation heights is assumed to be constant, which may not be a reasonable approximation for the similar reasons.
These additional sources of uncertainty may be significant for solar observational data and further investigation is required to quantify how these issues may affect the results produced by the methods tested in this article.

In the interest of further development, testing and implementation, the software for the global minimisation methods and the hybrid method is available at \url{http://www.cora.nwra.com/~ash/ambig2.tar.gz}.

\begin{acks}
The author thanks K.D. Leka and Graham Barnes for providing the synthetic datasets used in this investigation.
The author thanks the referee for helpful suggestions.
This work was supported by funding from NASA under contracts NNH05CC49C/NNH05CC75C and NNH09CE60C.
This work used the Extreme Science and Engineering Discovery Environment (XSEDE), which is supported by National Science Foundation grant number OCI-1053575.
\end{acks}


\begin{thebibliography}{36}
\ifx \bisbn   \undefined \def \bisbn  #1{ISBN #1}\fi
\ifx \binits  \undefined \def \binits#1{#1} \fi
\ifx \bauthor  \undefined \def \bauthor#1{#1} \fi
\ifx \batitle  \undefined \def \batitle#1{#1} \fi
\ifx \bjtitle  \undefined \def \bjtitle#1{\textit{#1}}\fi
\ifx \bvolume  \undefined \def \bvolume#1{\textbf{#1}}\fi
\ifx \byear  \undefined \def \byear#1{#1} \fi
\ifx \bissue  \undefined \def \bissue#1{#1} \fi
\ifx \bfpage  \undefined \def \bfpage#1{#1} \fi
\ifx \blpage  \undefined \def \blpage #1{#1} \fi
\ifx \burl  \undefined \def \burl#1{\textsf{#1}} \fi
\ifx \doiurl  \undefined \def \doiurl#1{\textsf{#1}} \fi
\ifx \betal  \undefined \def \betal{\textit{et al.}} \fi
\ifx \binstitute  \undefined \def \binstitute#1{#1} \fi
\ifx \bctitle  \undefined \def \bctitle#1{#1} \fi
\ifx \beditor  \undefined \def \beditor#1{#1} \fi
\ifx \bpublisher  \undefined \def \bpublisher#1{#1} \fi
\ifx \bbtitle  \undefined \def \bbtitle#1{\textit{#1}} \fi
\ifx \bedition  \undefined \def \bedition#1{#1} \fi
\ifx \bseriesno  \undefined \def \bseriesno#1{\textbf{#1}} \fi
\ifx \blocation  \undefined \def \blocation#1{#1} \fi
\ifx \bsertitle  \undefined \def \bsertitle#1{\textit{#1}} \fi
\ifx \bsnm \undefined \def \bsnm#1{#1} \fi
\ifx \bsuffix \undefined \def \bsuffix#1{#1} \fi
\ifx \bparticle \undefined \def \bparticle#1{#1} \fi
\ifx \barticle \undefined \def \barticle#1{#1} \fi
\ifx \botherref \undefined \def \botherref #1{#1} \fi
\ifx \url \undefined \def \url#1{\textsf{#1}} \fi
\ifx \bchapter \undefined \def \bchapter#1{#1} \fi
\ifx \bbook \undefined \def \bbook#1{#1} \fi
\ifx \bcomment \undefined \def \bcomment#1{#1} \fi
\ifx \oauthor \undefined \def \oauthor#1{#1} \fi
\ifx \citeauthoryear \undefined \def \citeauthoryear#1{#1} \fi
\def \endbibitem {}

\bibitem[\protect\citeauthoryear{{Ahnert} and
  {Abel}}{2007}]{2007CoPhC.177..764A}
\begin{barticle}
\bauthor{\bsnm{{Ahnert}}, \binits{K.}}, \bauthor{\bsnm{{Abel}}, \binits{M.}}:
\byear{2007},
\bjtitle{\cpc}
\bvolume{177},
\bfpage{764}.
doi:\doiurl{10.1016/j.cpc.2007.03.009}.
\end{barticle}
\endbibitem

\bibitem[\protect\citeauthoryear{{Boulmezaoud} and
  {Amari}}{1999}]{1999A+A...347.1005B}
\begin{barticle}
\bauthor{\bsnm{{Boulmezaoud}}, \binits{T.Z.}}, \bauthor{\bsnm{{Amari}},
  \binits{T.}}:
\byear{1999},
\bjtitle{\aap}
\bvolume{347},
\bfpage{1005}.
\end{barticle}
\endbibitem

\bibitem[\protect\citeauthoryear{{Chiu} and
  {Hilton}}{1977}]{1977ApJ...212..873C}
\begin{barticle}
\bauthor{\bsnm{{Chiu}}, \binits{Y.T.}}, \bauthor{\bsnm{{Hilton}},
  \binits{H.H.}}:
\byear{1977},
\bjtitle{\apj}
\bvolume{212},
\bfpage{873}.
doi:\doiurl{10.1086/155111}.
\end{barticle}
\endbibitem

\bibitem[\protect\citeauthoryear{{Collados}
  \textit{et~al.}}{1994}]{1994A+A...291..622C}
\begin{barticle}
\bauthor{\bsnm{{Collados}}, \binits{M.}}, \bauthor{\bsnm{{Martinez Pillet}},
  \binits{V.}}, \bauthor{\bsnm{{Ruiz Cobo}}, \binits{B.}}, \bauthor{\bsnm{{del
  Toro Iniesta}}, \binits{J.C.}}, \bauthor{\bsnm{{Vazquez}}, \binits{M.}}:
\byear{1994},
\bjtitle{\aap}
\bvolume{291},
\bfpage{622}.
\end{barticle}
\endbibitem

\bibitem[\protect\citeauthoryear{{Crouch} and
  {Barnes}}{2008}]{2008SoPh..247...25C}
\begin{barticle}
\bauthor{\bsnm{{Crouch}}, \binits{A.D.}}, \bauthor{\bsnm{{Barnes}},
  \binits{G.}}:
\byear{2008},
\bjtitle{\solphys}
\bvolume{247},
\bfpage{25}.
doi:\doiurl{10.1007/s11207-007-9096-1}.
\end{barticle}
\endbibitem

\bibitem[\protect\citeauthoryear{{Crouch}, {Barnes}, and
  {Leka}}{2009}]{2009SoPh..260..271C}
\begin{barticle}
\bauthor{\bsnm{{Crouch}}, \binits{A.D.}}, \bauthor{\bsnm{{Barnes}},
  \binits{G.}}, \bauthor{\bsnm{{Leka}}, \binits{K.D.}}:
\byear{2009},
\bjtitle{\solphys}
\bvolume{260},
\bfpage{271} (\cbl{}).
doi:\doiurl{10.1007/s11207-009-9454-2}.
\end{barticle}
\endbibitem

\bibitem[\protect\citeauthoryear{{Cullum}}{1971}]{1971SJNA....8..254C}
\begin{barticle}
\bauthor{\bsnm{{Cullum}}, \binits{J.}}:
\byear{1971},
\bjtitle{\sjna}
\bvolume{8},
\bfpage{254}.
doi:\doiurl{10.1137/0708026}.
\end{barticle}
\endbibitem

\bibitem[\protect\citeauthoryear{{Cuperman}, {Li}, and
  {Semel}}{1993}]{1993A+A...278..279C}
\begin{barticle}
\bauthor{\bsnm{{Cuperman}}, \binits{S.}}, \bauthor{\bsnm{{Li}}, \binits{J.}},
  \bauthor{\bsnm{{Semel}}, \binits{M.}}:
\byear{1993},
\bjtitle{\aap}
\bvolume{278},
\bfpage{279}.
\end{barticle}
\endbibitem

\bibitem[\protect\citeauthoryear{{del Toro Iniesta} and {Ruiz
  Cobo}}{1996}]{1996SoPh..164..169D}
\begin{barticle}
\bauthor{\bsnm{{del Toro Iniesta}}, \binits{J.C.}}, \bauthor{\bsnm{{Ruiz
  Cobo}}, \binits{B.}}:
\byear{1996},
\bjtitle{\solphys}
\bvolume{164},
\bfpage{169}.
doi:\doiurl{10.1007/BF00146631}.
\end{barticle}
\endbibitem

\bibitem[\protect\citeauthoryear{{Eibe}
  \textit{et~al.}}{2002}]{2002A+A...381..290E}
\begin{barticle}
\bauthor{\bsnm{{Eibe}}, \binits{M.T.}}, \bauthor{\bsnm{{Aulanier}},
  \binits{G.}}, \bauthor{\bsnm{{Faurobert}}, \binits{M.}},
  \bauthor{\bsnm{{Mein}}, \binits{P.}}, \bauthor{\bsnm{{Malherbe}},
  \binits{J.M.}}:
\byear{2002},
\bjtitle{\aap}
\bvolume{381},
\bfpage{290}.
doi:\doiurl{10.1051/0004-6361:20011495}.
\end{barticle}
\endbibitem

\bibitem[\protect\citeauthoryear{{Gary} and
  {Hagyard}}{1990}]{1990SoPh..126...21G}
\begin{barticle}
\bauthor{\bsnm{{Gary}}, \binits{G.A.}}, \bauthor{\bsnm{{Hagyard}},
  \binits{M.J.}}:
\byear{1990},
\bjtitle{\solphys}
\bvolume{126},
\bfpage{21}.
doi:\doiurl{10.1007/BF00158295}.
\end{barticle}
\endbibitem

\bibitem[\protect\citeauthoryear{{Georgoulis}}{2012}]{2012SoPh..276..423G}
\begin{barticle}
\bauthor{\bsnm{{Georgoulis}}, \binits{M.K.}}:
\byear{2012},
\bjtitle{\solphys}
\bvolume{276},
\bfpage{423}.
doi:\doiurl{10.1007/s11207-011-9819-1}.
\end{barticle}
\endbibitem

\bibitem[\protect\citeauthoryear{{Harvey}}{1969}]{1969PhDT.........3H}
\begin{botherref}
\oauthor{\bsnm{{Harvey}}, \binits{J.W.}}:
1969,
{Magnetic Fields Associated with Solar Active-Region Prominences.}
Ph.D. thesis,
University of Colorado, Boulder.
\end{botherref}
\endbibitem

\bibitem[\protect\citeauthoryear{{Kirkpatrick}, {Gelatt}, and
  {Vecchi}}{1983}]{1983Sci...220..671K}
\begin{barticle}
\bauthor{\bsnm{{Kirkpatrick}}, \binits{S.}}, \bauthor{\bsnm{{Gelatt}},
  \binits{C.D.}}, \bauthor{\bsnm{{Vecchi}}, \binits{M.P.}}:
\byear{1983},
\bjtitle{Science}
\bvolume{220},
\bfpage{671}.
doi:\doiurl{10.1126/science.220.4598.671}.
\end{barticle}
\endbibitem

\bibitem[\protect\citeauthoryear{{Leka} and
  {Barnes}}{2012}]{2012SoPh..277...89L}
\begin{barticle}
\bauthor{\bsnm{{Leka}}, \binits{K.D.}}, \bauthor{\bsnm{{Barnes}}, \binits{G.}}:
\byear{2012},
\bjtitle{\solphys}
\bvolume{277},
\bfpage{89}.
doi:\doiurl{10.1007/s11207-011-9821-7}.
\end{barticle}
\endbibitem

\bibitem[\protect\citeauthoryear{{Leka} and
  {Metcalf}}{2003}]{2003SoPh..212..361L}
\begin{barticle}
\bauthor{\bsnm{{Leka}}, \binits{K.D.}}, \bauthor{\bsnm{{Metcalf}},
  \binits{T.R.}}:
\byear{2003},
\bjtitle{\solphys}
\bvolume{212},
\bfpage{361}.
doi:\doiurl{10.1023/A:1022996404064}.
\end{barticle}
\endbibitem

\bibitem[\protect\citeauthoryear{{Leka}, {Barnes}, and
  {Crouch}}{2009}]{2009ASPC..415..365L}
\begin{botherref}
\oauthor{\bsnm{{Leka}}, \binits{K.D.}}, \oauthor{\bsnm{{Barnes}}, \binits{G.}},
  \oauthor{\bsnm{{Crouch}}, \binits{A.}}:
2009,
In: {Lites, B., Cheung, M., Magara, T., Mariska, J., Reeves, K.} (eds.)
\textit{The Second Hinode Science Meeting: Beyond Discovery-Toward
  Understanding},
\textit{ASP Conf. Ser.}
\textbf{415},
365.
\end{botherref}
\endbibitem

\bibitem[\protect\citeauthoryear{{Leka}
  \textit{et~al.}}{2009}]{2009SoPh..260...83L}
\begin{barticle}
\bauthor{\bsnm{{Leka}}, \binits{K.D.}}, \bauthor{\bsnm{{Barnes}}, \binits{G.}},
  \bauthor{\bsnm{{Crouch}}, \binits{A.D.}}, \bauthor{\bsnm{{Metcalf}},
  \binits{T.R.}}, \bauthor{\bsnm{{Gary}}, \binits{G.A.}},
  \bauthor{\bsnm{{Jing}}, \binits{J.}}, \bauthor{\bsnm{{Liu}}, \binits{Y.}}:
\byear{2009},
\bjtitle{\solphys}
\bvolume{260},
\bfpage{83}.
doi:\doiurl{10.1007/s11207-009-9440-8}.
\end{barticle}
\endbibitem

\bibitem[\protect\citeauthoryear{{Leka}
  \textit{et~al.}}{2012}]{2012SoPh..276..441L}
\begin{barticle}
\bauthor{\bsnm{{Leka}}, \binits{K.D.}}, \bauthor{\bsnm{{Barnes}}, \binits{G.}},
  \bauthor{\bsnm{{Gary}}, \binits{G.A.}}, \bauthor{\bsnm{{Crouch}},
  \binits{A.D.}}, \bauthor{\bsnm{{Liu}}, \binits{Y.}}:
\byear{2012},
\bjtitle{\solphys}
\bvolume{276},
\bfpage{441}.
doi:\doiurl{10.1007/s11207-011-9879-2}.
\end{barticle}
\endbibitem

\bibitem[\protect\citeauthoryear{{Li}, {Amari}, and
  {Fan}}{2007}]{2007ApJ...654..675L}
\begin{barticle}
\bauthor{\bsnm{{Li}}, \binits{J.}}, \bauthor{\bsnm{{Amari}}, \binits{T.}},
  \bauthor{\bsnm{{Fan}}, \binits{Y.}}:
\byear{2007},
\bjtitle{\apj}
\bvolume{654},
\bfpage{675}.
doi:\doiurl{10.1086/509062}.
\end{barticle}
\endbibitem

\bibitem[\protect\citeauthoryear{{Li}, {Cuperman}, and
  {Semel}}{1993}]{1993A+A...279..214L}
\begin{barticle}
\bauthor{\bsnm{{Li}}, \binits{J.}}, \bauthor{\bsnm{{Cuperman}}, \binits{S.}},
  \bauthor{\bsnm{{Semel}}, \binits{M.}}:
\byear{1993},
\bjtitle{\aap}
\bvolume{279},
\bfpage{214}.
\end{barticle}
\endbibitem

\bibitem[\protect\citeauthoryear{{Liu}
  \textit{et~al.}}{1996}]{1996SoPh..169...79L}
\begin{barticle}
\bauthor{\bsnm{{Liu}}, \binits{Y.}}, \bauthor{\bsnm{{Wang}}, \binits{J.}},
  \bauthor{\bsnm{{Yan}}, \binits{Y.}}, \bauthor{\bsnm{{Ai}}, \binits{G.}}:
\byear{1996},
\bjtitle{\solphys}
\bvolume{169},
\bfpage{79}.
doi:\doiurl{10.1007/BF00153835}.
\end{barticle}
\endbibitem

\bibitem[\protect\citeauthoryear{{Maltby}
  \textit{et~al.}}{1986}]{1986ApJ...306..284M}
\begin{barticle}
\bauthor{\bsnm{{Maltby}}, \binits{P.}}, \bauthor{\bsnm{{Avrett}},
  \binits{E.H.}}, \bauthor{\bsnm{{Carlsson}}, \binits{M.}},
  \bauthor{\bsnm{{Kjeldseth-Moe}}, \binits{O.}}, \bauthor{\bsnm{{Kurucz}},
  \binits{R.L.}}, \bauthor{\bsnm{{Loeser}}, \binits{R.}}:
\byear{1986},
\bjtitle{\apj}
\bvolume{306},
\bfpage{284}.
doi:\doiurl{10.1086/164342}.
\end{barticle}
\endbibitem

\bibitem[\protect\citeauthoryear{{Metcalf}}{1994}]{1994SoPh..155..235M}
\begin{barticle}
\bauthor{\bsnm{{Metcalf}}, \binits{T.R.}}:
\byear{1994},
\bjtitle{\solphys}
\bvolume{155},
\bfpage{235}.
doi:\doiurl{10.1007/BF00680593}.
\end{barticle}
\endbibitem

\bibitem[\protect\citeauthoryear{{Metcalf}
  \textit{et~al.}}{1995}]{1995ApJ...439..474M}
\begin{barticle}
\bauthor{\bsnm{{Metcalf}}, \binits{T.R.}}, \bauthor{\bsnm{{Jiao}},
  \binits{L.}}, \bauthor{\bsnm{{McClymont}}, \binits{A.N.}},
  \bauthor{\bsnm{{Canfield}}, \binits{R.C.}}, \bauthor{\bsnm{{Uitenbroek}},
  \binits{H.}}:
\byear{1995},
\bjtitle{\apj}
\bvolume{439},
\bfpage{474}.
doi:\doiurl{10.1086/175188}.
\end{barticle}
\endbibitem

\bibitem[\protect\citeauthoryear{{Metcalf}
  \textit{et~al.}}{2006}]{2006SoPh..237..267M}
\begin{barticle}
\bauthor{\bsnm{{Metcalf}}, \binits{T.R.}}, \bauthor{\bsnm{{Leka}},
  \binits{K.D.}}, \bauthor{\bsnm{{Barnes}}, \binits{G.}},
  \bauthor{\bsnm{{Lites}}, \binits{B.W.}}, \bauthor{\bsnm{{Georgoulis}},
  \binits{M.K.}}, \bauthor{\bsnm{{Pevtsov}}, \binits{A.A.}},
  \bauthor{\bsnm{{Balasubramaniam}}, \binits{K.S.}}, \bauthor{\bsnm{{Gary}},
  \binits{G.A.}}, \bauthor{\bsnm{{Jing}}, \binits{J.}}, \bauthor{\bsnm{{Li}},
  \binits{J.}}, \bauthor{\bsnm{{Liu}}, \binits{Y.}}, \bauthor{\bsnm{{Wang}},
  \binits{H.N.}}, \bauthor{\bsnm{{Abramenko}}, \binits{V.}},
  \bauthor{\bsnm{{Yurchyshyn}}, \binits{V.}}, \bauthor{\bsnm{{Moon}},
  \binits{Y.J.}}:
\byear{2006},
\bjtitle{\solphys}
\bvolume{237},
\bfpage{267}.
doi:\doiurl{10.1007/s11207-006-0170-x}.
\end{barticle}
\endbibitem

\bibitem[\protect\citeauthoryear{{Metropolis}
  \textit{et~al.}}{1953}]{1953JChPh..21.1087M}
\begin{barticle}
\bauthor{\bsnm{{Metropolis}}, \binits{N.}}, \bauthor{\bsnm{{Rosenbluth}},
  \binits{A.W.}}, \bauthor{\bsnm{{Rosenbluth}}, \binits{M.N.}},
  \bauthor{\bsnm{{Teller}}, \binits{A.H.}}, \bauthor{\bsnm{{Teller}},
  \binits{E.}}:
\byear{1953},
\bjtitle{\jcp}
\bvolume{21},
\bfpage{1087}.
doi:\doiurl{10.1063/1.1699114}.
\end{barticle}
\endbibitem

\bibitem[\protect\citeauthoryear{{Press}
  \textit{et~al.}}{1992}]{1992nrfa.book.....P}
\begin{bbook}
\bauthor{\bsnm{{Press}}, \binits{W.H.}}, \bauthor{\bsnm{{Teukolsky}},
  \binits{S.A.}}, \bauthor{\bsnm{{Vetterling}}, \binits{W.T.}},
  \bauthor{\bsnm{{Flannery}}, \binits{B.P.}}:
\byear{1992},
\bbtitle{{Numerical Recipes in FORTRAN. The Art of Scientific Computing}},
\bpublisher{Cambridge University Press}, \blocation{Cambridge},
\bfpage{436}.
\end{bbook}
\endbibitem

\bibitem[\protect\citeauthoryear{{Ruiz Cobo} and {del Toro
  Iniesta}}{1992}]{1992ApJ...398..375R}
\begin{barticle}
\bauthor{\bsnm{{Ruiz Cobo}}, \binits{B.}}, \bauthor{\bsnm{{del Toro Iniesta}},
  \binits{J.C.}}:
\byear{1992},
\bjtitle{\apj}
\bvolume{398},
\bfpage{375}.
doi:\doiurl{10.1086/171862}.
\end{barticle}
\endbibitem

\bibitem[\protect\citeauthoryear{{Socas-Navarro}}{2005}]{2005ApJ...631L.167S}
\begin{barticle}
\bauthor{\bsnm{{Socas-Navarro}}, \binits{H.}}:
\byear{2005},
\bjtitle{\apjl}
\bvolume{631},
\bfpage{L167}.
doi:\doiurl{10.1086/497334}.
\end{barticle}
\endbibitem

\bibitem[\protect\citeauthoryear{{Socas-Navarro}}{2007}]{2007ApJS..169..439S}
\begin{barticle}
\bauthor{\bsnm{{Socas-Navarro}}, \binits{H.}}:
\byear{2007},
\bjtitle{\apjs}
\bvolume{169},
\bfpage{439}.
doi:\doiurl{10.1086/510336}.
\end{barticle}
\endbibitem

\bibitem[\protect\citeauthoryear{{Socas-Navarro}, {Trujillo Bueno}, and {Ruiz
  Cobo}}{2000}]{2000ApJ...530..977S}
\begin{barticle}
\bauthor{\bsnm{{Socas-Navarro}}, \binits{H.}}, \bauthor{\bsnm{{Trujillo
  Bueno}}, \binits{J.}}, \bauthor{\bsnm{{Ruiz Cobo}}, \binits{B.}}:
\byear{2000},
\bjtitle{\apj}
\bvolume{530},
\bfpage{977}.
doi:\doiurl{10.1086/308414}.
\end{barticle}
\endbibitem

\bibitem[\protect\citeauthoryear{{Vernazza}, {Avrett}, and
  {Loeser}}{1981}]{1981ApJS...45..635V}
\begin{barticle}
\bauthor{\bsnm{{Vernazza}}, \binits{J.E.}}, \bauthor{\bsnm{{Avrett}},
  \binits{E.H.}}, \bauthor{\bsnm{{Loeser}}, \binits{R.}}:
\byear{1981},
\bjtitle{\apjs}
\bvolume{45},
\bfpage{635}.
doi:\doiurl{10.1086/190731}.
\end{barticle}
\endbibitem

\bibitem[\protect\citeauthoryear{{Westendorp Plaza}
  \textit{et~al.}}{1998}]{1998ApJ...494..453W}
\begin{barticle}
\bauthor{\bsnm{{Westendorp Plaza}}, \binits{C.}}, \bauthor{\bsnm{{del Toro
  Iniesta}}, \binits{J.C.}}, \bauthor{\bsnm{{Ruiz Cobo}}, \binits{B.}},
  \bauthor{\bsnm{{Martinez Pillet}}, \binits{V.}}, \bauthor{\bsnm{{Lites}},
  \binits{B.W.}}, \bauthor{\bsnm{{Skumanich}}, \binits{A.}}:
\byear{1998},
\bjtitle{\apj}
\bvolume{494},
\bfpage{453}.
doi:\doiurl{10.1086/305192}.
\end{barticle}
\endbibitem

\bibitem[\protect\citeauthoryear{{Westendorp Plaza}
  \textit{et~al.}}{2001}]{2001ApJ...547.1130W}
\begin{barticle}
\bauthor{\bsnm{{Westendorp Plaza}}, \binits{C.}}, \bauthor{\bsnm{{del Toro
  Iniesta}}, \binits{J.C.}}, \bauthor{\bsnm{{Ruiz Cobo}}, \binits{B.}},
  \bauthor{\bsnm{{Mart{\'{\i}}nez Pillet}}, \binits{V.}},
  \bauthor{\bsnm{{Lites}}, \binits{B.W.}}, \bauthor{\bsnm{{Skumanich}},
  \binits{A.}}:
\byear{2001},
\bjtitle{\apj}
\bvolume{547},
\bfpage{1130}.
doi:\doiurl{10.1086/318376}.
\end{barticle}
\endbibitem

\bibitem[\protect\citeauthoryear{{Wu} and {Ai}}{1990}]{1990AcApS..10..371W}
\begin{barticle}
\bauthor{\bsnm{{Wu}}, \binits{L.X.}}, \bauthor{\bsnm{{Ai}}, \binits{G.X.}}:
\byear{1990},
\bjtitle{Acta Astrophys. Sinica}
\bvolume{10},
\bfpage{371}.
\end{barticle}
\endbibitem

\end{thebibliography}

\end{article}
\end{document}